\documentclass[prb,showpacs, twocolumn, aps,superscriptaddress,floatfix]{revtex4-1}
\usepackage{amsmath}
\usepackage{amssymb}
\usepackage{bm}
\usepackage{graphicx}
\usepackage{color}
\usepackage[svgnames]{xcolor}
\usepackage{soul}
\usepackage{hyperref}
\usepackage{verbatim}% Long comments

\begin{document}

\title {Ordered states of undoped AB bilayer graphene: bias induced cascade of transitions}

\author{A.V. Rozhkov}

\author{A.O. Sboychakov}

\author{A.L. Rakhmanov}

\affiliation{Institute for Theoretical and Applied Electrodynamics, Russian
Academy of Sciences, 125412 Moscow, Russia}

\begin{abstract}
Using mean-field theory, we determine the electronic phase diagram of
undoped AB-stacked bilayer graphene in the presence of a transverse
electric field. In addition to multiple competing electronic instabilities
characterized by excitonic order parameters, our framework incorporates the
long-range Coulomb energy associated with interlayer polarization. This
long-range interaction plays a crucial role, as it significantly influences
both the structure and the relative energies of the competing ordered
states. We derive a set of self-consistency equations and solve them both
numerically and analytically. Our findings reveal that, as the bias field is
varied, the bilayer undergoes a cascade of first-order transitions between
several ordered insulating phases for which order-parameter structures are
explicitly identified. Some of these phases are characterized by two
inequivalent single-particle gaps, whose magnitudes depend on the valley
and spin quantum numbers. Field-driven transitions are accompanied by
discontinuous and non-monotonic variations of the single-electron gap. We
relate our results to Hartree-Fock numerical calculations and to
experimental research, including observations of fractional metallic phases
that emerge upon doping the bilayer system.
\end{abstract}

\date{\today}
\maketitle

\section{Introduction}

Ordered phases of electronic liquid in bilayer and multilayer graphene
structures are
studied~\cite{bilayer_review2016, Kotov2012RevModPhys}
for more than a decade, both
experimentally~\cite{Bao2012, Martin2010, Weitz2010, Mayorov2011,
Freitag2012, Freitag20122053, veligura2012, Velasco2012, freitag2013,
trilayer_quarter2021exper, ab_frac2022exper,
Seiler2022, zhou2022isospin}
and
theoretically~\cite{MCCANN2007110, Nandkishore2010, Nandkishore2010b,
vafek_rg2010, vafek_nemat_rg2010, Lemonik2010, Jung2011,
cvetkovic_multi2012, aa_graph_prl2012, haritonov_afm2012,
bilayer_half-metal2013numeric_MF,
baima2018dft_half_met_graphene,
aa_graph_BreyFertig2013, sboychakov2013AA, aa_quarter_met2021}.
Recent renewed interest in this subject is motivated by experimental
observations of fractional metallic phases in doped multilayer
samples~\cite{trilayer_quarter2021exper, Seiler2022},
for which fractional metallicity had been theoretically predicted earlier,
including
half-metal~\cite{bilayer_half-metal2013numeric_MF,
baima2018dft_half_met_graphene}
and quarter-metal
states~\cite{aa_quarter_met2021}.

It is reasonable to expect that the fractional metallicity of doped phases
may be linked to the properties of their undoped parent states.
Accordingly, gaining insight into the physics of the undoped phases may be
regarded as an important priority in this context. Motivated by these
considerations, we formulate a simple, analytically tractable mean field
scheme that describes competition of various ordered states in an undoped
sample of AB-stacked bilayer graphene (AB-BLG) placed in transverse
electric field.

Our work builds on a substantial body of prior theoretical
research, such as
Refs.~\onlinecite{MCCANN2007110, Nandkishore2010, Jung2011,
bilayer_half-metal2013numeric_MF}.
Extending these studies, we aim for a framework that balances
methodological clarity with comprehensive integration of the key elements
governing electronic order in AB-BLG. Beside fairly common two-band
Hamiltonian in the presence of transverse electric field, augmented by
exchange electron-electron interaction, we incorporate charging (Hartree)
energy associated with inter-layer polarization. Also, four distinct order
parameters (one per valley, per spin projection) are explicitly accounted
for in our calculations. This order parameters quadruplet enables us to
distinguish and keep track of several, rather dissimilar, ordered states of
undoped AB-BLG.

The model is studied within the variational formulation of the mean-field
approximation. The derived self-consistency equations are solved
analytically and numerically, allowing the phase diagram to be constructed.
It reveals a bias-driven cascade of first-order transitions among ordered
insulating phases, accompanied by a non-monotonic behavior of the
single-particle gap. Another notable result of this study is the
identification of several universal features of the phase diagram.

The four order parameters respond differently to the bias: at finite field,
their absolute values may be non-identical, signifying that the gaps in
different fermion sectors are non-identical. This spectral feature
has important consequences for the formation of the fractional-metal states
at finite doping.

Our analytical results are in good agreement with numerical studies, such
as
Refs.~\onlinecite{Jung2011, bilayer_half-metal2013numeric_MF}.
In particular, we show that data obtained from computationally intensive
Hartree-Fock calculations can be effectively reproduced using relatively
straightforward analytical methods. Encouragingly, previously unknown
universal features, which we identified in our study, are clearly visible
in the Hartree-Fock phase diagram.

We compare our results with experiment of
Ref.~\onlinecite{Geisenhof2022}.
It appears that the data partially consistent with the calculations
presented below. The relation between the model and experiment is
discussed.

The paper is organized as follows. In
Sec.~\ref{sec::model}
we explain our model's structure.
In. Sec.~\ref{var_MF}
the mean field theory is formulated, and the self-consistency equations are
derived. Thermodynamic phases compatible with the self-consistency
equations are discussed in
Sec.~\ref{MF_phases}.
The phase diagram is mapped in
Sec.~\ref{phase_diagram}.
Discussion of the results is in
Sec.~\ref{sec::discussion}.
Conclusions are in
Sec.~\ref{conclusion_section}.
Several calculations are relegated to Appendices.

\section{Model}
%%%%%%%%%%%%%%%%%%%%%%%%%%%%%%%%%%%%%%%%%%%%%%%%%%
\label{sec::model}
%%%%%%%%%%%%%%%%%%%%%%%%%%%%%%%%%%%%%%%%%%%%%%%%%%

\subsection{Tight-binding model}
%%%%%%%%%%%%%%%%%%%%%%%%%%%%%%%%%%%%%%%%%%%%%%%%%%
\label{subsec::TB_model}
%%%%%%%%%%%%%%%%%%%%%%%%%%%%%%%%%%%%%%%%%%%%%%%%%%

The unit cell of the Bernal-type or AB bilayer graphene (AB-BLG) includes
four carbon atoms on two sublattices. The carbon atoms in the sublattice B
of the top layer are located right above the atoms of the sublattice A of
the lower layer (such atoms are called dimers). The atoms in the sublattice
A of the top layer (sublattice B of the lower layer) are located above
(below) centers of the hexagons formed by the atoms of the lower layer (top
layer). These are nondimer sites. The elementary translation vectors for
the AB-BLG can be chosen as
$\mathbf{a}_{1,2}=a(\sqrt{3},\mp1)/2$,
where
$a=2.46$\AA\,
is the elementary unit length. The inter-layer distance for AB-BLG is
$d=3.35$\,\AA.

Below we consider an undoped gated sample of AB-BLG. The model Hamiltonian
can be presented as a sum of two terms
\begin{eqnarray}
H=H_0+H_{\text{int}},
\end{eqnarray}
where the first term is the single-particle Hamiltonian, while the second
term represents the Coulomb electron-electron interaction. We start with a
short discussion of the single-electron term
\begin{eqnarray}
%%%%%%%%%%%%%%%%%%%%%%%%%%%%%%%%%%%%%%%%%%%%%%%%%%
\label{Hkin}
%%%%%%%%%%%%%%%%%%%%%%%%%%%%%%%%%%%%%%%%%%%%%%%%%%
H_0
=
\sum_{\mathbf{k}\sigma}
	\psi^{\dag}_{\mathbf{k}\sigma}
	{\cal H}_{\mathbf{k}}
	\psi^{\phantom{\dag}}_{\mathbf{k}\sigma},
\end{eqnarray}
where the four-component operator-valued spinor
$\psi^{\dag}_{\mathbf{k}\sigma}$
is defined as
\begin{eqnarray}
\psi^{\dag}_{\mathbf{k}\sigma}
=
(d^{\dag}_{\mathbf{k}1A\sigma},\,d^{\dag}_{\mathbf{k}1B\sigma},\,
d^{\dag}_{\mathbf{k}2A\sigma},\,d^{\dag}_{\mathbf{k}2B\sigma}),
\end{eqnarray}
$d^{\dag}_{\mathbf{k}ia\sigma}$
and
$d^{\phantom{\dag}}_{\mathbf{k}ia\sigma}$
are the creation and annihilation operators of the electrons with momentum
$\mathbf{k}$ in the layer
$i$(=$1,\,2$),
in the sublattice
$a$(=$A\,,B$)
with spin projection
$\sigma(=\uparrow,\downarrow)$,
and the 4$\times$4 matrix
${\cal H}_{\mathbf{k}}$
equals
\begin{equation}
%%%%%%%%%%%%%%%%%%%%%%%%%%%%%%%%%%%%%%%%%%%%%%%%%%
\label{H0}
%%%%%%%%%%%%%%%%%%%%%%%%%%%%%%%%%%%%%%%%%%%%%%%%%%
{\cal H}_{\mathbf{k}}
=
\left(\begin{array}{cccc}
	{e\Phi}/{2}&-tf_{\mathbf{k}}&0&t_0\\
	-tf_{\mathbf{k}}^{*}&{e\Phi}/{2}&0&0\\
	0&0&-{e\Phi}/{2}&-tf_{\mathbf{k}}\\
	t_0&0&-tf_{\mathbf{k}}^{*}&-{e\Phi}/{2}
\end{array}\right).
\end{equation}
Here
$e>0$
is the elementary charge, $\Phi$ is the bias voltage, and the
function
$f_{\mathbf{k}}$
is
\begin{equation}
f_{\mathbf{k}}
=
e^{i\mathbf{k}\bm{\delta}}
\left[1+e^{-i\mathbf{ka}_1}+e^{-i\mathbf{ka}_2}\right],
\end{equation}
where
$\bm{\delta}=(\mathbf{a}_1+\mathbf{a}_2)/3$,
parameter
$t=2.7$\,eV
is the in-plane nearest-neighbor hopping amplitude, and
$t_0=0.4$\,eV
is the out-of-plane hopping amplitude between nearest-neighbor dimer
sites~\cite{bilayer_review2016}.
Here the hoping terms connecting non-dimer sites are omitted, therefore,
the effects of the trigonal warping are neglected.

The spectrum of the single-particle
Hamiltonian~\eqref{H0}
consists of four bands
\begin{equation}\label{4_bands}
\!\!\!\!\varepsilon^{(S)}_{\bf k}
=
\mp\sqrt{
	t_\mathbf{k}^2+\frac{e^2\Phi^2}{4}+\frac{t_0^2}{2}
	\mp
	\sqrt{t_\mathbf{k}^2\left(e^2\Phi^2 +t_0^2 \right)
		+
		\frac{t_0^4}{4}
	}
},
\end{equation}
two hole bands
$\varepsilon^{(1,2)}_{\bf k}$,
and two electron bands
$\varepsilon^{(3,4)}_{\bf k}$.
Here
$S=1,...,4$
and
$t_\mathbf{k}=t|f_\mathbf{k}|$.
When
$e\Phi =0$,
the bands
$\varepsilon^{(2)}_{\bf k}$
and
$\varepsilon^{(3)}_{\bf k}$
touch each other at the Dirac points
$\mathbf{K}_1=(0, 4\pi/3a)$
and
$\mathbf{K}_{-1}=-\mathbf{K}_1$.
At non-zero
$e\Phi$
a single-particle gap opens, and the undoped AB-BLG becomes an insulator.

The bispinor wave functions
\begin{equation}\label{bispinor}
\Psi^{(S)}_\mathbf{k}
=
\left(\Psi^{(S)}_{\mathbf{k}1A},\Psi^{(S)}_{\mathbf{k}1B},
\Psi^{(S)}_{\mathbf{k}2A},\Psi^{(S)}_{\mathbf{k}2B}\right)
\end{equation}
correspond to the eigenvalues
$\varepsilon^{(S)}_{\bf k}$.
The analytical formulas for
$\Psi^{(S)}_\mathbf{k}$
are quite cumbersome, and it is often more practical to evaluate them
numerically.

At a fixed
$e \Phi$,
it is convenient to introduce a new operator basis according to
\begin{equation}
%%%%%%%%%%%%%%%%%%%%%%%%%%%%%%%%%%%%%%%%%%%%%%%%%%
\label{gamma_spinors}
%%%%%%%%%%%%%%%%%%%%%%%%%%%%%%%%%%%%%%%%%%%%%%%%%%
d_{\mathbf{k}ia\sigma}^{\vphantom{\dag}}
=
\sum_S
	\Psi_{\mathbf{k}ia}^{(S)}
	\tilde \gamma^{\vphantom{\dag}}_{\mathbf{k}S\sigma},
\end{equation}
where operator
$\tilde \gamma^{\vphantom{\dag}}_{\mathbf{k}S\sigma}$
(operator
$\tilde \gamma^\dag_{\mathbf{k}S\sigma}$)
destroys (creates) an electron in the eigenstate $S$ with momentum
$\mathbf{k}$ and spin $\sigma$. The single-particle Hamiltonian
\begin{equation}
%%%%%%%%%%%%%%%%%%%%%%%%%%%%%%%%%%%%%%%%%%%%%%%%%%
\label{single_particle_H}
%%%%%%%%%%%%%%%%%%%%%%%%%%%%%%%%%%%%%%%%%%%%%%%%%%
H_0
=
\sum_{\mathbf{k}S\sigma}
	\varepsilon^{(S)}_\mathbf{k}
	\tilde \gamma^\dag_{\mathbf{k}S\sigma}
	\tilde \gamma^{\vphantom{\dag}}_{\mathbf{k}S\sigma}
\end{equation}
is diagonal in this operator basis.

Here we are interested in the low-temperature orderings of the AB-BLG. The
characteristic energy scale for such phenomena does not exceed several meV,
which is much smaller not only $t$, but
$t_0$
as well. We additionally assume that
\begin{eqnarray}
%%%%%%%%%%%%%%%%%%%%%%%%%%%%%%%%%%%%%%%%%%%%%%%%%%
\label{weak_field_requirement}
%%%%%%%%%%%%%%%%%%%%%%%%%%%%%%%%%%%%%%%%%%%%%%%%%%
|e \Phi| \lesssim t_0,
\end{eqnarray}
which is always fulfilled in experiment. In this regime, the two-band
approximation is appropriate: the bands
$\varepsilon^{(1,4)}_{\bf k}$
are ignored, and only bands
$\varepsilon^{(2,3)}_{\bf k}$,
which are the closest to the Fermi level, are kept in the formalism.

As a further simplification, only single-electron states sufficiently close
to the Dirac points are retained. For proper enumeration of such states, it
is convenient to introduce reciprocal space cutoff
$q_0=2t_0/\sqrt{3}ta<|\mathbf{K}_1-\mathbf{K}_{-1}|/2$,
and valley index
$\xi = \pm 1$.
If
${\bf k}$
satisfies the inequality
$|{\bf k} - {\bf K}_\xi| < q_0$,
a single-electron state with such a momentum is assigned to valley $\xi$.
In valley $\xi$, we count the electron momentum from the corresponding
Dirac point
$\mathbf{k}\rightarrow \mathbf{k}-\mathbf{K}_\xi$.

In the two-band framework, the operator basis introduced by
Eq.~(\ref{gamma_spinors})
is not always convenient: operators
$\tilde \gamma_{\mathbf{k}S\sigma}$
depend on transverse potential
$e \Phi$,
which may lead to awkward intermediate formulas, and they do not have
explicit valley designations. We address these matters by defining new
operators in the following manner. Observe that in the case of
$e\Phi=0$,
in the low-energy approximation we have simple analytical formulas for the
eigenvectors
$\Psi^{(S)}_{\mathbf{k}ia}$
[see also Eq.~(21) in
Ref.~\onlinecite{ab_supercond2023sboychakov}]
\begin{eqnarray}\label{eq::eigenvecs}
\nonumber
\Psi^{(2)}_{\mathbf{k}ia}=\frac{1}{\sqrt{2}}\left(0,\psi^*_\mathbf{k},\psi_\mathbf{k},0\right)^T,\\
\Psi^{(3)}_{\mathbf{k}ia}=\frac{1}{\sqrt{2}}\left(0,-\psi^*_\mathbf{k},\psi_\mathbf{k},0\right)^T,
\end{eqnarray}
where
$\psi_\mathbf{k}=\exp(i\pi/2-i\xi\phi_\mathbf{k})$
and
$\phi_\mathbf{k}$
is the polar angle for vector
${\bf k}$.
Then, using
Eq.~\eqref{gamma_spinors}
we obtain for non-dimer sites
\begin{eqnarray}
%%%%%%%%%%%%%%%%%%%%%%%%%%%%%%%%%%%%%%%%%%%%%%%%%%
\label{gamma_def_1B}
%%%%%%%%%%%%%%%%%%%%%%%%%%%%%%%%%%%%%%%%%%%%%%%%%%
d_{{\bf k} 1B \xi \sigma}
=
\frac{ -i e^{ i \xi \phi_{\bf k}} }{\sqrt{2}}
\left(
	\gamma_{{\bf k} 2 \xi \sigma}
	-
	\gamma_{{\bf k} 3 \xi \sigma}
\right),
\\
%%%%%%%%%%%%%%%%%%%%%%%%%%%%%%%%%%%%%%%%%%%%%%%%%%
\label{gamma_def_2A}
%%%%%%%%%%%%%%%%%%%%%%%%%%%%%%%%%%%%%%%%%%%%%%%%%%
d_{{\bf k} 2A \xi \sigma}
=
\frac{ i e^{ - i \xi \phi_{\bf k} } }{\sqrt{2}}
\left(
	\gamma_{{\bf k} 2 \xi \sigma}
	+
	\gamma_{{\bf k} 3 \xi \sigma}
\right),
\end{eqnarray}
where
$\gamma_{{\bf k} S \xi \sigma}$
are the operators
$\tilde \gamma_{{\bf k} S \sigma}$
at
$e\Phi=0$,
localized at specific valley $\xi$. Thus, at
$e\Phi = 0$,
the Hamiltonian can be approximated as ($\hbar=1$)
\begin{eqnarray}
H_0=\sum_{\mathbf{k} \xi \sigma}
	\varepsilon_\mathbf{k}
	\left(
		\gamma^{\dag}_{\mathbf{k}3 \xi \sigma}
		\gamma^{\phantom{\dag}}_{\mathbf{k}3 \xi \sigma}
		-
		\gamma^{\dag}_{\mathbf{k}2 \xi \sigma}
		\gamma^{\phantom{\dag}}_{\mathbf{k}2 \xi \sigma}
	\right),
\end{eqnarray}
where
$\varepsilon_\mathbf{k}=v_{\rm F}^2{\bf k}^2/t_0$
and
$v_\textrm{F}=\sqrt{3}at/2$.

To account for a non-zero
$e\Phi$,
we should add the term
$e\Phi ( \rho_{10} - \rho_{20})/2$
to the Hamiltonian, where
$\rho_{i0}$
is the density operator in the layer
$i=1,2$
\begin{eqnarray}\label{eq::rhodefinition}
\rho_{i {\bf k} } =
\sum_{\mathbf{q} \xi \sigma}\left(
	d^\dag_{\mathbf{q + k}i A \xi \sigma}
	d^{\vphantom{\dagger}}_{\mathbf{q}i A \xi \sigma}
	+
	d^\dag_{\mathbf{q + k}i B \xi \sigma}
	d^{\vphantom{\dagger}}_{\mathbf{q}i B \xi \sigma}\right),
\end{eqnarray}
evaluated at small wave vector
${\bf k} \rightarrow 0$.
In terms of the band operators one can write
\begin{eqnarray}
%%%%%%%%%%%%%%%%%%%%%%%%%%%%%%%%%%%%%%%%%%%%%%%%%%
\label{diff_rho}
%%%%%%%%%%%%%%%%%%%%%%%%%%%%%%%%%%%%%%%%%%%%%%%%%%
\rho_{10} - \rho_{20}
=
- \sum_{{\bf k} \xi \sigma}
	\left(\gamma_{{\bf k} 2 \xi \sigma}^\dag
	\gamma_{{\bf k} 3 \xi \sigma}^{\vphantom{\dag}}
	+
	\gamma_{{\bf k} 3 \xi \sigma}^\dag
	\gamma_{{\bf k} 2 \xi \sigma}^{\vphantom{\dag}}\right).
\end{eqnarray}
Therefore, the Hamiltonian
$H_0$
becomes
\begin{eqnarray}
%%%%%%%%%%%%%%%%%%%%%%%%%%%%%%%%%%%%%%%%%%%%%%%%%%
\label{2band_H}
%%%%%%%%%%%%%%%%%%%%%%%%%%%%%%%%%%%%%%%%%%%%%%%%%%
H_0
=
\sum_{{\bf k} \xi \sigma}
	\left(
		\gamma_{{\bf k} 2 \xi \sigma}^\dag,
		\gamma_{{\bf k} 3 \xi \sigma}^\dag
	\right)
	\left(
		\begin{matrix}
			- \varepsilon_{\bf k} & - \frac{e\Phi}{2} \\
			- \frac{e\Phi}{2} & \varepsilon_{\bf k} \\
		\end{matrix}
	\right)
	\left(
		\begin{matrix}
			\gamma_{{\bf k} 2 \xi \sigma}^{\vphantom{\dag}} \\
			\gamma_{{\bf k} 3 \xi \sigma}^{\vphantom{\dag}} \\
		\end{matrix}
	\right).
\end{eqnarray}
The spectrum of the
Hamiltonian~\eqref{2band_H} is
\begin{equation}
%%%%%%%%%%%%%%%%%%%%%%%%%%%%%%%%%%%%%%%%%%%%%%%%%%
\label{spectrum_2bands}
%%%%%%%%%%%%%%%%%%%%%%%%%%%%%%%%%%%%%%%%%%%%%%%%%%
\varepsilon_{\bf k}^{(2,3)}=\mp\sqrt{\varepsilon_{\bf k}^2+\frac{e^2\Phi^2}{4}}.
\end{equation}
This Hamiltonian is invariant under action of the group
\begin{eqnarray}
%%%%%%%%%%%%%%%%%%%%%%%%%%%%%%%%%%%%%%%%%%%%%%%%%%
\label{invar_group}
%%%%%%%%%%%%%%%%%%%%%%%%%%%%%%%%%%%%%%%%%%%%%%%%%%
G = {\rm U}(2)\times {\rm U}(2)
\end{eqnarray}
that transforms the band operators according to
\begin{eqnarray}
%%%%%%%%%%%%%%%%%%%%%%%%%%%%%%%%%%%%%%%%%%%%%%%%%%
\label{Bobolyubov}
%%%%%%%%%%%%%%%%%%%%%%%%%%%%%%%%%%%%%%%%%%%%%%%%%%
\gamma_{{\bf k} s \xi \sigma}
\rightarrow
\sum_{\sigma'}
	u^\xi_{\sigma \sigma'} \gamma_{{\bf k} s \xi \sigma'},
\end{eqnarray}
where
$u^\xi_{\sigma \sigma'}$
are matrix elements of a
$2\times 2$
unitary matrix
$\hat U_\xi$.
This transformation is more general than the usual spin-rotation symmetry: it allows for an independent spin rotation in each valley.

\subsection{Interaction Hamiltonian}
%%%%%%%%%%%%%%%%%%%%%%%%%%%%%%%%%%%%%%%%%%%%%%%%%%
\label{subsec::interaction_Hamiltonian}
%%%%%%%%%%%%%%%%%%%%%%%%%%%%%%%%%%%%%%%%%%%%%%%%%%

We model the interaction Hamiltonian as follows
\begin{eqnarray}
%%%%%%%%%%%%%%%%%%%%%%%%%%%%%%%%%%%%%%%%%%%%%%%%%%
\label{H_int}
%%%%%%%%%%%%%%%%%%%%%%%%%%%%%%%%%%%%%%%%%%%%%%%%%%
H_{\rm int} = \frac{1}{2 N_c}\sum_{\bf k}
	\left(V^{11}_{\bf k} \rho_{1 {\bf k}} \rho_{1 -{\bf k}}
	+
	V^{22}_{\bf k} \rho_{2 {\bf k}} \rho_{2 -{\bf k}}\right.
	\\
\nonumber
\left.
	+V^{12}_{\bf k} \rho_{1 {\bf k}} \rho_{2 -{\bf k}}
	+
	V^{21}_{\bf k} \rho_{2 {\bf k}} \rho_{1 -{\bf k}}\right),
\end{eqnarray}
where
$N_c$
is the number of unit cells in a sample, and
$V_\mathbf{k}^{ij}$
is the screened Coulomb potential that accounts for interaction between
electrons in $i$-th and $j$-th layers. Note also that we neglect the
electron scattering with large momentum transfer and below is assumed that
\begin{equation}\label{eq::k_condition}
|\mathbf{k}|\ll |\mathbf{K}_1|.
\end{equation}

%When doping is absent, we have $\rho_{1\mathbf{k}}=-\rho_{2\mathbf{k}} $ and for ${\bf k} \rightarrow 0$ we readily obtain

At small $\mathbf{k}$, one can write for the ``bare'' Coulomb interaction:
\begin{eqnarray}
%%%%%%%%%%%%%%%%%%%%%%%%%%%%%%%%%%%%%%%%%%%%%%%%%%
\label{electostatic}
%%%%%%%%%%%%%%%%%%%%%%%%%%%%%%%%%%%%%%%%%%%%%%%%%%
V^{11}_{\bf k}
&=&
V^{22}_{\bf k}
=
\frac{2\pi e^2}{\epsilon S_{0} |{\bf k}|},
\\
%%%%%%%%%%%%%%%%%%%%%%%%%%%%%%%%%%%%%%%%%%%%%%%%%%
\label{Vdiff}
%%%%%%%%%%%%%%%%%%%%%%%%%%%%%%%%%%%%%%%%%%%%%%%%%%
V^{12}_{\bf k}
&=&
V^{21}_{\bf k}
=
\frac{2\pi e^2}{\epsilon S_{0} |{\bf k}|} e^{- |{\bf k}| d}
\approx
V^{11}_{\bf k} - \frac{2\pi e^2 d}{\epsilon S_{0} },
\end{eqnarray}
where $\epsilon$ is an effective permittivity and
$S_0=\sqrt{3}a^2/2$
is the area of the graphene unit cell.

Now we re-write the Hamiltonian~(\ref{H_int}) in the form
\begin{eqnarray}
%%%%%%%%%%%%%%%%%%%%%%%%%%%%%%%%%%%%%%%%%%%%%%%%%%
\label{eq::interaction_decomposition}
%%%%%%%%%%%%%%%%%%%%%%%%%%%%%%%%%%%%%%%%%%%%%%%%%%
H_{\rm int} = H_{\rm int}^{\rm cap} + \Tilde H_{\rm int},
\end{eqnarray}
splitting the interaction in "Hartree" term,
$H_{\rm int}^{\rm cap}$
at
$\mathbf{k}=0$
and  ''Hartree-Fock'' term
\begin{eqnarray}
%%%%%%%%%%%%%%%%%%%%%%%%%%%%%%%%%%%%%%%%%%%%%%%%%%
\label{H_nonHartree_def}
%%%%%%%%%%%%%%%%%%%%%%%%%%%%%%%%%%%%%%%%%%%%%%%%%%
\Tilde H_{\rm int}
=
\frac{1}{2N_c} \sideset{}{'}\sum_{{\bf k} }
	\left(
		V^{11}_{\bf k} \rho_{1 {\bf k}} \rho_{1 -{\bf k}}
		+
		V^{22}_{\bf k} \rho_{2 {\bf k}} \rho_{2 -{\bf k}}
\right.
\\
\nonumber
	\left.
		+
		V^{12}_{\bf k} \rho_{1 {\bf k}} \rho_{2 -{\bf k}}
		+
		V^{21}_{\bf k} \rho_{2 {\bf k}} \rho_{1 -{\bf k}}
	\right),
\end{eqnarray}
where the prime at the summation sign implies that
${\bf k} \ne 0$.
The contribution at
${\bf k} = 0$
is simply a capacitive electrostatic energy
\begin{eqnarray}
%%%%%%%%%%%%%%%%%%%%%%%%%%%%%%%%%%%%%%%%%%%%%%%%%%
\label{capacitance_Eo}
%%%%%%%%%%%%%%%%%%%%%%%%%%%%%%%%%%%%%%%%%%%%%%%%%%
H_{\rm int}^{\rm cap}
=
- \frac{{\cal E}_0}{2 N_c} \rho_{1 0} \rho_{2 0},
\end{eqnarray}
where
\begin{eqnarray}
%%%%%%%%%%%%%%%%%%%%%%%%%%%%%%%%%%%%%%%%%%%%%%%%%%
\label{capacitance_Eo_value}
%%%%%%%%%%%%%%%%%%%%%%%%%%%%%%%%%%%%%%%%%%%%%%%%%%
{\cal E}_0 = \frac{4 \pi e^2 d }{S_0\epsilon }
\approx
\frac{116}{\epsilon}\,\text{eV}
\approx
\frac{43 t}{\epsilon}.
\end{eqnarray}

Beside the capacitance term, there is, of course, another Hartree contribution associated with the long-range Coulomb force [see Eqs.~\eqref{electostatic}]. It is proportional to
$V^{11}_{\bf k}$
in the limit
${\bf k} \rightarrow 0$.
Clearly, this is a divergent quantity whose role is to enforce total electroneutrality of the sample. Thus, we can write
\begin{eqnarray}
%%%%%%%%%%%%%%%%%%%%%%%%%%%%%%%%%%%%%%%%%%%%%%%%%%
\label{neutrality}
%%%%%%%%%%%%%%%%%%%%%%%%%%%%%%%%%%%%%%%%%%%%%%%%%%
(\rho_{1 0 } + \rho_{2 0 } ) = 4 N_c.
\end{eqnarray}
In other words, we always work within the subspace where the operator
$(\rho_{1 0 } + \rho_{2 0 } )$
has eigenvalue
$4N_c$.
This allows us to re-write
$H_{\rm int}^{\rm cap}$
as follows
\begin{eqnarray}
%%%%%%%%%%%%%%%%%%%%%%%%%%%%%%%%%%%%%%%%%%%%%%%%%%
\label{capacitance_Eo2}
%%%%%%%%%%%%%%%%%%%%%%%%%%%%%%%%%%%%%%%%%%%%%%%%%%
H_{\rm int}^{\rm cap}
=
- 2 {\cal E}_0 N_c+\frac{{\cal E}_0}{8 N_c} (\rho_{1 0} - \rho_{2 0})^2.
\end{eqnarray}

It is necessary to express
$\tilde H_{\rm int}^{\vphantom{\dag}}$
and
$H_{\rm int}^{\rm cap}$
in terms of the band operators. The term
$\tilde H_{\rm int}^{\vphantom{\dag}}$
is transformed first. We start by calculating
$V^{ij}_{\bf k}$
at
${\bf k} \ne 0$
using the random phase approximation. This allows us to account for
many-body screening
effects~\cite{ab_supercond2023sboychakov}.
The potentials
$V^{ij}_{\bf k}$
decrease at large transferred momentum. This justifies us restricting
ourselves to small transferred momentum
limit~(\ref{eq::k_condition}).

According to definitions~\eqref{gamma_spinors} and \eqref{eq::rhodefinition} we introduce projected coupling constants as
\begin{eqnarray}
%%%%%%%%%%%%%%%%%%%%%%%%%%%%%%%%%%%%%%%%%%%%%%%%%%
\label{Gamma12}
%%%%%%%%%%%%%%%%%%%%%%%%%%%%%%%%%%%%%%%%%%%%%%%%%%
\Gamma^{(1)}_{\mathbf{kk}'}\!&=&\!
\sum_{ij}\!
        \Big(\!
                \sum_{a}
                        \Psi^{(2)*}_{\mathbf{k}ia}
                        \Psi^{(2)}_{\mathbf{k}'ia}
        \Big)\!
        V^{ij}_{\mathbf{k}-\mathbf{k}'}\!
        \Big(\!
		\sum_{b}
                        \Psi^{(3)}_{\mathbf{k}jb}
                        \Psi^{(3)*}_{\mathbf{k}'jb}
        \Big),
\;\;\;\;
\nonumber\\
\Gamma^{(2)}_{\mathbf{kk}'}\!&=&\!
\sum_{ij}\!
        \Big(\!
		\sum_{a}
			\Psi^{(2)*}_{\mathbf{k}ia}
                        \Psi^{(3)}_{\mathbf{k}'ia}
        \Big)\!
        V^{ij}_{\mathbf{k}-\mathbf{k}'}\!
        \Big(\!
                \sum_{b}
                        \Psi^{(3)}_{\mathbf{k}jb}
                        \Psi^{(2)*}_{\mathbf{k}'jb}
        \Big),\;\;\;\;\nonumber\\
\end{eqnarray}
and write down the Hamiltonian
$\tilde{H}_{\textrm{int}}$
in terms of the band operators
$\tilde \gamma_{\mathbf{k}S\sigma}$,
$\tilde \gamma^\dag_{\mathbf{k}S\sigma}$:
\begin{eqnarray}
%%%%%%%%%%%%%%%%%%%%%%%%%%%%%%%%%%%%%%%%%%%%%%%%%%
\label{HintPsi}
%%%%%%%%%%%%%%%%%%%%%%%%%%%%%%%%%%%%%%%%%%%%%%%%%%
\!\!\!\!\!\Tilde H_{\rm int}
=
-\frac{1}{2 N_c}
\sum_{\mathbf{kk}'\sigma \sigma'}\left(
        \tilde \gamma^{\dag}_{\mathbf{k}2\sigma}
        \tilde \gamma^{\phantom{\dag}}_{\mathbf{k}3{\sigma'}}
        \Gamma^{(1)}_{\mathbf{kk}'}
        \tilde \gamma^{\dag}_{\mathbf{k}'3{\sigma'}}
        \tilde \gamma^{\phantom{\dag}}_{\mathbf{k}'2\sigma}+
\right.
\nonumber
\\
\left.
\tilde \gamma^{\dag}_{\mathbf{k}2\sigma}
        \tilde \gamma^{\phantom{\dag}}_{\mathbf{k}3{\sigma'}}
        \Gamma^{(2)}_{\mathbf{kk}'}
        \tilde \gamma^{\dag}_{\mathbf{k}'2{\sigma'}}
        \tilde \gamma^{\phantom{\dag}}_{\mathbf{k}'3\sigma}\right)
        +
        {\rm H.c.}
\end{eqnarray}
Here we only kept the terms responsible for the gap opening (at $\Phi=0$).
The coupling constants satisfy the relations
\begin{eqnarray}
%%%%%%%%%%%%%%%%%%%%%%%%%%%%%%%%%%%%%%%%%%%%%%%%%%
\label{Gamma_relations}
%%%%%%%%%%%%%%%%%%%%%%%%%%%%%%%%%%%%%%%%%%%%%%%%%%
\Gamma^{(1)}_{\mathbf{kk}'} = \Gamma^{(1)*}_{\mathbf{k'k}},
\quad
\Gamma^{(2)}_{\mathbf{kk}'} = \Gamma^{(2)}_{\mathbf{k'k}}.
\end{eqnarray}
Among these two,
$\Gamma^{(1)}_{\mathbf{kk}'}$
is responsible for the direct coupling, while
$\Gamma^{(2)}_{\mathbf{kk}'}$
represents the umklapp scattering.

Note that the definition~(\ref{H_nonHartree_def}) of
$\Tilde H_{\rm int}$
excludes zero-momentum terms. In the context of Eq.~(\ref{HintPsi}) this constraint is equivalent to the condition
${\bf k} \ne {\bf k}'$.
However, there is no singularity at
${\bf k} = {\bf k}'$
in the case of the screened interaction. Thus, we can ignore this restriction, and assume everywhere that summations over
${\bf k}$
and
${\bf k}'$
in Eq.~(\ref{HintPsi}) (and related expressions) are independent of each other.

The condition~\eqref{eq::k_condition} allows us to introduce valleys in our
description of the interaction Hamiltonian, as it was done for
$H_0$.
As a result, we have
\begin{eqnarray}
%%%%%%%%%%%%%%%%%%%%%%%%%%%%%%%%%%%%%%%%%%%%%%%%%%
\label{HintPsi_valley}
%%%%%%%%%%%%%%%%%%%%%%%%%%%%%%%%%%%%%%%%%%%%%%%%%%
\Tilde H_{\rm int}
&\approx&
\sum_\xi\Tilde H_{\rm int}^\xi,
\\
%%%%%%%%%%%%%%%%%%%%%%%%%%%%%%%%%%%%%%%%%%%%%%%%%%
\label{Hint_gamma}
%%%%%%%%%%%%%%%%%%%%%%%%%%%%%%%%%%%%%%%%%%%%%%%%%%
\Tilde H_{\text{int}}^\xi
&=&
-\frac{1}{2 N_c}
\sum_{\mathbf{kk}'\sigma \sigma'}\left(
        \gamma^{\dag}_{\mathbf{k}2\xi \sigma}
        \gamma^{\phantom{\dag}}_{\mathbf{k}3{\xi \sigma'}}
        \Gamma^{(1)}_{\mathbf{kk}'}
        \gamma^{\dag}_{\mathbf{k}'3{\xi \sigma'}}
        \gamma^{\phantom{\dag}}_{\mathbf{k}'2\xi \sigma}
\right.
+
\nonumber
\\
&&
\left.
        \gamma^{\dag}_{\mathbf{k}2\xi \sigma}
        \gamma^{\phantom{\dag}}_{\mathbf{k}3{\xi \sigma'}}
        \Gamma^{(2)}_{\mathbf{kk}'}
        \gamma^{\dag}_{\mathbf{k}'2{\xi \sigma'}}
        \gamma^{\phantom{\dag}}_{\mathbf{k}'3\xi \sigma}\right)
        +
        {\rm H.c.}
\end{eqnarray}
Here
${\bf k}$
and
${\bf k}'$
are limited to a vicinity of
${\bf K}_\xi$. Note that Hamiltonian~\eqref{Hint_gamma} is written in terms of operators $\gamma_{\mathbf{k}S\xi\sigma}$ instead of operators $\tilde{\gamma}_{\mathbf{k}S\sigma}$ [as in Eq.~\eqref{HintPsi}]. In this case, the wave functions
$\Psi^{(S)}_{\mathbf{k}ia}$
in
Eq.~\eqref{Gamma12} are given by Eq.~\eqref{eq::eigenvecs}. Our $\tilde H_{\rm int}$
has same-valley terms only. Such a simplification excludes some more exotic
order parameters (the so-called Kekul{\'{e}}-like orders), which we believe
is justified. Indeed, while various Kekul{\'{e}}-like phases are considered
in theoretical
literature~\cite{Nandkishore2010b, Lemonik2010, cvetkovic_multi2012,
lemonic_rg_nemat_long2012, min_pseudo_fm2008,rozhkov2025ab_su4}
as a mathematical possibility, detailed calculations often demonstrate that
they do not become ground state under realistic
conditions~\cite{lemonic_rg_nemat_long2012, cvetkovic_multi2012}.

It is convenient to define an operator-valued 2$\times$2 matrix
$\hat\Xi_{{\bf k} \xi}$
whose matrix elements are
\begin{eqnarray}
\Xi_{{\bf k} \xi}^{\sigma \sigma'}
=
\gamma^{\dag}_{\mathbf{k}3{\xi \sigma}}
\gamma^{\phantom{\dag}}_{\mathbf{k} 2\xi \sigma'},
\;\;
[\Xi_{{\bf k} \xi}^\dag]^{\sigma \sigma'}
=
\gamma^{\dag}_{\mathbf{k}2{\xi \sigma}}
\gamma^{\phantom{\dag}}_{\mathbf{k} 3\xi \sigma'}.
\end{eqnarray}
Then Eq.~(\ref{Hint_gamma}) can be re-written as
\begin{equation}\label{Hint1}
\!\!\Tilde H_{\rm int}^\xi
\!=\!
-\!\frac{1}{2N_c}
\sum_{\mathbf{kk}'}
{\rm Tr}\/
\left(
        \hat\Xi_{{\bf k} \xi}^\dag
        \Gamma^{(1)}_{\mathbf{kk}'}
	\hat\Xi_{{\bf k}' \xi}^{\vphantom{\dag}}
	\!+\!
        \hat\Xi_{{\bf k} \xi}^\dag
        \Gamma^{(2)}_{\mathbf{kk}'}
	\hat\Xi_{{\bf k}' \xi}^\dag
\right)
\!+\!
{\rm H.c.}
\end{equation}
We introduce averaged coupling constants
\begin{eqnarray}\label{Gammaav}
\bar\Gamma_{1,2}
=
\int \frac{d^2 {\bf k}}{\pi q_0^2} \frac{d^2 {\bf k}'}{\pi q_0^2}
	\Gamma^{(1,2)}_{\mathbf{kk}'},
\end{eqnarray}
where integration is performed over
$|\mathbf{k}|,|\mathbf{k}'|<q_0$.
The averaged coupling constants
$\bar \Gamma_{1,2}$
were estimated in
Ref.~\onlinecite{rozhkov2025ab_su4},
where they were denoted as
$\bar V_{\rm dir, um}$.
We will use for estimates the values obtained in
Ref.~\onlinecite{rozhkov2025ab_su4}:
\begin{eqnarray}
%%%%%%%%%%%%%%%%%%%%%%%%%%%%%%%%%%%%%%%%%%%%%%%%%%
\label{Gamma_def}
%%%%%%%%%%%%%%%%%%%%%%%%%%%%%%%%%%%%%%%%%%%%%%%%%%
\bar \Gamma_{1} = 9.37 t,
\quad
\bar \Gamma_{2} = 8.93 t.
\end{eqnarray}
Using these constants, we simplify the interaction Hamiltonian
\begin{equation}\label{Hint2}
\Tilde H_{\rm int}^\xi
\!=\!
\!\frac{-1}{2N_c}\!\!
\sum_{\mathbf{kk}'}
{\!\rm Tr}\!
\left[
        2 \bar \Gamma_{1}
        \hat\Xi_{{\bf k} \xi}^\dag
	\hat\Xi_{{\bf k}' \xi}^{\vphantom{\dag}}
\!+\!
%\right.
%\\
%\left.
\bar \Gamma_{2}
	\left(
        	\hat\Xi_{{\bf k} \xi}^\dag
		\hat\Xi_{{\bf k}' \xi}^\dag
		\!+\!
        	\hat\Xi_{{\bf k} \xi}^{\vphantom{\dag}}
		\hat\Xi_{{\bf k}' \xi}^{\vphantom{\dag}}
	\right)
\!\right].
\end{equation}
The transition from Eq.~\eqref{Hint1} to Eq.~\eqref{Hint2} is an approximation. However, there is certain physical reason standing behind it. Using the momentum dependent $\Gamma^{(1,2)}_{\mathbf{kk}'}$ one can prove that the most stable order parameters (introduced below) are of the $s$-wave type. Assuming that they are slightly depend on $\mathbf{k}$ in the range $k<q_0$ and decay fast outside this region, one can use the averaged coupling constants in the form of Eq.~\eqref{Gammaav}.

According to Eq.~(\ref{diff_rho}), we have
\begin{eqnarray}
%%%%%%%%%%%%%%%%%%%%%%%%%%%%%%%%%%%%%%%%%%%%%%%%%%
\label{diff_rho_Xi}
%%%%%%%%%%%%%%%%%%%%%%%%%%%%%%%%%%%%%%%%%%%%%%%%%%
\rho_{10} - \rho_{20}
=
- \sum_{{\bf k} \xi}
	{\rm Tr}\! \left(
			\hat\Xi_{{\bf k} \xi}^{\vphantom{\dag}}
			+
			\hat\Xi_{{\bf k} \xi}^\dag
	\right).
\end{eqnarray}
Then, we can express
$H_{\rm int}^{\rm cap}$
in terms of
$\hat\Xi$
as
\begin{eqnarray}
%%%%%%%%%%%%%%%%%%%%%%%%%%%%%%%%%%%%%%%%%%%%%%%%%%
\label{capacitance_Eo_Xi}
%%%%%%%%%%%%%%%%%%%%%%%%%%%%%%%%%%%%%%%%%%%%%%%%%%
H_{\rm int}^{\rm cap}
&=&-2 {\cal E}_0N_c
+
\\
\nonumber
&&\frac{{\cal E}_0}{8 N_c}
\sum_{{\bf k} {\bf k}' \xi \xi' }
	{\rm Tr}\! \left(
			\hat\Xi_{{\bf k} \xi}^{\vphantom{\dag}}
			+
			\hat\Xi_{{\bf k} \xi}^\dag
	\right)
	{\rm Tr}\! \left(
			\hat\Xi_{{\bf k}' \xi'}^{\vphantom{\dag}}
			+
			\hat\Xi_{{\bf k}' \xi'}^\dag
	\right).
\end{eqnarray}
Collecting all these terms, we write the interaction Hamiltonian in the
form
\begin{eqnarray}
%%%%%%%%%%%%%%%%%%%%%%%%%%%%%%%%%%%%%%%%%%%%%%%%%%
\label{interaction_Eo_Xi}
%%%%%%%%%%%%%%%%%%%%%%%%%%%%%%%%%%%%%%%%%%%%%%%%%%
&&H_{\rm int}
=
\frac{{\cal E}_0}{8 N_c}
\sum_{{\bf k} {\bf k}' \xi \xi' }
	{\rm Tr}\! \left(
			\hat\Xi_{{\bf k} \xi}^{\vphantom{\dag}}
			+
			\hat\Xi_{{\bf k} \xi}^\dag
	\right)
	{\rm Tr}\! \left(
			\hat\Xi_{{\bf k}' \xi'}^{\vphantom{\dag}}
			+
			\hat\Xi_{{\bf k}' \xi'}^\dag
	\right)
\\
&&-
\nonumber
\frac{1}{2N_c}
\sum_{\mathbf{kk}' \xi}
	{\rm Tr}\!
	\left[
		2 \bar \Gamma_{1}
	        \hat\Xi_{{\bf k} \xi}^\dag
		\hat\Xi_{{\bf k}' \xi}^{\vphantom{\dag}}
		+
       	 	\bar \Gamma_{2}
		\left(
        		\hat\Xi_{{\bf k} \xi}^\dag
			\hat\Xi_{{\bf k}' \xi}^\dag
			+
       		 	\hat\Xi_{{\bf k} \xi}^{\vphantom{\dag}}
			\hat\Xi_{{\bf k}' \xi}^{\vphantom{\dag}}
		\right)
	\right].
\end{eqnarray}
Here, the
$c$-number contribution
$- 2{\cal E}_0N_c$
is ignored.

The transformation group
$G$
introduced by Eqs.~(\ref{invar_group}) and~(\ref{Bobolyubov}) keeps the  interaction Hamiltonian invariant. Such an invariance is a consequence of two features: (i) operator
$\Tilde H_{\rm int}$
splits into two valley-specific terms
$\Tilde H_{\rm int}^\xi$,
each being invariant under spin rotation
$\hat U_\xi$,
and (ii) operator
$H^{\rm cap}_{\rm int}$
depends on the electrical polarization
$(\rho_{10} - \rho_{20})$
only, which is a true scalar under the action of the spin rotations in
either valleys, as
Eq.~(\ref{diff_rho_Xi})
readily demonstrates.

\section{Variation version of the mean field theory}
%%%%%%%%%%%%%%%%%%%%%%%%%%%%%%%%%%%%%%%%%%%%%%%%%%
\label{var_MF}
%%%%%%%%%%%%%%%%%%%%%%%%%%%%%%%%%%%%%%%%%%%%%%%%%%

\subsection{Mean field variational wave function}

Our ultimate goal is to map phase diagram of the many-body model formulated
in the previous section. Theoretical tool we use is a zero temperature mean
field approach. It is well-known that the mean field method can be
implemented using several (equivalent) frameworks: for example, one can
view it through lens of the Hubbard-Stratonovich decoupling, or,
alternatively, as a self-consistency scheme. Here, we choose variational
formulation of the mean field approximation for it allows a consistent
treatment of multiple competing ordered phases.

In the mean-field approach the main object is an order parameter. Here we
consider exciton-type (non-superconducting) order parameters that open
insulating gap. For AB-BLG, such order parameters can be organized into
$2\times2$
matrices
$\hat \Delta_\xi$
and
$\hat \Delta_{\bar\xi}$,
acting in spin space, with the valley labels $\xi$ and
$\bar\xi$,
respectively (bar over index invert its value,
$\bar{\xi}=-\xi$).
We denote the variational wave function $|\hat \Delta_\xi\rangle$ that depends on the proper order parameter.

To construct the desired variational wave function, consider the following
Hamiltonian
\begin{eqnarray}
%%%%%%%%%%%%%%%%%%%%%%%%%%%%%%%%%%%%%%%%%%%%%%%%%%
\label{MF_Hamiltonian_def}
%%%%%%%%%%%%%%%%%%%%%%%%%%%%%%%%%%%%%%%%%%%%%%%%%%
H_{\rm MF}
=
H_0
- \sum_{{\bf k} \xi}
	{\rm Tr}\!  \left(
		\hat{\Delta}_{ \xi}^{\vphantom{\dag}}
		\hat\Xi_{{\bf k} \xi}^\dag
		+
		\hat\Xi_{{\bf k} \xi}^{\vphantom{\dag}}
		\hat{\Delta}_{ \xi}^\dag
	\right).
\end{eqnarray}
Its ground state
$\left| \hat{\Delta}_\xi \right>$
satisfies
\begin{eqnarray}
H_{\rm MF} \left|\hat{\Delta}_\xi \right>
=
E_{\rm MF} \left|\hat{\Delta}_\xi \right>.
\end{eqnarray}
Here
$E_{\rm MF} = E_{\rm MF} (\hat \Delta_\xi)$
is the ground state energy of
$\hat H_{\rm MF} (\hat \Delta_\xi)$.
Obviously
$E_{\rm MF} = \langle \hat H_{\rm MF} \rangle$,
where
$\langle \ldots \rangle$
denotes averaging with respect to the wave function
$\left|\hat{\Delta}_\xi \right>$.
We use
$\left|\hat{\Delta}_\xi \right>$
as variational wave function, with
$\hat{\Delta}_\xi$
as an optimization parameter: we adjust
$\hat{\Delta}_\xi$
to minimize variational energy
\begin{eqnarray}
%%%%%%%%%%%%%%%%%%%%%%%%%%%%%%%%%%%%%%%%%%%%%%%%%%
\label{eq::Evar_def}
%%%%%%%%%%%%%%%%%%%%%%%%%%%%%%%%%%%%%%%%%%%%%%%%%%
E_{\rm var} = \langle H_0 + H_{\rm int} \rangle.
\end{eqnarray}

The considered mean field Hamiltonian changes under spin rotation and we
can use
$H_{\rm MF}$
to describe states with broken spin-rotation symmetry. Group $G$ is broader
than the uniform spin-rotation group U(2), therefore, the mean field
Hamiltonian changes under the action of $G$. However, it is easy to see
that the Bogolyubov
transformation~(\ref{Bobolyubov})
induces a unitary transformation of
$H_{\rm MF}$
that can be viewed as a unitary rotation of the order parameter belonging
to $G$. Consequently,  the function
$E_{\rm MF} = E_{\rm MF} (\hat \Delta_\xi)$
is invariant under the transformation $G$, since any eigenenergy remains
invariant under a unitary transformation of a Hamiltonian.

The value
$E_{\rm MF}$
is not the same as the variation energy
$E_{\rm var}$.
However,
$E_{\rm MF}$
is a useful technical object for further considerations. The utility of
$E_{\rm MF}$
stems specifically from the fact that
\begin{eqnarray}
%%%%%%%%%%%%%%%%%%%%%%%%%%%%%%%%%%%%%%%%%%%%%%%%%%
\label{Hellmann_Feyn_theorem_matrxi}
%%%%%%%%%%%%%%%%%%%%%%%%%%%%%%%%%%%%%%%%%%%%%%%%%%
\sum_{\bf k} \langle \hat\Xi_{{\bf k} \xi}^\dag \rangle
=
- \frac{\partial E_{\rm MF}}{\partial \hat \Delta_\xi},
\end{eqnarray}
as Hellmann-Feynman theorem guarantees. In this formula, the
differentiation is performed over matrix elements
$\left[ \hat \Delta_\xi \right]_{\sigma \sigma'}$.
The indices $\sigma$ and $\sigma'$ run from 1 to 2, enumerating all four
matrix elements of the
$2\times 2$
order parameter matrix.

To perform such a differentiation, one needs to determine how
$E_{\rm MF}$
depends on
$\hat \Delta_\xi$.
To find the desired expression, negative eigenvalues of the
$4\times4$
matrix
\begin{eqnarray}
{\cal H}^{\rm MF}_{{\bf k} \xi}
=
\left(
	\begin{matrix}
		- \varepsilon_{\bf k}^{\vphantom{\dag}} &
		- \hat\Delta_{\xi}^{\vphantom{\dag}} \\
		- \hat \Delta_{ \xi}^\dag &
		\varepsilon_{\bf k}^{\vphantom{\dag}} \\
	\end{matrix}
\right)
\end{eqnarray}
must be found.
Introducing spinors
$a_i$
and
$b_i$,
we write the following relation for eigenvalues
$\varepsilon_{i \xi}$
\begin{eqnarray}
\left(
	\begin{matrix}
		- \varepsilon_{\bf k}^{\vphantom{\dag}} &
		- \hat\Delta_\xi^{\vphantom{\dag}} \\
		- \hat \Delta_\xi^\dag &
		\varepsilon_{\bf k}^{\vphantom{\dag}} \\
	\end{matrix}
\right)
\left(
	\begin{matrix}
		a_i\\
		b_i\\
	\end{matrix}
\right)
=
\varepsilon_{i\xi}
\left(
	\begin{matrix}
		a_i\\
		b_i\\
	\end{matrix}
\right).
\end{eqnarray}
Excluding
$b_i$,
one can derive
\begin{eqnarray}
( \varepsilon_{i\xi}^2 - \varepsilon_{\bf k}^2 )
a_i
=
\hat\Delta_{\xi}^{\vphantom{\dag}} \hat\Delta_{\xi}^\dag a_i.
\end{eqnarray}
This reduces the
$4\times4$
eigenvalue problem to
$2\times2$
one: we need to diagonalize the Hermitian matrix
$\hat\Delta_{\xi}^{\vphantom{\dag}} \hat\Delta_{\xi}^\dag$.
Once its eigenvalues
${\cal{D}}_{1\xi} \geqslant 0$,
${\cal{D}}_{2\xi} \geqslant 0$
are known, the eigenvalues
$\varepsilon_{i \xi}$
can be determined according to
\begin{eqnarray}
\varepsilon_{i\xi} = \pm \sqrt{\varepsilon_{\bf k}^2 + {\cal{D}}_{1,2\xi}}.
\end{eqnarray}
Now, we can express the ground state energy as a sum of negative
$\varepsilon_{i\xi}$
\begin{eqnarray}
E_{\rm MF}
=
- \sum_{{\bf k} \xi}
	\left(
		\sqrt{\varepsilon_{\bf k}^2 + {\cal{D}}_{1\xi}}
		+
		\sqrt{\varepsilon_{\bf k}^2 + {\cal{D}}_{2\xi}}
	\right).
\end{eqnarray}
Since
$\sqrt{\varepsilon_{\bf k}^2 + {\cal{D}}_{1\xi}} + \sqrt{\varepsilon_{\bf k}^2 + {\cal{D}}_{2\xi}} = {\rm Tr} \sqrt{\varepsilon_{\bf k}^2 + \hat\Delta_{\xi}^{\vphantom{\dag}} \hat\Delta_{\xi}^\dag}$,
we can evaluate
$E_{\rm MF}$
without explicit diagonalization of
$\hat\Delta_{\xi}^{\vphantom{\dag}} \hat\Delta_{\xi}^\dag$:
\begin{eqnarray}
%%%%%%%%%%%%%%%%%%%%%%%%%%%%%%%%%%%%%%%%%%%%%%%%%%
\label{E_MF_VS_Delta}
%%%%%%%%%%%%%%%%%%%%%%%%%%%%%%%%%%%%%%%%%%%%%%%%%%
E_{\rm MF}
=
- \sum_{{\bf k} \xi}
	{\rm Tr}
	\sqrt{\varepsilon_{\bf k}^2
	+
	\hat\Delta_{\xi}^{\vphantom{\dag}}
	\hat\Delta_\xi^\dag}.
\end{eqnarray}
As discussed above, this expression is invariant under unitary
transformations of
$\hat \Delta_\xi$.

Applying
Eq.~\eqref{Hellmann_Feyn_theorem_matrxi}
to this formula for
$E_{\rm MF}$,
we derive
\begin{eqnarray}
%%%%%%%%%%%%%%%%%%%%%%%%%%%%%%%%%%%%%%%%%%%%%%%%%%
\label{Hellmann_Feyn_result}
%%%%%%%%%%%%%%%%%%%%%%%%%%%%%%%%%%%%%%%%%%%%%%%%%%
\sum_{\bf k} \langle \hat\Xi_{{\bf k} \xi}^{\vphantom{\dag}} \rangle
=
\frac{1}{2} \sum_{\bf k}
	\frac{1}{\sqrt{
		\varepsilon_{\bf k}^2
		+ 		
		\hat \Delta_{ \xi}^\dag
		\hat \Delta_{ \xi}^{\vphantom{\dag}}
	}}
	\hat \Delta_{ \xi}^{\vphantom{\dag}},
\end{eqnarray}
connecting the order parameters with anomalous expectation values.

Let us assume that the order parameter is Hermitian,
$\hat \Delta_\xi^\dag = \hat \Delta_\xi^{\vphantom{\dag}}$.
In principle, the order parameter needs not be Hermitian: a very general
self-consistency condition, derived in
Appendix~\ref{app_Delta_VS_Xi},
admits both Hermitian and non-Hermitian solutions. However, the discussions
in
Refs.~\onlinecite{rozhkov2023aa_su4, rozhkov2025ab_su4}
indicate that non-Hermitian (current-carrying) order parameters are
energetically less favorable under realistic conditions. Restricting the
analysis to Hermitian matrices therefore represents a substantial
simplification of the calculations, while remaining well justified
physically.

Hermitian
$\hat{\Delta}_\xi$
is diagonalizable
\begin{eqnarray}
\!\!\!\!\!\hat \Delta_\xi^{\vphantom{\dag}}
\!=\!
\hat U_\xi^{\vphantom{\dag}}
{\rm diag}\, \left( D_{\uparrow \xi}, D_{\downarrow \xi} \right)
\hat U_\xi^\dag,
\,\,\,
\hat U_\xi \!\in\! {\rm U}(2),
\,\,\,
D_{\sigma\xi}\! \in\! \mathbb{R}.
\end{eqnarray}
This allows us to derive
\begin{eqnarray}
%%%%%%%%%%%%%%%%%%%%%%%%%%%%%%%%%%%%%%%%%%%%%%%%%%
\label{eq::Xi_vs_D}
%%%%%%%%%%%%%%%%%%%%%%%%%%%%%%%%%%%%%%%%%%%%%%%%%%
&&\sum_{\bf k} \langle \hat\Xi_{{\bf k} \xi}^{\vphantom{\dag}} \rangle
=
\sum_{\bf k} \langle \hat\Xi_{{\bf k} \xi}^\dag \rangle
=
\\
\nonumber
&&\frac{N_c \nu_0}{2}
\hat U_\xi^{\vphantom{\dag}}
{\rm diag}\!\left[\!
	D_{\uparrow \xi}
	\ln\left(\frac{2t_0}{|D_{\uparrow \xi}|}\right)\!,
	D_{\downarrow \xi}
	\ln\left(\frac{2t_0}{|D_{\downarrow \xi}|}\right)
\right]
\hat U_\xi^\dag,
\end{eqnarray}
where
$\nu_0=t_0/2\sqrt{3}\pi t^2$
is the density of states of undoped AB-BLG, and
$D_{\sigma\xi}\ll t_0$
is assumed. The details of the derivation can be found in
Appendix~\ref{appendix_sum}.

For Hermitian order parameter
$\hat \Delta_\xi$,
the energy
$E_{\rm MF}$
is equal to
\begin{eqnarray}
%%%%%%%%%%%%%%%%%%%%%%%%%%%%%%%%%%%%%%%%%%%%%%%%%%
\label{E_MF_Herm}
%%%%%%%%%%%%%%%%%%%%%%%%%%%%%%%%%%%%%%%%%%%%%%%%%%
E_{\rm MF} = - \sum_{{\bf k} m} \sqrt{\varepsilon_{\bf k}^2 + D_m^2}.
\end{eqnarray}
Here multi-index
$m=(\sigma, \xi)$
is introduced: it enumerates four fermion sectors. Mathematically,
$D_m = D_{\sigma \xi}$
are (real) eigenvalues of
$\hat \Delta_\xi$,
two eigenvalues,
$D_{\uparrow \xi}$
and
$D_{\downarrow \xi}$,
per each valley. Physically,
$|D_m|$
represent four single-electron gaps in four fermion sectors of AB-BLG.

The latter expression for
$E_{\rm MF}$
implies that this energy depends only on the Hermitian order parameters'
eigenvalues, and is independent of eigenvectors. The sum over
${\bf k}$
in
Eq.~(\ref{E_MF_Herm})
is calculated in
Appendix~\ref{appendix_sum}.
Namely,
\begin{eqnarray}\label{eq::EMF_sum}
\!\!\!\!\!\!E_{\rm MF}
\!=\!
E_{\rm MF} (0)
\!-\!
\frac{1}{2} N_c \sum_m
	\nu_0 D_m^2
	\left[
		\!\ln \left( \frac{2t_0}{|D_m|} \right)
		\!+\!
		\frac{1}{2}
	\right],
\end{eqnarray}
where
$E_{\rm MF} (0)$
is a constant independent of the order parameters.

\subsection{Variation energy}

Here we evaluate the variation
energy~(\ref{eq::Evar_def})
as a function of
$\hat \Delta_\xi$.
We can write using
Eq.~\eqref{MF_Hamiltonian_def}
\begin{eqnarray}
%%%%%%%%%%%%%%%%%%%%%%%%%%%%%%%%%%%%%%%%%%%%%%%%%%
\label{E_MF_VS_OP}
%%%%%%%%%%%%%%%%%%%%%%%%%%%%%%%%%%%%%%%%%%%%%%%%%%
\!\!\!\!\!\!\!E_{\rm var}
\!=\!
E_{\rm MF}
\!+\!
\sum_{{\bf k} \xi}
	{\rm Tr}\!  \left(
		\hat{\Delta}_{ \xi}^{\vphantom{\dag}}
		\langle \hat\Xi_{{\bf k} \xi}^\dag \rangle
		\!+\!
		\langle \hat\Xi_{{\bf k} \xi}^{\vphantom{\dag}}  \rangle
		\hat{\Delta}_{ \xi}^\dag
	\right)
\!+\!
\langle H_{\rm int} \rangle.
\end{eqnarray}
To calculate
$\langle H_{\rm int} \rangle$,
Eq.~\eqref{interaction_Eo_Xi},
we apply the Wick theorem and obtain
\begin{widetext}
\begin{eqnarray}
%%%%%%%%%%%%%%%%%%%%%%%%%%%%%%%%%%%%%%%%%%%%%%%%%%
\label{interaction_Wick}
%%%%%%%%%%%%%%%%%%%%%%%%%%%%%%%%%%%%%%%%%%%%%%%%%%
%\!\!\!&&
\langle H_{\rm int} \rangle
=
\frac{{\cal E}_0}{8 N_c}
\left(
	\sum_{{\bf k} \xi }
		{\rm Tr} \langle
				\hat\Xi_{{\bf k} \xi}^{\vphantom{\dag}}
			\rangle
			+
		{\rm Tr} \langle \hat\Xi_{{\bf k} \xi}^\dag \rangle
\right)^2
-
%\\
%\nonumber
%\!\!\!&&
\frac{1}{2N_c}
\sum_{\mathbf{kk}' \xi}\!\!
	{\rm Tr}\!
	\left[
		2 \bar \Gamma_{1}
	        \langle \hat\Xi_{{\bf k} \xi}^\dag \rangle
		\langle \hat\Xi_{{\bf k}' \xi}^{\vphantom{\dag}} \rangle
		\!+\!
       	 	\bar \Gamma_{2}\!
		\left(
			\langle \hat\Xi_{{\bf k} \xi}^\dag \rangle
			\langle \hat\Xi_{{\bf k}' \xi}^\dag \rangle
			\!+\!
       		 	\langle
				\hat\Xi_{{\bf k} \xi}^{\vphantom{\dag}}
			\rangle
			\langle
				\hat\Xi_{{\bf k}' \xi}^{\vphantom{\dag}}
			\rangle
		\right)
	\!\right].
\end{eqnarray}
Therefore, the full expression for the variation energy reads as
\begin{eqnarray}
%%%%%%%%%%%%%%%%%%%%%%%%%%%%%%%%%%%%%%%%%%%%%%%%%%
\label{E_var_full}
%%%%%%%%%%%%%%%%%%%%%%%%%%%%%%%%%%%%%%%%%%%%%%%%%%
E_{\rm var} (\hat \Delta_\xi)
&=&
E_{\rm MF}
+
\sum_{{\bf k} \xi}
	{\rm Tr}\!  \left(
		\hat{\Delta}_{ \xi}^{\vphantom{\dag}}
		\langle \hat\Xi_{{\bf k} \xi}^\dag \rangle
		+
		\langle \hat\Xi_{{\bf k} \xi}^{\vphantom{\dag}}  \rangle
		\hat{\Delta}_{ \xi}^\dag
	\right)
-
%\\
%\nonumber
%&&
\frac{1}{2N_c}
\sum_{\mathbf{kk}' \xi}
	\!{\rm Tr}\!
	\left[
		2 \bar \Gamma_{1}
	        \langle \hat\Xi_{{\bf k} \xi}^\dag \rangle
		\langle \hat\Xi_{{\bf k}' \xi}^{\vphantom{\dag}} \rangle
		\!+\!
       	 	\bar \Gamma_{2}
		\left(
			\langle \hat\Xi_{{\bf k} \xi}^\dag \rangle
			\langle \hat\Xi_{{\bf k}' \xi}^\dag \rangle
			\!+\!
       		 	\langle
				\hat\Xi_{{\bf k} \xi}^{\vphantom{\dag}}
			\rangle
			\langle
				\hat\Xi_{{\bf k}' \xi}^{\vphantom{\dag}}
			\rangle
		\right)
	\right]\nonumber
\\
&&+
\frac{{\cal E}_0}{8 N_c}
\left(
	\sum_{{\bf k} \xi }
		{\rm Tr} \langle
				\hat\Xi_{{\bf k} \xi}^{\vphantom{\dag}}
			\rangle
			+
		{\rm Tr} \langle \hat\Xi_{{\bf k} \xi}^\dag \rangle
\right)^2.
\end{eqnarray}
For Hermitian order parameter, we perform summation over
$\mathbf{k}$
in Eq.~\eqref{E_var_full} similar to Eq.~\eqref{eq::EMF_sum} and obtain
\begin{eqnarray}
%%%%%%%%%%%%%%%%%%%%%%%%%%%%%%%%%%%%%%%%%%%%%%%%%%
\label{Evar_total_herm}
%%%%%%%%%%%%%%%%%%%%%%%%%%%%%%%%%%%%%%%%%%%%%%%%%%
\frac{E_{\rm var}}{N_c}
%&=&
=
\frac{E_{\rm MF} (0)}{N_c}
+
\frac{\nu_0}{2}
\sum_m
	D_m^2
	\left[
		\ln \left(\frac{2t_0}{|D_m|} \right) - \frac{1}{2}
%\\
%&-&
	-
	\frac{\nu_0 (\bar \Gamma_1 + \bar \Gamma_2)}{2}
	\ln^2 \left( \frac{2t_0}{|D_m|} \right)
\right]
%\\
%\nonumber
%&+&
+
\frac{{\cal E}_0}{8}
\left[
	\sum_m \nu_0 D_m \ln \left( \frac{2t_0}{|D_m|} \right)
\right]^2.
\end{eqnarray}
\end{widetext}
In this formula, the last term is the contribution due to capacitance
electrostatic energy, the term proportional to
$(\bar \Gamma_1 + \bar \Gamma_2)$
corresponds to
$\langle \Tilde H_{\rm int} \rangle$.
The remaining terms come from
$\langle H_0 \rangle$.

Analyzing
Eq.~(\ref{Evar_total_herm}),
one can notice that the exchange interaction energy
$\langle \Tilde H_{\rm int} \rangle$
is non-zero as long as at least one
$D_m \ne 0$.
At the same time, the electrostatic energy can be nullified even when all
$D_m$'s
are finite. This suggests that the system can maintain symmetry broken
state even at finite (and, possibly, large)
${\cal E}_0$
energy.

At the same time, the electrostatic charging effects do affect the
ordering, as we will see below. Fortunately, we can control, to some
extent, these effects by applied bias voltage
$e \Phi$.
It contributes the following term to the variational energy
\begin{eqnarray}
%%%%%%%%%%%%%%%%%%%%%%%%%%%%%%%%%%%%%%%%%%%%%%%%%%
\label{ephi_energy_def}
%%%%%%%%%%%%%%%%%%%%%%%%%%%%%%%%%%%%%%%%%%%%%%%%%%
\delta E_{\rm var}
\!=\!
\frac{e \Phi}{2}\langle\rho_{10}\!-\!\rho_{20}\rangle
\!=\!
- \frac{e \Phi}{2}
\sum_{{\bf k} \xi}
	{\rm Tr}\! \left(
			\hat\Xi_{{\bf k} \xi}^{\vphantom{\dag}}
			\!+\!
			\hat\Xi_{{\bf k} \xi}^\dag
	\right),
\end{eqnarray}
where
Eq.~(\ref{diff_rho_Xi})
was used. Expressing traces of the order parameter matrices in terms of
$D_m$'s, see
Eq.~(\ref{eq::Xi_vs_D}),
it is possible to demonstrate that
\begin{eqnarray}
%%%%%%%%%%%%%%%%%%%%%%%%%%%%%%%%%%%%%%%%%%%%%%%%%%
\label{ephi_energy}
%%%%%%%%%%%%%%%%%%%%%%%%%%%%%%%%%%%%%%%%%%%%%%%%%%
\frac{\delta E_{\rm var}}{N_c} = - e \Phi P,
\end{eqnarray}
where
\begin{eqnarray}
%%%%%%%%%%%%%%%%%%%%%%%%%%%%%%%%%%%%%%%%%%%%%%%%%%
\label{polarization_def}
%%%%%%%%%%%%%%%%%%%%%%%%%%%%%%%%%%%%%%%%%%%%%%%%%%
P =
- \frac{1}{N_c e}
\frac{\partial E_{\rm var}}{ \partial \Phi}
=
\frac{1}{2}
	\sum_m \nu_0 D_m \ln\! \left( \frac{2t_0}{|D_m|} \right)
\end{eqnarray}
is transverse (inter-layer) polarization. Once
$\delta E_{\rm var}$
is added to the right-hand side of
Eq.~(\ref{Evar_total_herm}),
our variational energy is complete.

\subsection{Self-consistency equations}

The self-consistency equation can be obtained in general case by
minimization of the variational
energy~(\ref{E_var_full})
over the order parameters
$\hat \Delta_\xi$.
However, such a procedure is quite cumbersome and we placed it in the
Appendix~\ref{app_Delta_VS_Xi}.
Here we limit ourselves to the case of the Hermitian order parameters.
Therefore, our task here is to minimize the energy, expressed by
Eqs.~\eqref{Evar_total_herm}
and~\eqref{ephi_energy},
over
$D_m$'s.
Differentiating over
$D_m$,
we obtain
\begin{eqnarray}
%%%%%%%%%%%%%%%%%%%%%%%%%%%%%%%%%%%%%%%%%%%%%%%%%%
\label{sefl_cons_gaps_log}
%%%%%%%%%%%%%%%%%%%%%%%%%%%%%%%%%%%%%%%%%%%%%%%%%%
\frac{\nu_0}{2} D_m \ln \left( \frac{2t_0}{|D_m|} \right)
=
\frac{D_m}{\bar \Gamma_1\! +\! \bar \Gamma_2}
-
\quad
\\
\nonumber
\frac{{\cal E}_0}{
	2(\bar \Gamma_1\! +\! \bar \Gamma_2)
	(2{\cal E}_0 \!-\! \bar \Gamma_1 \!-\! \bar \Gamma_2)
}
\sum_{m'} \! D_{m'}
\!+\!
\frac{e \Phi }{2(2{\cal E}_0 - \bar \Gamma_1 - \bar \Gamma_2)}.
\end{eqnarray}
Our
Eqs.~(\ref{capacitance_Eo_value})
and~(\ref{Gamma_def})
allow us to establish the following inequality
\begin{eqnarray}
%%%%%%%%%%%%%%%%%%%%%%%%%%%%%%%%%%%%%%%%%%%%%%%%%%
\label{large_E0}
%%%%%%%%%%%%%%%%%%%%%%%%%%%%%%%%%%%%%%%%%%%%%%%%%%
2 {\cal E}_0 > (\bar{\Gamma}_1 + \bar{\Gamma}_2).
\end{eqnarray}
It implies that all material-specific coefficients in
Eq.~(\ref{sefl_cons_gaps_log})
are positive. Our analysis will be mostly focused on this regime. Yet,
inequality~(\ref{large_E0})
is not fundamental and may be violated. While not identical, these two
limits shares a lot of features, as we will see below.

It is convenient to introduce a BCS-like order parameter as
\begin{eqnarray}
%%%%%%%%%%%%%%%%%%%%%%%%%%%%%%%%%%%%%%%%%%%%%%%%%%
\label{Delta0_def}
%%%%%%%%%%%%%%%%%%%%%%%%%%%%%%%%%%%%%%%%%%%%%%%%%%
\Delta_0
=
2t_0 \exp \left(
	- \frac{2}{\nu_0 (\bar \Gamma_1 + \bar \Gamma_2)}
\right),
\end{eqnarray}
After a simple algebra we derive the self-consistency equations in the final form
\begin{eqnarray}
%%%%%%%%%%%%%%%%%%%%%%%%%%%%%%%%%%%%%%%%%%%%%%%%%%
\label{sefl_cons_gaps_simp}
%%%%%%%%%%%%%%%%%%%%%%%%%%%%%%%%%%%%%%%%%%%%%%%%%%
&&D_m \!\ln \!\left( \!\frac{\Delta_0}{|D_m|} \!\right)\!
\!=\!
-
\frac{{\cal E}_0}{
	\nu_0(\bar \Gamma_1\! +\! \bar \Gamma_2)
	(2{\cal E}_0 \!-\! \bar \Gamma_1\! -\! \bar \Gamma_2)
}
\sum_{m'}\! D_{m'}
\quad\nonumber\\
&&+
\frac{e \Phi }{\nu_0 (2{\cal E}_0 - \bar \Gamma_1 - \bar \Gamma_2)},
\quad
m = 1, \ldots, 4.
\end{eqnarray}
Here, the multi-index $m$ enumerate all four possible choices of the spin
$\sigma = \uparrow \downarrow$,
and valley
$\xi =\pm 1$.
From the formal mathematical point of view, the exact order of this
enumeration does not affect the solution of this set of equations. However,
choosing different order of enumeration we will obtain different physical
phases.

Expression~(\ref{sefl_cons_gaps_simp})
is the central result of the paper. Together with the variational energy
formulas, it allows us to analyze the mean-field phases of biased AB-BLG.

\section{Mean field phases}
%%%%%%%%%%%%%%%%%%%%%%%%%%%%%%%%%%%%%%%%%%%%%%%%%%
\label{MF_phases}
%%%%%%%%%%%%%%%%%%%%%%%%%%%%%%%%%%%%%%%%%%%%%%%%%%

\subsection{Ferroelectric state}
%%%%%%%%%%%%%%%%%%%%%%%%%%%%%%%%%%%%%%%%%%%%%%%%%%
\label{FE_state_stability}
%%%%%%%%%%%%%%%%%%%%%%%%%%%%%%%%%%%%%%%%%%%%%%%%%%

First we discuss the simple ferroelectric (FE) state in the AB-BLG. Namely,
we assess the possibility of spontaneous inter-layer polarization of AB-BLG
at
$e\Phi = 0$.
This question caused some controversy back in early days of bilayer
research. Specifically,
Ref.~\onlinecite{MCCANN2007110}
argued that, unless external electric field is applied, the FE phase
stability is destroyed by large electrostatic energy cost. This was
challenged in
Ref.~\onlinecite{Nandkishore2010},
whose authors proposed that the contributions due to exchange interaction
$\tilde H_{\rm int}$
is sufficient to overcome the electrostatic energy, making the FE phase
stable. Our calculations presented below demonstrate that for realistic
conditions the FE state is likely to be unstable.

We begin our investigation with the observation that in the FE phase all
four
$D_m$'s
have identical absolute values and identical signs. That is
\begin{eqnarray}
%%%%%%%%%%%%%%%%%%%%%%%%%%%%%%%%%%%%%%%%%%%%%%%%%%
\label{FE_ansatz}
%%%%%%%%%%%%%%%%%%%%%%%%%%%%%%%%%%%%%%%%%%%%%%%%%%
D_m = D_{\rm FE} \quad \text{for all }m.
\end{eqnarray}
Using
$\sum_{m'} D_{m'} = 4 D_{\rm FE}$
in
Eqs.~(\ref{sefl_cons_gaps_log}),
one can derive an equation for
$D_{\rm FE}$
\begin{eqnarray}
%%%%%%%%%%%%%%%%%%%%%%%%%%%%%%%%%%%%%%%%%%%%%%%%%%
\label{self_cons_FE_lambda}
%%%%%%%%%%%%%%%%%%%%%%%%%%%%%%%%%%%%%%%%%%%%%%%%%%
&&D_{\rm FE}
\left[
	\lambda_{\rm FE}\ln \left( \frac{2t_0}{|D_{\rm FE} |} \right) + 1
\right]
=
\frac{e \Phi}{2},\\
\nonumber
&&\lambda_{\rm FE}
=
\frac{ \nu_0 (2 {\cal E}_0 \!-\! \bar \Gamma_1 \!-\! \bar \Gamma_2)}{2}
\approx 0.461.
\end{eqnarray}
When no bias voltage is applied, we have
\begin{eqnarray}
%%%%%%%%%%%%%%%%%%%%%%%%%%%%%%%%%%%%%%%%%%%%%%%%%%
\label{self_cons_FE_nofield}
%%%%%%%%%%%%%%%%%%%%%%%%%%%%%%%%%%%%%%%%%%%%%%%%%%
\ln \left( \frac{2t_0}{|D_{\rm FE} |} \right)
=
- \frac{1}{\lambda_{\rm FE}}.
\end{eqnarray}
This equation has formal solution
$|D_{\rm FE} |=\frak{E}$,
where the energy scale
\begin{eqnarray}
%%%%%%%%%%%%%%%%%%%%%%%%%%%%%%%%%%%%%%%%%%%%%%%%%%
\label{energy_scale_def}
%%%%%%%%%%%%%%%%%%%%%%%%%%%%%%%%%%%%%%%%%%%%%%%%%%
\frak{E} = 2t_0 e^{1/\lambda_{\rm FE}}
\end{eqnarray}
is introduced. Yet this root
$\frak{E} \approx 7$\,{\rm eV}
violates the condition
$|D_{\rm FE}|<2t_0$
for our two-band model. One must conclude that, unless bias field is
applied, only trivial solution
$D_{\rm FE} = 0$
for
Eq.~(\ref{self_cons_FE_lambda})
is valid, in agreement with results of
Ref.~\onlinecite{MCCANN2007110}.

What would happen when the external field is applied? In the limit of high
bias voltage when we can neglect interaction,
Eq.~(\ref{self_cons_FE_lambda})
has an obvious solution
$D_{\rm FE} = e \Phi/2$.
That is, the application of the bias voltage opens a gap in the spectrum
proportional to the electric field. This case corresponds to the limit
$\lambda_{\rm FE}\ln{(4t_0/e\Phi)}\ll 1$,
or, equivalently,
\begin{eqnarray}
%%%%%%%%%%%%%%%%%%%%%%%%%%%%%%%%%%%%%%%%%%%%%%%%%%
\label{large_field}
%%%%%%%%%%%%%%%%%%%%%%%%%%%%%%%%%%%%%%%%%%%%%%%%%%
\frac{e\Phi}{2}\gg \frak{E} > 2t_0.
\end{eqnarray}
Inequality~(\ref{large_field})
is inconsistent with the two-band applicability
criterion~(\ref{weak_field_requirement}).
Additionally, for
relation~(\ref{large_field})
to be valid, the electric field between graphene layers
must be at least
$2 \frak{E}/d \approx 4\times 10^8$\,V/cm,
which is quite large.

\begin{figure}
  \includegraphics[width=8.5 cm]{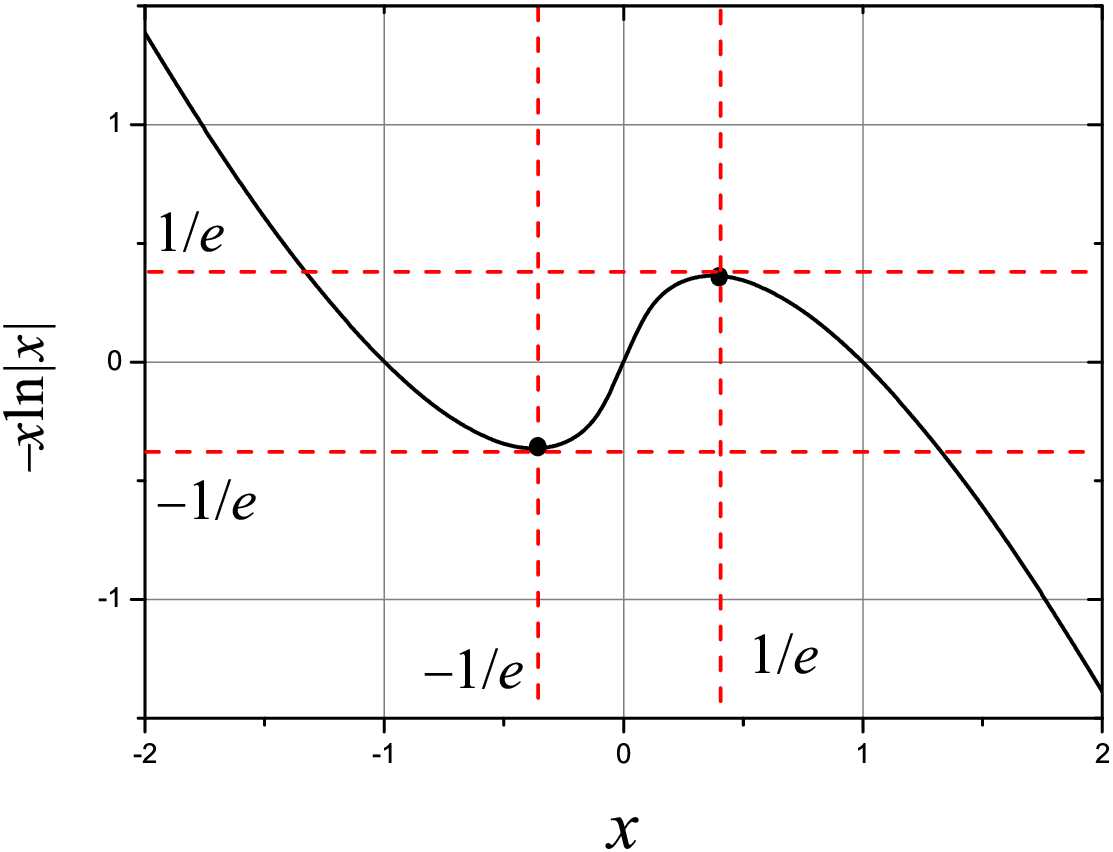}\\
  \caption{Function $y = x \ln (1/|x|)$.
Formally, the derivative of the function diverges at $x = 0$ as $\sim \ln (1/|x|)$.
Yet, due to weakness of this divergence, vertical tangent is essentially unobservable on the plot.
%We see a minimum (maximum) at $y=-1/e$ (at $y=1/e$).
%These extrema separate three types of roots of
%Eq.~(\ref{equation_C}):
%the roots to the left of the minimum (to the right of the maximum) is root
%${\sl x}_-$ (root ${\sl x}_+$). Between the extrema is
%${\sl x}_s$. The equation has three roots only when
%$|C| < 1/e$. When $|C| > 1/e$,
%only one root exists.
%%%%%%%%%%%%%%%%%%%%%%%%%%%%%%%%%%%%%%%%%%%%%%%%%%
\label{Fig1}
%%%%%%%%%%%%%%%%%%%%%%%%%%%%%%%%%%%%%%%%%%%%%%%%%%
}
\end{figure}

At smaller fields, the interaction effects cannot be ignored. It is useful
to rewrite
Eq.~\eqref{self_cons_FE_lambda}
in the form
\begin{eqnarray}
%%%%%%%%%%%%%%%%%%%%%%%%%%%%%%%%%%%%%%%%%%%%%%%%%%
\label{self_cons_FE_lambda2}
%%%%%%%%%%%%%%%%%%%%%%%%%%%%%%%%%%%%%%%%%%%%%%%%%%
\quad
\frac{D_{\rm FE}}{\frak{E}}
\ln \left( \frac{\frak{E}}{|D_{\rm FE} |} \right)
=
\frac{ e \Phi}{2\lambda_{\rm FE} \frak{E}}.
\end{eqnarray}
Thus, we need to analyze the equation
\begin{eqnarray}
%%%%%%%%%%%%%%%%%%%%%%%%%%%%%%%%%%%%%%%%%%%%%%%%%%
\label{log_equation}
%%%%%%%%%%%%%%%%%%%%%%%%%%%%%%%%%%%%%%%%%%%%%%%%%%
x \ln \left( \frac{1}{|x|} \right) = y,
\end{eqnarray}
in the limit
$|y| \ll 1$,
where
$x = D_{\rm FE}/\frak{E}$
and
$y = e\Phi/(2\lambda_{\rm FE} \frak{E})$.
The dependence
$x = x(y)$
can be learned from the plot in
Fig.~\ref{Fig1}.
There we observe three branches of the function
$y = y(x)$:
the first starts from
$x=0$,
the second and third branches pass through
$x= \pm 1$.
The branches merge at
$|x|=1/e$.
It is evident that only the first one is physical and obeys the condition
$|y| \ll 1$.
The first branch corresponds to the required FE state induced by the gate
voltage. As one might expect, the gap
$D_{\rm FE}$
grows monotonically when the voltage grows.

We can see that, for
$|y| > 1/e$,
there is no solution of
Eq.~\eqref{log_equation}
on the physical branch. One can speculate, if this implies that at
$|y|=1/e$
an electric breakdown of the bilayer occurs. However, we must remember that
our model is valid only in the regime
$e\Phi/2,|D_m|\ll 2t_0$,
which does not cover the fields near the extrema.

We already pointed out that, in our model, no spontaneous FE gap
$D_{\rm FE}$
exists. This absence of spontaneous FE order may be traced back to
inequality~(\ref{large_E0}).
Basically, the inequality tells us that the macroscopic electric
polarization of the sample is very costly in terms of energy (large
${\cal E}_0$),
and any potential energy minimization due to exchange contributions (Fock
terms) cannot compensate large electrostatic energy.

At the same time, if
inequality~(\ref{large_E0})
is violated,
$\lambda_{\rm FE}$
is negative, and the energy scale
$\frak{E} = 2 t_0 e^{-1/|\lambda_{\rm FE}|}$
becomes less than
$2t_0$.
This makes the zero-bias FE solution
\begin{eqnarray}
%%%%%%%%%%%%%%%%%%%%%%%%%%%%%%%%%%%%%%%%%%%%%%%%%%
\label{metastable_FE_OP}
%%%%%%%%%%%%%%%%%%%%%%%%%%%%%%%%%%%%%%%%%%%%%%%%%%
|D_{\rm FE}| = \frak{E} = \Delta_0\exp\left[-\frac{2{\cal E}_0}{|\lambda_{\rm FE}|( \bar\Gamma_1 + \bar\Gamma_2)} \right]
\end{eqnarray}
compatible with the two-band model assumption
$|D_{\rm FE}| < 2t_0$,
in agreement with
Ref.~\onlinecite{Nandkishore2010}.
It is easy to check that, as long as
${\cal E}_0$
remains finite, one has
$|D_{\rm FE}| < \Delta_0$.
Thus, violation
of~(\ref{large_E0})
turns the FE phase metastable at
$e \Phi = 0$.
However, the FE cannot become zero-bias ground state as solutions with
zero spontaneous electrostatic polarization have larger order parameter
($\sim \Delta_0$),
and lower energy. It is likely that
$\lambda_{\rm FE} < 0$
regime was modeled numerically in
Ref.~\onlinecite{Jung2011}.

\subsection{Layered antiferromagnetic and other possible states}

Large positive contribution due to the electrostatic energy, which
destabilized the FE phase, can be drastically reduced if the FE phase
requirement~(\ref{FE_ansatz})
is abandoned, and
$D_m$'s
are allowed to vary from one fermion sector to another. Such a situation
corresponds to several orderings, including, in particular, the layered
antiferromagnetic state, as well as other phases with non-zero spin or
valley polarization. To describe them, it is convenient to re-write the set
of
equations~(\ref{sefl_cons_gaps_simp})
as follows
\begin{eqnarray}
%%%%%%%%%%%%%%%%%%%%%%%%%%%%%%%%%%%%%%%%%%%%%%%%%%
\label{self_cons_Hermit_again}
%%%%%%%%%%%%%%%%%%%%%%%%%%%%%%%%%%%%%%%%%%%%%%%%%%
&&\left( \frac{D_{m}}{\Delta_0} \right)
\ln \left(
	\frac{\Delta_0}{|D_{m} |}
\right)
=
\frac{e \Phi} {2 \lambda_{\rm FE} \Delta_0}
-
\\
\nonumber
&&\frac{ {\cal E}_0} {2 \lambda_{\rm FE} ( \bar\Gamma_1 + \bar\Gamma_2) }
\sum_{m'} \left( \frac{D_{m'}}{\Delta_0} \right),
\text{  for  }
m \!=\! 1, \ldots, 4.
\end{eqnarray}
To solve this set of equations, let us introduce dimensionless constant
$\Lambda =
{\cal E}_0/ [2 \lambda_{\rm FE} ( \bar\Gamma_1 + \bar\Gamma_2) ]$.
It can be equivalently expressed as
\begin{eqnarray}
%%%%%%%%%%%%%%%%%%%%%%%%%%%%%%%%%%%%%%%%%%%%%%%%%%
\label{Lambda_definition}
%%%%%%%%%%%%%%%%%%%%%%%%%%%%%%%%%%%%%%%%%%%%%%%%%%
\Lambda =
\frac{ {\cal E}_0}
	{ \nu_0 (2 {\cal E}_0 \!-\! \bar \Gamma_1 \!-\! \bar \Gamma_2)
	( \bar\Gamma_1 + \bar\Gamma_2) }
\approx 2.55.
\end{eqnarray}
This is a very important parameter that controls various physical
properties of AB-BLG. It satisfies
\begin{eqnarray}
%%%%%%%%%%%%%%%%%%%%%%%%%%%%%%%%%%%%%%%%%%%%%%%%%%
\label{Lambda_range}
%%%%%%%%%%%%%%%%%%%%%%%%%%%%%%%%%%%%%%%%%%%%%%%%%%
\frac{1}{2 \gamma} < \Lambda ,
\text{ where}
\quad
\gamma=(\bar{\Gamma}_1+\bar{\Gamma}_2)\nu_0 \approx 0.249.
\end{eqnarray}
Clearly, our
estimate~(\ref{Lambda_definition})
is consistent with
inequality~(\ref{Lambda_range}).
It is also worth keeping in mind that, for the mean field to be valid,
$\gamma/2$
must be smaller than unity. Consequently,
$\Lambda > 1/(2 \gamma) > 0.25$,
which may be viewed as the lowest possible bound on $\Lambda$.

We further define dimensionless applied voltage $V$ and dimensionless gap
parameters
${\sl x}_m$
according to
\begin{eqnarray}
{\sl x}_m = \frac{D_m}{\Delta_0},
\quad
V = \frac{e \Phi} {2 \lambda_{\rm FE} \Delta_0}.
\end{eqnarray}
Using these notations, we rewrite
Eq.~(\ref{self_cons_Hermit_again})
as
\begin{eqnarray}
%%%%%%%%%%%%%%%%%%%%%%%%%%%%%%%%%%%%%%%%%%%%%%%%%%
\label{self_consist_dimensionless}
%%%%%%%%%%%%%%%%%%%%%%%%%%%%%%%%%%%%%%%%%%%%%%%%%%
{\sl x}_m\! \ln \!\left(\! \frac{1}{|{\sl x}_m|}\! \right)
\!\!=\!
V \!\!-\! \Lambda \left(
	{\sl x}_1\! +\! {\sl x}_2\! +\! {\sl x}_3\! +\! {\sl x}_4
\right)\!.
\end{eqnarray}
Here all variables
${\sl x}_m$
can be either positive, or negative, or zero. From the physical point of
view, the term proportional to $\Lambda$ in
Eq.~(\ref{self_consist_dimensionless})
describes the suppression of the inter-layer charge polarization due to
significant electrostatic energy. If the applied potential $V$ is zero,
this energy contribution is the smallest at
$\sum_{m'} {\sl x}_{m'} = 0$.
A finite applied potential $V$ shifts this minimum from zero of
$\,\sum_{m'} {\sl x}_{m'}$
to non-zero value.

The system of
Eqs.~(\ref{self_consist_dimensionless})
includes four unknowns. Mathematical search for its roots is
greatly simplified due to the following feature: all four equations have
identical right-hand side, which is independent of $m$. Thus, for any
$m \ne m'$
\begin{eqnarray}
%%%%%%%%%%%%%%%%%%%%%%%%%%%%%%%%%%%%%%%%%%%%%%%%%%
\label{relation_mm}
%%%%%%%%%%%%%%%%%%%%%%%%%%%%%%%%%%%%%%%%%%%%%%%%%%
{\sl x}_m \ln \left( \frac{1}{|{\sl x}_m|} \right)
=
{\sl x}_{m'} \ln \left( \frac{1}{|{\sl x}_{m'}|} \right).
\end{eqnarray}
An evident solution of the latter equation is
${\sl x}_m = {\sl x}_{m'}$.
Yet, this is not the only possibility. We already discussed function
$y(x) = x \ln (1/|x|)$
in connection with solutions of
Eq.~(\ref{log_equation}),
see
Fig.~\ref{Fig1}:
clearly, the equation
\begin{eqnarray}
%%%%%%%%%%%%%%%%%%%%%%%%%%%%%%%%%%%%%%%%%%%%%%%%%%
\label{equation_C}
%%%%%%%%%%%%%%%%%%%%%%%%%%%%%%%%%%%%%%%%%%%%%%%%%%
{\sl x} \ln \left( \frac{1}{|{\sl x}|} \right) = C,
\end{eqnarray}
may have one, two, or three roots, depending on
$|C|$.
It is convenient to denote the three possible solutions of
Eq.~(\ref{equation_C})
as
${\sl x}_\pm$
and
${\sl x}_s$
according to the following rule:
\begin{eqnarray}
%%%%%%%%%%%%%%%%%%%%%%%%%%%%%%%%%%%%%%%%%%%%%%%%%%
\label{root_interval}
%%%%%%%%%%%%%%%%%%%%%%%%%%%%%%%%%%%%%%%%%%%%%%%%%%
{\sl x}_- \!<\! - 1/e,
\,\,\,
{\sl x}_+ \!>\! 1/e,
\,\,\,
- 1/e < {\sl x}_s < 1/e,
\end{eqnarray}
one root per branch of
$y(x)$.
With this definition, the set of
equations~(\ref{self_consist_dimensionless})
can be reformulated in a parametric representation
\begin{eqnarray}
V = C + \Lambda \left[ n_s {\sl x}_s (C) + n_+ {\sl x}_+ (C) + n_- {\sl x}_- (C) \right],
\end{eqnarray}
where
${\sl x}_\pm = {\sl x}_\pm (C)$
and
${\sl x}_s = {\sl x}_s (C)$
satisfy Eq.~(\ref{equation_C}) and inequalities~(\ref{root_interval}). Non-negative integers
$0 \leqslant n_\pm \leqslant 4$,
$0 \leqslant n_s \leqslant 4$
satisfy the relation
\begin{eqnarray}
n_+ + n_- + n_s = 4.
\end{eqnarray}
The integer
$n_s$
(integer
$n_\pm$)
show how many roots
${\sl x}_s$
(roots
${\sl x}_\pm$)
among four variables
${\sl x}_{1,\ldots,4}$.

There are 15 possible partitions of four into three parts. This specifies
15 solutions of
Eq.~(\ref{self_consist_dimensionless}),
one solution for a given partition. For example, the FE state corresponds
to
$n_s = 4$,
$n_\pm = 0$.
It is easy to check that the layered antiferromagnet is represented by the
choice
$n_s = 0$,
$n_\pm = 2$.
This guarantees that electrostatic effects represented by
$\Lambda \sum_{m'} {\sl x}_{m'}$
are small in the layered antiferromagnet state: in this sum, two terms are
positive and two terms are negative.
However, among these solutions some are physically equivalent. For example,
the transition between phases with
$n_s=4$, $n_{\pm}=0$
and $n_s=0$, $n_{-}=0$, $n_{+}=4$,
occurs at some
$V=4.122$ (for $\Lambda=2.55$)
without any thermodynamic changes and is not a phase transition.
In
addition, the system properties are evidently symmetric with respect to the
sign of the applied voltage. Thus, in what follows we consider only the
case of
$V>0$.

\subsection{Ordered state energy}

Among these solutions, a single ground state must be
identified for every $V$. To find the ground state, we need to calculate the variational
energies of the competing phases, and compare them against each other. To
this end, we re-write the variational
energy~\eqref{Evar_total_herm}
and~\eqref{ephi_energy} in
the form
\begin{eqnarray}
%%%%%%%%%%%%%%%%%%%%%%%%%%%%%%%%%%%%%%%%%%%%%%%%%%
\label{Evar_dim}
%%%%%%%%%%%%%%%%%%%%%%%%%%%%%%%%%%%%%%%%%%%%%%%%%%
\bar{E}
=
\frac{1}{2}\sum_m	
	{\sl x}_m^2 \!\left[
		\ln\! \left(\frac{1}{\delta_0|{\sl x}_m|} \right)
		\!-\!
		\frac{\gamma}{2}
		\ln^2\!\! \left( \frac{1}{\delta_0|{\sl x}_m|} \right)
		\!-\!
		\frac{1}{2}
	\right]
+
\quad
\\
\nonumber
\frac{\bar{\varepsilon}}{8}
\left[	
	\sum_m {\sl x}_m \ln \left( \frac{1}{\delta_0|{\sl x}_m|} \right)
\right]^2
\!\!-\!
V\lambda_{\textrm{FE}}
\sum_m
	{\sl x}_m \!\ln\!\! \left( \frac{1}{\delta_0|{\sl x}_m|} \right)\!,
\end{eqnarray}
where
\begin{eqnarray}
%%%%%%%%%%%%%%%%%%%%%%%%%%%%%%%%%%%%%%%%%%%%%%%%%%
\label{dimensionless}
%%%%%%%%%%%%%%%%%%%%%%%%%%%%%%%%%%%%%%%%%%%%%%%%%%
\bar{E}
=
\frac{E_{\rm var}-E_{\rm MF}(0)}{N_c\nu_0\Delta_0^2},
\quad
\delta_0=\frac{\Delta_0}{2t_0},
\\
%%%%%%%%%%%%%%%%%%%%%%%%%%%%%%%%%%%%%%%%%%%%%%%%%%
\label{epsilon_bar}
%%%%%%%%%%%%%%%%%%%%%%%%%%%%%%%%%%%%%%%%%%%%%%%%%%
\bar{\varepsilon} = {\cal E}_0\nu_0 \approx 0.585.
\end{eqnarray}
When
${\sl x}_m$'s
in
Eq.~\eqref{Evar_dim}
satisfy the self-consistency
equations~\eqref{self_consist_dimensionless},
the expression for the system's energy can be simplified: after rather
straightforward calculations, one can derive
\begin{eqnarray}
%%%%%%%%%%%%%%%%%%%%%%%%%%%%%%%%%%%%%%%%%%%%%%%%%%
\label{bar_E}
%%%%%%%%%%%%%%%%%%%%%%%%%%%%%%%%%%%%%%%%%%%%%%%%%%
\bar{E}
=
-\frac{1}{4}\sum_m {\sl x}_m^2 + \frac{V}{2} \sum_m {\sl x}_m - 2 \lambda_{\rm FE} V^2.
\end{eqnarray}
The first term in this formula represents the mean field condensation energy,
the others are induced by the electrostatic interaction between the external
field and the inter-layer polarization.

\section{Phase diagram}
%%%%%%%%%%%%%%%%%%%%%%%%%%%%%%%%%%%%%%%%%%%%%%%%%%
\label{phase_diagram}
%%%%%%%%%%%%%%%%%%%%%%%%%%%%%%%%%%%%%%%%%%%%%%%%%%
%
\begin{figure}
    \includegraphics[width=8.5 cm]{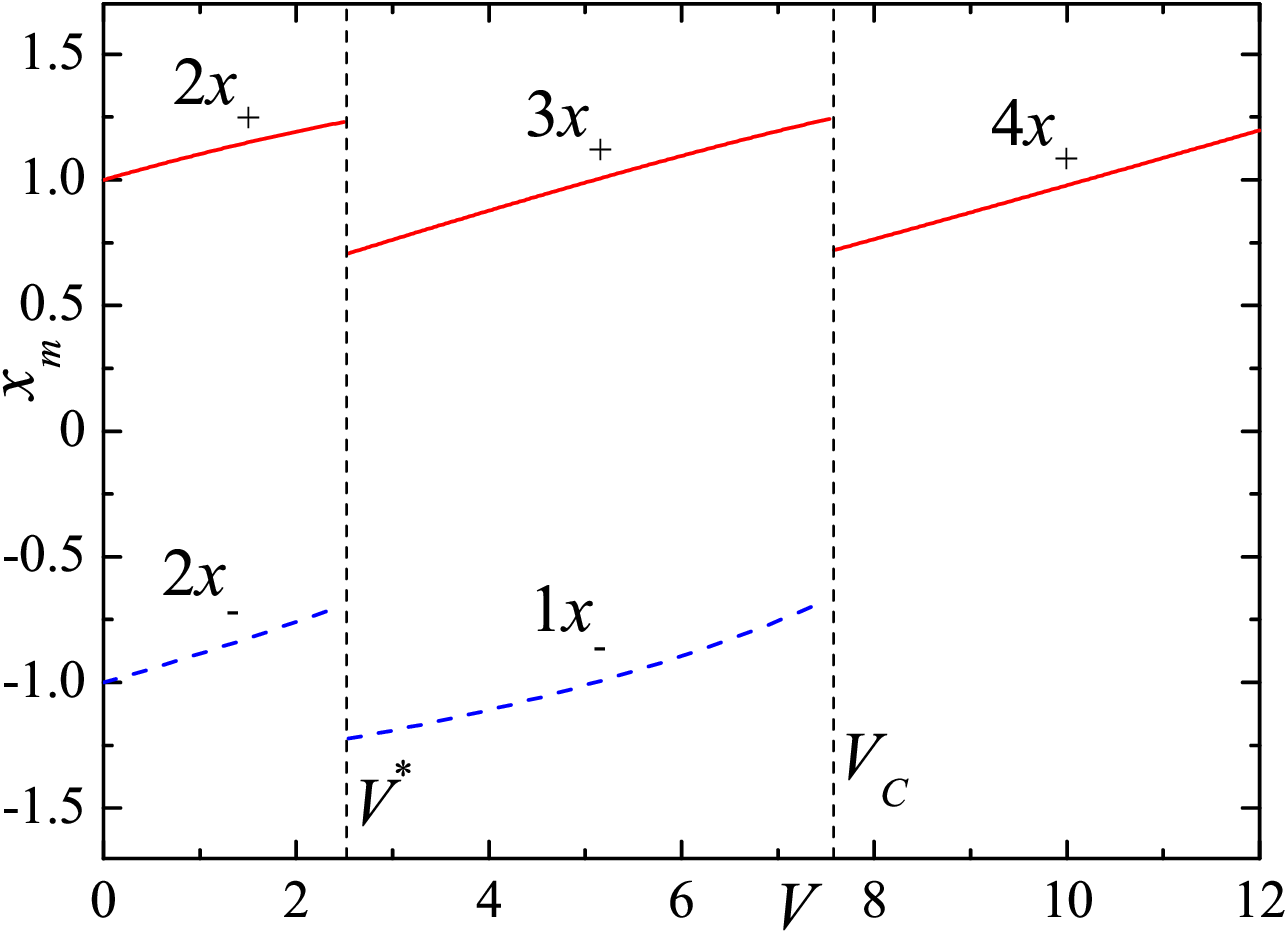}\\
  \caption{The phase diagram of the system at
$V>0$
and the dependence of the order parameters
${\sl x}_m=D_m/\Delta_0$
on the applied voltage. Solid (red) lines represent
${\sl x}_+$,
dashed (blue) lines represent
${\sl x}_-$.
The voltage
$V^*\approx 2.3$
marks the transition between the "antiferromagnetic states" with
$(n_+=2,n_-=2)$
and
$(n_+=3,n_-=1)$.
When
$V>V_C\approx 8.8$
the ``antiferromagnetic orders" disappear and the gap in the spectrum is
due to the layers polarization only.
%%%%%%%%%%%%%%%%%%%%%%%%%%%%%%%%%%%%%%%%%%%%%%%%%%
\label{Figure_OP}
%%%%%%%%%%%%%%%%%%%%%%%%%%%%%%%%%%%%%%%%%%%%%%%%%%
}
\end{figure}
\begin{figure}
    \includegraphics[width=8.5 cm]{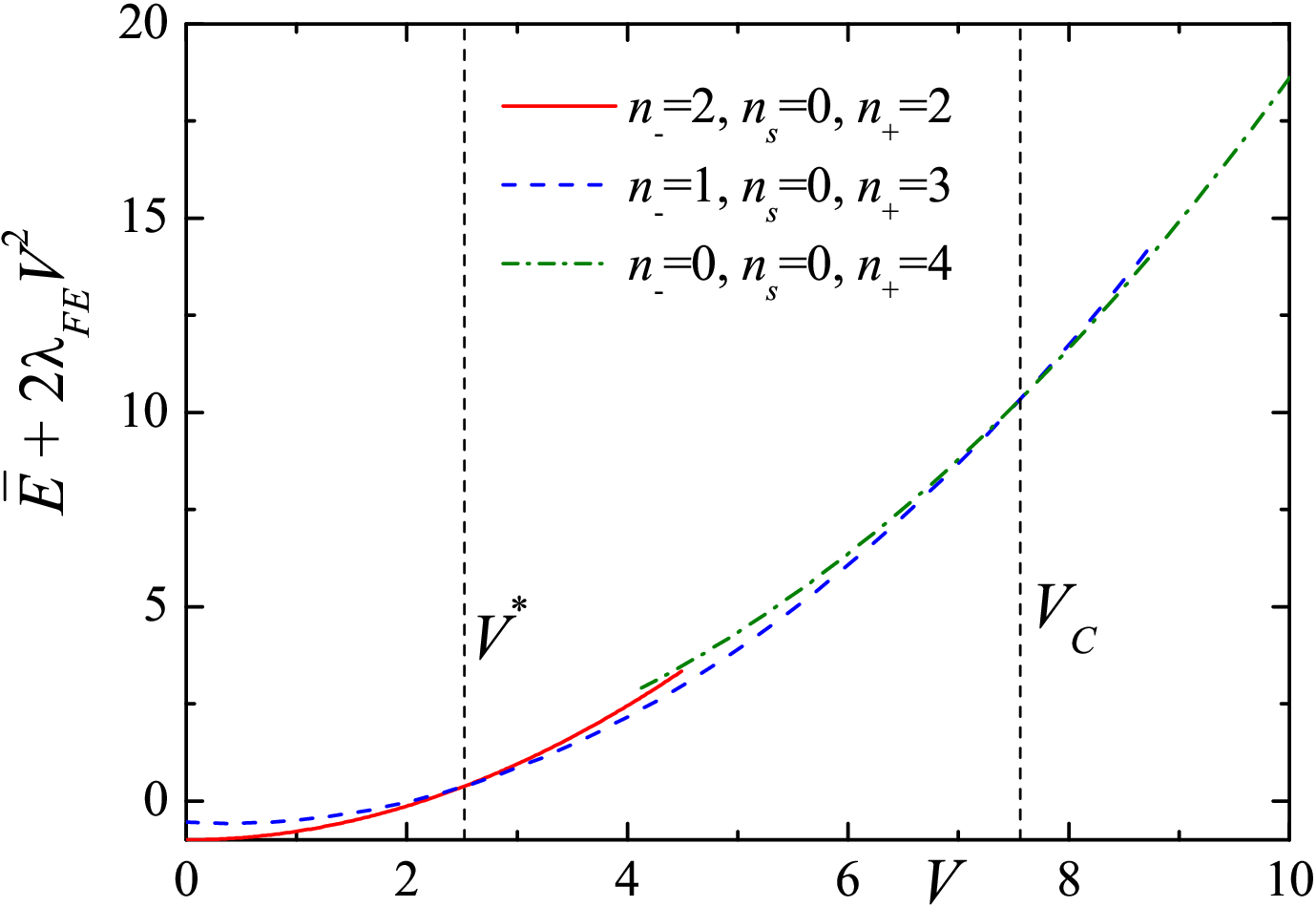}\\
\caption{The system energies as functions of the applied voltage. For
$V \geqslant 0$,
only three ground states are possible, see the legend. The plotted energy is
the sum of the first and the second terms in
definition~(\ref{bar_E}).
The third term in
Eq.~(\ref{bar_E}), while contributing a large offset to the total energy,
does not influence relative stability of phases. It is therefore omitted
for clarity. For
$V<V^*$,
the solid (red) curve is the lowest. Thus, the solution
$n_\pm = 2$
is the ground state. As $V$ increases, a transition into
$n_+ = 3$,
$n_- = 1$
type, followed by the second transition into the FE state
$n_+ = 4$,
$n_- = 0$,
occurs. Note that in the interval
$V^* < V < V_{\rm C}$
the solution
$n_\pm = 2$
disappears, while the FE solution emerges. When $V$ exceeds
$V_{\rm C}$
sufficiently, the solution
$n_+ = 3$,
$n_- = 1$
disappears, leaving the FE solution the only one at elevated $V$.
%%%%%%%%%%%%%%%%%%%%%%%%%%%%%%%%%%%%%%%%%%%%%%%%%%
\label{Fig_energy}
%%%%%%%%%%%%%%%%%%%%%%%%%%%%%%%%%%%%%%%%%%%%%%%%%%
}
\end{figure}

\subsection{General remarks}

In our model, the ordered states depend only on two dimensionless
parameters, $V$ and $\Lambda$. Indeed, of all model parameters, only $V$
and $\Lambda$ enter the dimensionless self-consistency
equations~(\ref{self_consist_dimensionless}).
Competition between ordered states is controlled by
$\bar E$,
the latter being a sum of three terms, of which two (the first and the
second terms) depend explicitly on $V$, and implicitly (through
${\sl x}_m$)
on $V$ and $\Lambda$. One can suggest superficially that
$\lambda_{\rm FE}$
might affect the phase diagram as it enters the third term in
Eq.~(\ref{bar_E}).
Yet, this contribution is the same for all ordered phases, and,
consequently, does not affect their competition at fixed $V$ and $\Lambda$.
Therefore, we conclude that the values of two dimensionless parameters, $V$
and $\Lambda$, determine the ground state order in our model.

\begin{figure}
    \includegraphics[width=8.5 cm]{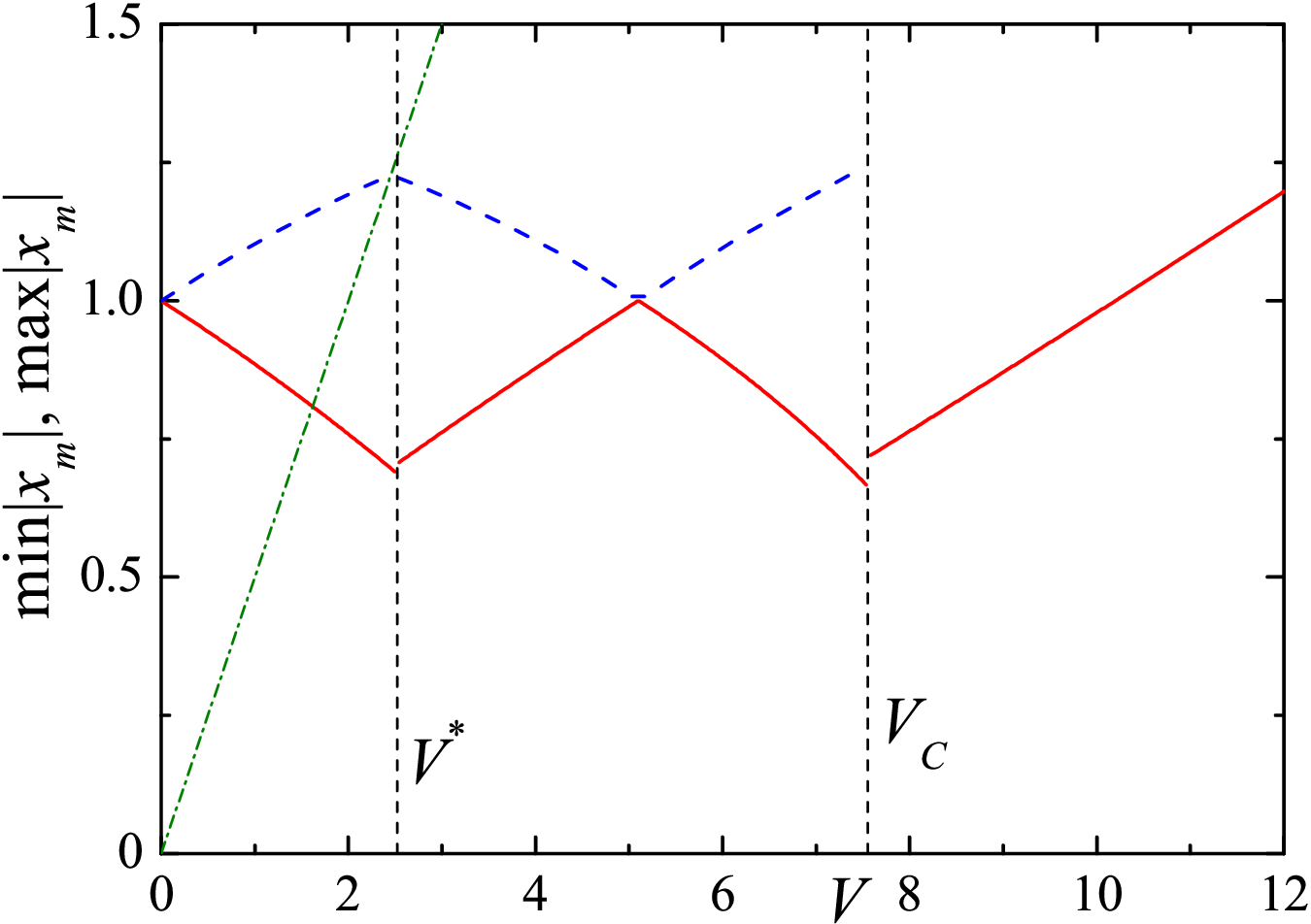}\\
  \caption{Dependence of dimensionless single-electron gap
$\min_m |{\sl x}_m|$
(red solid line) versus dimensionless bias $V$. In the interval
$0 < V < V_{\rm C}$,
both
$n_\pm > 0$,
thus, positive and negative
${\sl x}_m$'s
may have unequal
$|{\sl x}_m|$,
see
Fig.~\ref{Figure_OP}.
For such values of $V$, dashed (blue) line shows
$\max_m |{\sl x}_m|$.
The gap is a non-monotonic function of $V$, with discontinuities at
$V = V^*, V_{\rm C}$.
At
$V = 2 \Lambda$,
the gap has local maximum
${\sl x} = 1$,
see
Sec.~\ref{subsection_Vstar}.
Dash-dotted (green) line is
$\Delta = e\Phi/2$,
or, in dimensionless units,
${\sl x}=\lambda_{\textrm{FE}}V$.
It represents bias-induced gap in the model with no interaction. We see
that interaction effects screen the bare field significantly.
%%%%%%%%%%%%%%%%%%%%%%%%%%%%%%%%%%%%%%%%%%%%%%%%%%
  \label{Fig_gap}
%%%%%%%%%%%%%%%%%%%%%%%%%%%%%%%%%%%%%%%%%%%%%%%%%%
}
\end{figure}

For value of $\Lambda$ as in
Eq.~(\ref{Lambda_definition}),
numerically calculated phase diagram is shown in
Fig.~\ref{Figure_OP}.
The figure also presents the dependence of the order parameters on the
applied voltage
$V>0$.
Figure~\ref{Fig_energy}
shows the energies of the states that become ground state at certain $V$.

We see that, at zero or low bias, the ground state corresponds to the
solution
$n_+=n_-=2$.
As $V$ grows, two phase transitions occur. At
$V^* \approx 2.51$
the solution
$n_+=n_-=2$
is replaced by
$n_+=3$,
$n_-=1$
solution. The transition takes place when the energies of these solutions,
expressed by
Eq.~(\ref{bar_E}),
become equal each other, see
Fig.~\ref{Fig_energy}.

As $V$ exceeds
$V_{\rm C} \approx 7.6$,
solution
$n_+ = 4$,
$n_- = 0$
becomes the ground state. It remains the ground state for even higher
voltages, as long as the model remains valid.

To extrapolate
Figs.~\ref{Figure_OP}
and~\ref{Fig_energy}
to
$V<0$,
we can use the symmetry of the phase diagram relative to
$V = 0$
point: simultaneous substitutions
$V \rightleftarrows -V$
and
$n_+\rightleftarrows n_-$
keep the phase diagram unchanged.

For electronic states in sector
$m=(\xi, \sigma)$,
the value of
$|D_m|$
represents the single-electron gap in this fermion sector. Clearly, the
overall spectral gap, visible in, e.g., transport experiments, is the
smallest of
$|D_m|$'s,
or, in dimensionless units,
$\min_m |{\sl x}_m|$,
see
Fig.~\ref{Fig_gap}.
We observe that the gap is non-monotonic function of the bias. For
comparison, the latter figure plots the non-interacting model gap
$e\Phi/2$.
We see that interaction effects reduce this ``bare'' gap quite
dramatically (except low-bias regime, where spontaneous gap exceeds the
bare one).

\subsection{Linearized self-consistency equations}

Curiously, the phase diagram may be recovered analytically. If we
assume that
$|{\sl x}_m|$
is close to unity, we can approximate
${\sl x}_m \ln (1/|{\sl x}_m|) = - \zeta_m + O(\zeta_m^2)$
for
${\sl x}_m = \pm 1 + \zeta_m$.
Using
Eq.~(\ref{relation_mm}),
one can conclude that all
$\zeta_m$'s
are identical for any $m$. Thus, we can drop subscript and write
$\zeta = \zeta_{m}$.
In this approximation, the self-consistency
conditions~(\ref{self_consist_dimensionless})
become
\begin{eqnarray}
%%%%%%%%%%%%%%%%%%%%%%%%%%%%%%%%%%%%%%%%%%%%%%%%%%
\label{self_cons_linear}
%%%%%%%%%%%%%%%%%%%%%%%%%%%%%%%%%%%%%%%%%%%%%%%%%%
- \zeta \approx V - \Lambda (n_+ - n_-) - 4 \Lambda \zeta,
\end{eqnarray}
and can be solved
\begin{eqnarray}
%%%%%%%%%%%%%%%%%%%%%%%%%%%%%%%%%%%%%%%%%%%%%%%%%%
\label{self_cons_linear_solution}
%%%%%%%%%%%%%%%%%%%%%%%%%%%%%%%%%%%%%%%%%%%%%%%%%%
\zeta \approx \frac{V - \Lambda (n_+ - n_-)}{4 \Lambda - 1}.
\end{eqnarray}
This solution is valid as long as $\zeta$ is small. Also, because of
singularity at
${\sl x} =0$,
linearization is impossible for solutions with
$n_s > 0$.
Fortunately, our numerical data indicates that we can ignore them, as long
as $\Lambda$ is not too small.

Substituting approximate
solutions~(\ref{self_cons_linear_solution})
in
formula~(\ref{bar_E}),
we derive the following approximate expression
\begin{eqnarray}
%%%%%%%%%%%%%%%%%%%%%%%%%%%%%%%%%%%%%%%%%%%%%%%%%%
\label{energy_linearized_approximation}
%%%%%%%%%%%%%%%%%%%%%%%%%%%%%%%%%%%%%%%%%%%%%%%%%%
\bar{E}
\approx
-1 - \frac{2 \lambda_{\rm FE}^2}{\bar \varepsilon}  V^2
+ \left( 1 - \frac{1}{2 \Lambda} \right) \zeta^2.
\end{eqnarray}
This energy depends on the solution type through $\zeta$.

Using the latter formula, we establish
$V^* \approx \Lambda$,
and
$V_{\rm C} \approx 3 \Lambda$.
Both estimates are consistent with our numerical results. If
$|V|$
does not exceed
$V_{\rm C}$
too much, the value of
$|\zeta|$
remains below 0.28, which guarantees good practical accuracy of the linear
approximation.

Alternatively, we can write that the transitions between the phases occur
at the following bias values
\begin{eqnarray}
%%%%%%%%%%%%%%%%%%%%%%%%%%%%%%%%%%%%%%%%%%%%%%%%%%
\label{ePhi_transitions}
%%%%%%%%%%%%%%%%%%%%%%%%%%%%%%%%%%%%%%%%%%%%%%%%%%
e\Phi^* =
\frac{{\cal E}_0}{ \bar\Gamma_1 + \bar\Gamma_2} \Delta_0,
\quad
e\Phi_{\rm C} =
\frac{3 {\cal E}_0}{ \bar\Gamma_1 + \bar\Gamma_2} \Delta_0,
\end{eqnarray}
where
$e\Phi^* = 2 \lambda_{\rm FE} \Delta_0 V^*$,
and
$e\Phi_{\rm C}
= 2 \lambda_{\rm FE} \Delta_0 V_{\rm C}$.

Obviously, there is a ``universal" connection between
$e\Phi^*$
and
$e\Phi_{\rm C}$
\begin{eqnarray}
%%%%%%%%%%%%%%%%%%%%%%%%%%%%%%%%%%%%%%%%%%%%%%%%%%
\label{universal_ratio}
%%%%%%%%%%%%%%%%%%%%%%%%%%%%%%%%%%%%%%%%%%%%%%%%%%
\frac{e\Phi_{\rm C}}{e\Phi^*} = 3,
\end{eqnarray}
which can be used for analysis of experimental or numerical data.

\subsection{Layered antiferromagnet}
%%%%%%%%%%%%%%%%%%%%%%%%%%%%%%%%%%%%%%%%%%%%%%%%%%
\label{layered_AFM_subsection}
%%%%%%%%%%%%%%%%%%%%%%%%%%%%%%%%%%%%%%%%%%%%%%%%%%
%
\begin{figure}
    \includegraphics[width=8.5 cm]{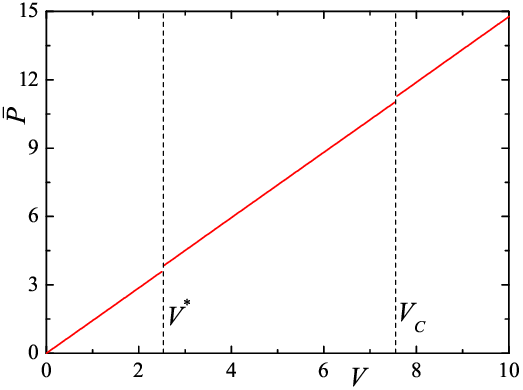}\\
  \caption{Numerically estimated dimensionless polarization
$\bar{P}$
versus the dimensionless bias voltage $V$. The plot for
$\bar{P}$
demonstrates discontinuities at the transition points
$V=V^*$
and
$V=V_{\rm C}$.
Between the transitions, the graph is virtually indistinguishable from
linear function. This behavior is in agreement
with
Eq.~(\ref{polarization_linear}),
attesting the quality of approximate
solution~(\ref{self_cons_linear_solution}).
%%%%%%%%%%%%%%%%%%%%%%%%%%%%%%%%%%%%%%%%%%%%%%%%%%
\label{Fig_polarization}
%%%%%%%%%%%%%%%%%%%%%%%%%%%%%%%%%%%%%%%%%%%%%%%%%%
}
\end{figure}

Both numerical and approximate analytical solutions demonstrate that, for
our choice of $\Lambda$, in the interval
$|V| < V^*$,
the solution
$n_+=n_-=2$
has the lowest energy, see
Fig.~\ref{Fig_energy}.
In this regime,
$\zeta = V/(4 \Lambda - 1)$.

There are
$C_4^2 = 6$
possible assignments of two pluses
($n_+ = 2$)
and two minuses
($n_- = 2$)
among four
${\sl x}_m$'s.
These six choices can be grouped into three pairs, each pair representing
either layered antiferromagnet (LA), or quantum anomalous Hall (QAH) state,
or quantum spin Hall (QSH)
state~\cite{zhand2011hall_state_classific}. %\cite{min_pseudo_fm2008, Jung2011}.
Every pair can be characterized by a certain spontaneous polarization
(spin, valley, or spin-valley), which breaks one of the model's
symmetries~\cite{rozhkov2025ab_su4}.
States within a pair are connected to each other by inversion of the
polarization's sign.

Due to our model's simplifications, all these states have the same energy,
and the ground state smoothly evolves as $V$ changes. Yet, it is necessary
to remember that a more complicated
framework~\cite{haritonov_afm2012, cvetkovic_multi2012, lemonic_rg_nemat_long2012}
can lift this degeneracy: previous calculations
indicate~\cite{haritonov_afm2012, lemonic_rg_nemat_long2012}
that LA is the ground state of AB-BLG at
$V=0$.
If the degeneracy is removed, a possibility of first-order transitions
between the LA/QAH/QSH states must be kept in mind as small energies
separating the latter phases might be voltage-dependent.

Clearly, the applied voltage makes polarization $P$, introduced by
Eq.~(\ref{polarization_def}),
finite.
%\begin{eqnarray}
%%%%%%%%%%%%%%%%%%%%%%%%%%%%%%%%%%%%%%%%%%%%%%%%%%
%\label{dimless_polarization_def}
%%%%%%%%%%%%%%%%%%%%%%%%%%%%%%%%%%%%%%%%%%%%%%%%%%%
%\bar{P} = - \frac{\partial \bar{E}}{\partial V}
%=
%\lambda_{\textrm{FE}} \sum_m  {\sl x}_m \!\ln\!\! \left( \frac{1}{\delta_0|{\sl x}_m|} \right).
%\end{eqnarray}
It is easy to demonstrate that,
for
${\sl x}_m$
satisfying the self-consistency
equations~(\ref{self_consist_dimensionless}),
dimensionless polarization
$\bar{P} = - \partial \bar{E}/\partial V$
is
\begin{eqnarray}
%%%%%%%%%%%%%%%%%%%%%%%%%%%%%%%%%%%%%%%%%%%%%%%%%%
\label{polarization}
%%%%%%%%%%%%%%%%%%%%%%%%%%%%%%%%%%%%%%%%%%%%%%%%%%
\bar{P} = 4 \lambda_{\textrm{FE}} V - \sum_m  {\sl x}_m.
\end{eqnarray}
Numerically estimated
$\bar{P}(V)$
is plotted in
Fig.~\ref{Fig_polarization}.
On the other hand, using
approximation~(\ref{self_cons_linear_solution}),
one can derive
\begin{eqnarray}
%%%%%%%%%%%%%%%%%%%%%%%%%%%%%%%%%%%%%%%%%%%%%%%%%%
\label{polarization_linear}
%%%%%%%%%%%%%%%%%%%%%%%%%%%%%%%%%%%%%%%%%%%%%%%%%
\bar{P}
\approx
4 \left( \lambda_{\textrm{FE}} - \frac{1}{4 \Lambda - 1} \right) V
+
\frac{n_+ - n_-}{4 \Lambda - 1}.
\end{eqnarray}
Substituting
$n_+ = n_- = 2$
into this formula, we find
$\bar{P} \approx 1.41 V$,
which agrees well with the data in
Fig.~\ref{Fig_polarization}.

At
$V = 0$,
we have
$|{\sl x}_m| = 1$
for all $m$. As $V$ deviates from zero, disparity emerges: if
${\sl x}_m > 0$,
its absolute value
$|{\sl x}_m |$
increases (decreases) for positive (negative) $V$. For
${\sl x}_m < 0$,
the trend is opposite. This feature has an important consequence.
Single-electron gap
$E_{\rm g} (V) = 2 \Delta_0 \min_m |{\sl x}_m (V)|$
becomes a function of voltage, see
Fig.~\ref{Fig_gap}.
In the LA phase, the gap is the largest at
$V=0$.
The gap modulation may be detected in transport experiments.

\subsection{Transition at $V = V^*$}
%%%%%%%%%%%%%%%%%%%%%%%%%%%%%%%%%%%%%%%%%%%%%%%%%%
\label{subsection_Vstar}
%%%%%%%%%%%%%%%%%%%%%%%%%%%%%%%%%%%%%%%%%%%%%%%%%%

When $V$ crosses
$V^*$,
the lowest energy solution is
$n_+=3$,
$n_-=1$,
see
Fig~\ref{Fig_energy}.
Likewise, solution
$n_+=1$,
$n_-=3$
becomes the ground state at
$V = - V^* < 0$.
These solutions are stable in the range
$V^*<|V|<V_C$.
Thus, condition
$|V| = V^*$
marks locations of first-order transitions.

There are
$C_4^1 = 4$
possible sign assignments for
$n_+=3$,
$n_-=1$
solution. Each assignment represents one of four QAH-type phases, which
form a degenerate quartet. Another quartet of symmetry-related QAH phases
emerge at
$V < -V^*$.
In terminology of
Ref.~\onlinecite{rozhkov2025ab_su4},
these are class~III states. When smoothly extrapolated to
$V = 0$,
all these phases have finite valley, spin, spin-valley, and charge
polarizations. Finite charge polarization undermines their stability at low
bias. Yet, at finite $V$, the same feature improves their stability
relative to the LA.

Both
$\bar P$
and
$|{\sl x}_m|$
exhibit jumps at the transition point, see
Fig.~\ref{Fig_polarization}
and
Fig.~\ref{Figure_OP}.
If we use
approximation~(\ref{polarization_linear}),
the polarization discontinuity is controlled by the second term:
$n_+ - n_-$
changes by 2 at the transition. Thus, the discontinuity can be estimated as
$2/(4 \Lambda - 1) \approx 0.22$,
in agreement with the numerical calculations.

Beside the discontinuities at the transitions points, the data in
Fig.~\ref{Fig_gap}
reveals that, at certain bias all
$|{\sl x}_m|$
are identical for both positive and negative
${\sl x}_m$.
It is easy to check that such a condition can be fulfilled when
$|{\sl x}_m| = 1$
for all $m$. This further implies that the right-hand side of
Eq.~(\ref{self_consist_dimensionless})
vanishes. For
$n_+ = 3$,
$n_- = 1$,
this is only possible when bias is equal to
$V_{\rm max} = 2 \Lambda$,
or, equivalently,
\begin{eqnarray}
%%%%%%%%%%%%%%%%%%%%%%%%%%%%%%%%%%%%%%%%%%%%%%%%%%
\label{maximum_gap}
%%%%%%%%%%%%%%%%%%%%%%%%%%%%%%%%%%%%%%%%%%%%%%%%%%
e\Phi_{\rm max} =
\frac{2 {\cal E}_0}{ \bar\Gamma_1 + \bar\Gamma_2} \Delta_0.
\end{eqnarray}
This is exact relation, and not a consequence of
approximation~(\ref{self_cons_linear_solution}).
A ``universal" ratio
\begin{eqnarray}
%%%%%%%%%%%%%%%%%%%%%%%%%%%%%%%%%%%%%%%%%%%%%%%%%%
\label{universal_ratio2}
%%%%%%%%%%%%%%%%%%%%%%%%%%%%%%%%%%%%%%%%%%%%%%%%%%
\frac{e\Phi_{\rm max}}{e\Phi^*} = 2,
\end{eqnarray}
on the other hand, is an approximate relation since it relies on
Eq.~(\ref{ePhi_transitions}),
the latter being derived using linearization.

Examining
Fig.~\ref{Fig_gap},
one can check that the single-electron gap is a non-monotonic
function of $V$ within this phase: the maximum gap is
$| {\sl x}_m| = 1$,
or, in dimensional units,
$\Delta_0$.
This value is reached at
$e \Phi = e \Phi_{\rm max}$.

\subsection{Finite-bias ferroelectric state}

When $V$ crosses $V_{\rm C}$,
solution $n_+ = 4$
becomes the ground state. It is non-degenerate and represents
the FE phase. It may also be
classified~\cite{zhand2011hall_state_classific}
as a quantum valley Hall state.

For our $\Lambda$ estimate, the FE is unstable at zero voltage, yet, it
becomes stable when external electric field is applied. As with the QAH
phases discussed in
subsection~\ref{subsection_Vstar},
finite polarization of the FE phase pushes its energy upward at low $V$,
while at large bias nonzero $P$ is the most important factor of the FE
state stabilization.

The polarization is discontinuous at
$V = V_{\rm C}$,
as the numerical data in
Fig.~\ref{Fig_polarization}
attest.
Formula~(\ref{polarization_linear})
shows that the discontinuity is 0.22, the same as the discontinuity at
$V = V^*$.

The gap changes discontinuously at the transition, see
Fig.~\ref{Fig_gap}.
Beyond
$V_{\rm C}$,
the gap monotonically increases when
$|V|$
grows.

\section{Discussion}
%%%%%%%%%%%%%%%%%%%%%%%%%%%%%%%%%%%%%%%%%%%%%%%%%%
\label{sec::discussion}
%%%%%%%%%%%%%%%%%%%%%%%%%%%%%%%%%%%%%%%%%%%%%%%%%%

\subsection{Comparison with experiment}
%%%%%%%%%%%%%%%%%%%%%%%%%%%%%%%%%%%%%%%%%%%%%%%%%%
\label{subsect_on_experiment}
%%%%%%%%%%%%%%%%%%%%%%%%%%%%%%%%%%%%%%%%%%%%%%%%%%

To compare our results with experiment, we need to determine the
zero-bias order parameter size
$\Delta_0$.
It is possible to use the
definition~(\ref{Delta0_def})
and evaluate
$\Delta_0 \approx 0.27$\,meV.
This is order of magnitude smaller than the
reported~\cite{Velasco2012,Bao2012,Freitag2012}
value of
$\sim 2-4$\,meV.
Such a difference between the experiment and theory is not surprising. Due to non-analytical dependence of
$\Delta_0$
on
$\gamma/2$,
even moderate errors in our knowledge of the coupling constant strength may
severely skew the value of
$\Delta_0$.
Additionally, the pre-exponential energy scale is not immediately known. In
the two-band theory we are limited by the pre-exponential being of order of
$t_0$,
which is fairly small energy. As we will see below, in the Hartree-Fock
calculations of
Ref.~\onlinecite{Jung2011}
the pre-exponential energy is drastically larger than
$t_0$,
leading to unrealistically large values of
$\Delta_0$.
In such a situation, it appears more prudent to use order parameter value
from experiment. Thus, below we will assume that
$\Delta_0 = 2$\,meV.

Our theory predicts several features that may be tested against experiment,
at least in principle. We know that the bias induces phase transitions
between various thermodynamic states of AB-BLG. Also, the field changes the
value of single-electron gap: the gap is a non-monotonic function of
$e\Phi$.
Such behavior have identifiable experimental consequences, as the gap is
often affects low-temperature transport measurements. In this regard, let
us mention experiment of
Ref.~\onlinecite{Geisenhof2022}
that does show non-monotonicity of resistance as a function of bias field,
see Fig.~2(b) of the latter paper.

Unfortunately, the measured pattern of resistance versus bias cannot be
completely understood within the framework of our theory. Specifically, the
variation of the resistance of AB-BLG sample does not trace the variation
of the gap in Fig.~\ref{Fig_gap}.
Yet, partial interpretation of the data is still possible.

On the experimental curve for device~A in Ref.~\onlinecite{Geisenhof2022}, one can see a well-defined maximum
at zero bias, which may be associated with the gap that our calculations find at
$e\Phi = 0$.
%Unfortunately, this is the only unambiguously identifiable element of that plot.

Away from the zero-bias state, the resistance data shows symmetric minima
at
$\sim \pm 20$\,mV/nm,
which correspond to
\begin{eqnarray}
%%%%%%%%%%%%%%%%%%%%%%%%%%%%%%%%%%%%%%%%%%%%%%%%%%
\label{experimental_min_ePhi}
%%%%%%%%%%%%%%%%%%%%%%%%%%%%%%%%%%%%%%%%%%%%%%%%%%
e\Phi_{\rm min}^{\rm exper} \approx \pm 6.7\,{\rm meV}.
\end{eqnarray}
Identifying the minimum of resistance with the gap minimum at
$e\Phi^*$,
that is, equating
$e\Phi^* = e\Phi_{\rm min}^{\rm exper}$,
and assuming that the low-temperature gap size is
$\Delta_0 = 2$\,meV,
as agreed above, it is possible to evaluate the ratio
\begin{eqnarray}
%%%%%%%%%%%%%%%%%%%%%%%%%%%%%%%%%%%%%%%%%%%%%%%%%%
\label{experimental_ratio}
%%%%%%%%%%%%%%%%%%%%%%%%%%%%%%%%%%%%%%%%%%%%%%%%%%
\left( \frac{{\cal E}_0}{ \bar\Gamma_1 + \bar\Gamma_2} \right)_{\rm exper}
\approx 3.4.
\end{eqnarray}
The latter number is roughly consistent with our theoretical estimate
$\sim 2.3$,
which follows from
Eqs.~(\ref{capacitance_Eo_value})
and~(\ref{Gamma_def}).
Additionally, since
Eq.~(\ref{experimental_ratio})
does not violate
inequality~(\ref{large_E0}),
we conclude that both $\Lambda$ and
$\lambda_{\rm FE}$
are positive, and the FE state is absolutely unstable at zero bias, as we
discussed in
Sec.~\ref{FE_state_stability}.

Using ``universal"
relations~(\ref{universal_ratio})
and~(\ref{universal_ratio2}),
it is possible to estimate the maximum gap bias as
$e \Phi_{\rm max}^{\rm exper}
= 2e \Phi_{\rm min}^{\rm exper} \approx \pm 13$\,meV,
which is equivalent to transverse field of
$\pm 40$\,mV/nm.
Further, the theory indicates that, at
$\pm 60$\,mV/nm,
the gap minimum corresponding to
$e\Phi_{\rm C}$
must be visible. Yet, instead of these features, the resistance plot shows
a shoulder (or broad maximum) for the field to the left of
-30\,mV/nm,
or, equivalently, for
$e\Phi < - 10$\,{\rm meV}.
%\begin{eqnarray}
%%%%%%%%%%%%%%%%%%%%%%%%%%%%%%%%%%%%%%%%%%%%%%%%%%
%\label{experimental_shoulder}
%%%%%%%%%%%%%%%%%%%%%%%%%%%%%%%%%%%%%%%%%%%%%%%%%%
%-20\,{\rm meV} < e\Phi < - 10\,{\rm meV}.
%\end{eqnarray}

Since no data at high bias
$|e \Phi| > 20$\,{\rm meV}
%$- 20\,{\rm meV} < e\Phi < 7\,{\rm meV}$
were reported, the properties of the high-bias FE state remain untested.
Specifically, the expected steady increase of single-electron gap with
increasing $V$ beyond
$V_{\rm C}$
is not apparent in the data of
Ref.~\onlinecite{Geisenhof2022}.

At the same time, we must remember that the energy differences between the
competing phases are very small. Indeed, examining
Fig.~\ref{Fig_energy},
we see that the competing solutions' energies differ by less than
$\sim \nu_0 \Delta_0^2 = 2 \times 10^{-8}$\,eV
per unit cell. In addition, both
$n_\pm = 2$
solution and
$n_+ = 3$,
$n_- = 1$
solution each represent several degenerate phases, see discussion in
subsections~\ref{layered_AFM_subsection}
and~\ref{subsection_Vstar}.

In this situation, various factors may alter the relative stability of the
competing phases and thereby modify the phase diagram. Furthermore, domain
walls between degenerate or nearly degenerate states can arise. Because
such domain walls may be conducting, transport measurements may constitute
an unreliable probe of the single-electron gap.

In our considerations we neglect the interlayer hopping
$t_3\approx 0.3$~eV
between the non-dimer sites. Taking into account this process introduces
the so-called trigonal warping: the parabolic dispersion near the Dirac
points is replaced at low energy by a structure consisting of four Dirac
cones. This effects can be characterized by an energy
scale~\cite{bilayer_review2016}
$\varepsilon_L=t_0(t_3/t)^2$.
The value $\varepsilon_L$ is about 1~meV or even lower. This energy scale
is significantly smaller than the gap $\Delta_0$, which is estimated to be
about 2-4~meV based on experimental
data~\cite{veligura2012}.
Thus, we can assume that the obtained results are robust against the effect
of the trigonal warping, at least qualitatively.

Taking into account the nearest neighbor hoping between dimer and non-dimer
sites breaks electron-hole symmetry. However, the value of the
corresponding hoping integral, 
$t_4$, 
is smaller than 
$t_3$ 
(see, e.g., the
review~\onlinecite{bilayer_review2016}
and references therein). Note that the existence of the exact electron-hole
symmetry is not crucial for our formalism, this effect is small, and can be
neglected.

\subsection{Order parameter identification}
%%%%%%%%%%%%%%%%%%%%%%%%%%%%%%%%%%%%%%%%%%%%%%%%%%
\label{subsection_OP}
%%%%%%%%%%%%%%%%%%%%%%%%%%%%%%%%%%%%%%%%%%%%%%%%%%

Theoretical studies of low-temperature electronic properties of AB-BLG
indicate that the system can host various ordered states, some of them
gapful, some of them have no gap. We additionally limit our investigation
to gapful order parameters, with no inter-valley coherence. Yet, even these
restrictions leave fairly extensive list of symmetry-breaking types.
Distinguishing them in experiment is a significant challenge. 

As previously noted, all order parameters in our model can be interpreted
as transverse polarizations in terms of one of four possible ``flavors''
(electric charge, spin, valley, and spin-valley). Thus, in principle,
knowledge of the inter-layer flavor imbalances can determine the nature of
the order. The necessary measurements must be (i)~spin-sensitive, and
(ii)~resolve valley-specific information. 

Between these two rather demanding criteria, condition~(ii) appears to be
particularly difficult since purely single-site measurement
$\propto \langle 
	d^\dag_{i a \sigma} ({\bf r}) d^{\vphantom{\dag}}_{ia\sigma'} ({\bf r})
\rangle$
is insufficient to discern symmetry violations at individual valleys.
Indeed, the symmetry-breaking part of this average is always a linear
combination of elements in matrix
$\sum_\xi \langle \hat\Xi_\xi \rangle$,
to which both valleys contribute equally. One can additionally fix a finite
lattice vector ${\bf a}$, and measure two-site average
$\langle d^\dag_{i a \sigma} ({\bf r}) d^{\vphantom{\dag}}_{ia\sigma'} ({\bf r} + {\bf a}) \rangle$,
which is a linear function of 
$\sum_\xi e^{i\xi ({\bf K}_\xi \cdot {\bf a})} \langle \hat\Xi_\xi \rangle$.
By combining these two measurements, valley-specific data can be derived.
Alternatively, measurements at several ${\bf a}$'s could be utilized for
the same purpose.

This summary highlights the significant difficulties encountered in the
identification of symmetry violations within AB-BLG models. It underscores
the necessity of relying on a diverse array of indirect evidence in such
contexts.

\subsection{Comparison with Hartree-Fock simulations}
%%%%%%%%%%%%%%%%%%%%%%%%%%%%%%%%%%%%%%%%%%%%%%%%%%
\label{subsection_Hartree_Fock}
%%%%%%%%%%%%%%%%%%%%%%%%%%%%%%%%%%%%%%%%%%%%%%%%%%

While exhaustive comparison with the experiment is challenging at this
point, our conclusions agree quite well with extensive numerical
Hartree-Fock simulations of
AB-BLG, presented in
Ref.~\onlinecite{Jung2011}.
Specifically, Fig.~5 of Ref.~\onlinecite{Jung2011} is analogous to Fig.~\ref{Fig_energy} of the present paper.
It shows that the low-bias AF phase (our
$n_\pm = 2$
solution) is replaced by Fi phase (represents our
$n_+ = 3$,
$n_- = 1$
solution). The latter is ultimately replaced by F phase (that is, the FE
phase in our terminology) at higher external fields. It is easy to
check that bias fields at which these transitions occur satisfy
``universality''
relation~(\ref{universal_ratio}).

Examining lower panel of Fig.~6 in
Ref.~\onlinecite{Jung2011},
we clearly see a gap maximum inside the Fi phase. In terms of the bias, the
maximum is located exactly in the middle of the Fi phase, precisely in
agreement with
relation~(\ref{universal_ratio2}).
Note that the gap value at this maximum is exactly the same as zero-bias
gap (both are
$\sim 33$\,meV).
This feature is clearly visible in our
Fig.~\ref{Fig_gap},
where the gap at
$V=0$
and the gap at
$V = 2 \Lambda \approx 5.1$
are equal to each other.

It is possible to extract model parameters from the data in Figs.~5
and~6 of
Ref.~\onlinecite{Jung2011}.
This is described in detail in Appendix~\ref{Hartree_Fock_appendix}. We would like to draw attention of a reader to two circumstances.
(i)~It is demonstrated that the Hartree-Fock data corresponds to the regime of
negative
$\lambda_{\rm FE}$
and $\Lambda$. This means that
inequality~(\ref{large_E0})
is violated. We already commented above that this inequality is not of the
fundamental nature. However, both our estimates and experimental data in
Ref.~\onlinecite{Geisenhof2022}
indicate that
relation~(\ref{large_E0})
does hold true. For some unclear reason, the Hartree-Fock simulation
underestimates
${\cal E}_0$
energy (elevated dielectric constant assumed for the simulations is not
large enough to account for the discrepancy). Smallness of
${\cal E}_0$
relative to
$\bar{\Gamma}_{1,2}$
makes the FE phase metastable at
$e \Phi = 0$,
which indeed what
Ref.~\onlinecite{Jung2011}
reported, see Fig.~5 there.

Additionally, (ii)~the order parameters
values in
Ref.~\onlinecite{Jung2011}
are very large: 16\,meV and even larger. On one hand, an ordering of such a
strength was never observed in AB-BLG. On the other hand, to fit this gap
value into our framework, we need to assume large
($\sim 8.5$\,eV)
pre-exponential factor in the expression for
$\Delta_0$.
This huge pre-exponential energy scale points toward certain weakness of a
Hartree-Fock approach, partially explaining why the Hartree-Fock scheme produces
unrealistically high order parameters values.

\subsection{Fractional metallicity at weak doping}

While a detailed study of doped AB-BLG is still in progress, several
relevant observations about doping can already be made. For
$0 < |V| < V_{\rm C}$,
the spectrum contains sectors with smaller and larger gaps (see
Fig.~\ref{Fig_gap}).
Upon doping, additional carriers enter the sector with the smallest gap
first. This immediately implies that weak doping of the
$n_+ = 3$,
$n_- = 1$
phases stabilizes a quarter-metal state -- that is, a metallic phase with
complete polarization of the Fermi surface in both
spin and valley quantum
numbers~\cite{aa_quarter_met2021, rakhmanov2023ab_FraM}.

At weak doping and finite $V$, the fate of the
$n_\pm = 2$
state can follow either of two routes, depending on microscopic details. If
the added electrons distribute evenly between the two (degenerate) low-gap
sectors, a half-metal
emerges~\cite{bilayer_half-metal2013numeric_MF}.
Alternatively, the doping itself may lift the sector
degeneracy~\cite{keldysh1965half_metal,
keldysh_kopaev_collection2024half_metal, rodionov2025half_metal},
resulting in a quarter-metal state.

This overall picture is partially consistent with the investigation of
Ref.~\onlinecite{bilayer_half-metal2013numeric_MF},
whose the numerical mean-field framework was restricted to
two independent order parameters, rather than four. As a result of this
limitation, only the half-metallic phase was captured.
Similar conclusions were drawn in
Ref.~\onlinecite{baima2018dft_half_met_graphene}
based on density-functional calculations.

\section{Conclusions}
%%%%%%%%%%%%%%%%%%%%%%%%%%%%%%%%%%%%%%%%%%%%%%%%%%
\label{conclusion_section}
%%%%%%%%%%%%%%%%%%%%%%%%%%%%%%%%%%%%%%%%%%%%%%%%%%

We formulated a simple mean field theory that describes electronic
properties of undoped AB-BLG placed in transverse electric field. The model
accounts not only for short-range screened Coulomb interaction between
electrons, but for the effects of the electrostatic energy associated with
inter-layer polarization of the bilayer. We additionally introduce four
independent order parameters, each order parameter is associated with a
specific values of valley and spin quantum numbers. (Current-carrying order
parameters are ignored from the start.) A set of four self-consistency
equations is derived and solved. Despite physical dissimilarity, many
ordered phases are found to be degenerate. We are able to classify these
phases in accordance with their spin and valley projections. The external
field induces transitions between various phases of the system.
Single-electron gap is a non-monotonic function of the bias, and, for fixed
bias, the gaps for different valleys and/or spin projections do not have to
be the same. We map the phase diagram: at low bias, the ground state is
likely to be layered antiferromagnet, at sufficiently large bias the
bilayer is in the ferroelectric state. Comparison with experiment and other
theoretical works is performed. The proposed framework may be relevant for
study of doped AB-BLG.

%%%%%%%%%%%%%%%%%%%%%%%%%%%%%%%%%%%%%%%%%%%%%%%%%%%%%%%%%%%%%%%%%%%%%%%%%%%%%%%%%%%%%%%%%%%%%%%%%%%%%%%%%%%%%%%%%%%%%%%%%%%%%%%%%%%%%%
\appendix
%%%%%%%%%%%%%%%%%%%%%%%%%%%%%%%%%%%%%%%%%%%%%%%%%%%%%%%%%%%%%%%%%%%%%%%%%%%%%%%%%%%%%%%%%%%%%%%%%%%%%%%%%%%%%%%%%%%%%%%%%%%%%%%%%%%%%%%

\section{Summation for MF energy}
%%%%%%%%%%%%%%%%%%%%%%%%%%%%%%%%%%%%%%%%%%%%%%%%%%
\label{appendix_sum}
%%%%%%%%%%%%%%%%%%%%%%%%%%%%%%%%%%%%%%%%%%%%%%%%%%

Let us evaluate the sum in Eq.~\eqref{E_MF_Herm}
\begin{eqnarray}
E (D) = - \sum_{\bf k} \sqrt{\varepsilon_{\bf k}^2 + D^2}.
\end{eqnarray}
We use the following trick: we write
$E (D) = E(0) + \int_0^D E' (\Tilde D) d \Tilde D$,
where
$E(0)$
is a constant independent of
$D$,
while the derivative
$E'(\Tilde D)$
is equal to
\begin{eqnarray}
%%%%%%%%%%%%%%%%%%%%%%%%%%%%%%%%%%%%%%%%%%%%%%%%%%
\label{app_E_derivative}
%%%%%%%%%%%%%%%%%%%%%%%%%%%%%%%%%%%%%%%%%%%%%%%%%%
\!\!\!\!\!\!\!\!E' (\Tilde D)
=
- \sum_{\bf k} \frac{\Tilde D}{\sqrt{\varepsilon_{\bf k}^2 + \Tilde D^2}}
\approx
- N_c \nu_0 \Tilde D \ln \left( \frac{2t_0}{|\Tilde D|} \right).
\end{eqnarray}
Indeed, if
$D\ll t_0$,
we obtain
\begin{eqnarray}
%%%%%%%%%%%%%%%%%%%%%%%%%%%%%%%%%%%%%%%%%%%%%%%%%%
\label{log_function}
%%%%%%%%%%%%%%%%%%%%%%%%%%%%%%%%%%%%%%%%%%%%%%%%%%
\nonumber
&&\frac{1}{N_c}\!
\sum_{\bf k}
	\frac{\Tilde D }{\sqrt{ \varepsilon_{\bf k}^2 + \Tilde D^2 }}
=
\int \frac{ S_0 d^2 {\bf k}}{(2\pi)^2}
	\frac{\Tilde D}{\sqrt{ \varepsilon_{\bf k}^2 + \Tilde D^2 }}
=
\\
&&\frac{t_0}{2 \sqrt{3} \pi t^2}
\int_0^{t_0} d \varepsilon
	\frac{\Tilde D }{\sqrt{ \varepsilon^2 + \Tilde D^2 }}
\approx
\nu_0 \Tilde D \ln \left( \frac{2t_0}{|\Tilde D|} \right),
\\
\nonumber
&&\nu_0 = \frac{t_0}{2 \sqrt{3} \pi t^2},
\end{eqnarray}
and we derive Eq.~(\ref{app_E_derivative}). If we then integrate over
$\Tilde D$,
we get
\begin{eqnarray}
E (D) &=& E(0)
-
N_c \nu_0 \int_0^D
	\Tilde D \ln \left( \frac{2t_0}{|\Tilde D|} \right) d \Tilde D
\\
&=&
\nonumber
E(0)
-
\frac{1}{2} N_c \nu_0 D^2
\left[
	\ln \left( \frac{2t_0}{|D|} \right)
	+
	\frac{1}{2}
\right].
\end{eqnarray}

\section{Variation energy minimization}
%%%%%%%%%%%%%%%%%%%%%%%%%%%%%%%%%%%%%%%%%%%%%%%%%%
\label{app_Delta_VS_Xi}
%%%%%%%%%%%%%%%%%%%%%%%%%%%%%%%%%%%%%%%%%%%%%%%%%%

We minimize the energy
$E_{\rm var}$,
Eq.~(\ref{E_var_full}), over
$\hat \Delta_\xi$
in a general case. Such a minimization allows us to derive corresponding self-consistency equation for the order parameters. In so doing, we differentiate
$E_{\rm var} = E_{\rm var} (\hat \Delta_\xi )$
over the matrix elements
$\hat \Delta_\xi$.
We start with
$\langle H_0 \rangle$,
Eq.~\eqref{MF_Hamiltonian_def}, and obtain
\begin{eqnarray}
%%%%%%%%%%%%%%%%%%%%%%%%%%%%%%%%%%%%%%%%%%%%%%%%%%
\label{H0_diff}
%%%%%%%%%%%%%%%%%%%%%%%%%%%%%%%%%%%%%%%%%%%%%%%%%%
\frac{\partial \langle H_0 \rangle }{
	\partial \left[\hat \Delta_\xi \right]_{\sigma \sigma' }}
=
\frac{\partial E_{\rm MF}}{
	\partial \left[\hat \Delta_\xi \right]_{\sigma \sigma' }}
+
\sum_{\bf k}
	\langle
		\left[ \hat\Xi_{{\bf k} \xi}^\dag \right]_{\sigma' \sigma}
	\rangle
+
\\
\nonumber
\sum_{\bf k}
{\rm Tr}\! \left(
	\frac{\partial
		\langle \hat\Xi_{{\bf k} \xi}^{\dag}  \rangle}
		{\partial \left[ \hat \Delta_\xi \right]_{\sigma \sigma'}}
	\hat{\Delta}_{ \xi}^{\vphantom{\dag}}
	+
	\hat{\Delta}_{ \xi}^\dag
	\frac{\partial
		\langle \hat\Xi_{{\bf k} \xi}^{\vphantom{\dag}}  \rangle}
		{\partial \left[ \hat \Delta_\xi \right]_{\sigma \sigma'}}
\right)\!.
\end{eqnarray}

The first two terms in the right-hand side of Eq.~(\ref{H0_diff}) cancel each other, as Eq.~(\ref{Hellmann_Feyn_theorem_matrxi}) indicates. Therefore, the following is true
\begin{eqnarray}\label{eq::difer_H0}
\partial \langle H_0 \rangle
=
\sum_{\bf k}
	{\rm Tr}\!
	\left(
		\hat{\Delta}_{ \xi}^{\vphantom{\dag}}
		\partial \langle \hat\Xi_{{\bf k} \xi}^{\dag}  \rangle
		+
		\hat{\Delta}_{ \xi}^\dag
		\partial \langle
			\hat\Xi_{{\bf k} \xi}^{\vphantom{\dag}}
		\rangle
	\right),
\end{eqnarray}
where
$\partial X$
means
$\partial X/\partial \left[ \hat \Delta_\xi \right]_{\sigma \sigma'}$.

Next we differentiate
$\langle H_{\rm int} \rangle$,
Eq.~\eqref{interaction_Wick}, and get
\begin{widetext}
\begin{eqnarray}
%%%%%%%%%%%%%%%%%%%%%%%%%%%%%%%%%%%%%%%%%%%%%%%%%%
\label{interaction_diff}
%%%%%%%%%%%%%%%%%%%%%%%%%%%%%%%%%%%%%%%%%%%%%%%%%%
\partial \langle H_{\rm int} \rangle
=
\frac{{\cal E}_0}{4N_c}
\left(
	\sum_{{\bf k}' \xi' }
		{\rm Tr} \langle
				\hat\Xi_{{\bf k}' \xi'}^{\vphantom{\dag}}
			\rangle
			+
		{\rm Tr} \langle \hat\Xi_{{\bf k}' \xi'}^\dag \rangle
\right)
\left(
	\sum_{\bf k}
		{\rm Tr}\,
		\partial \langle
				\hat\Xi_{{\bf k} \xi}^{\vphantom{\dag}}
			\rangle
		+
		{\rm Tr}
		\partial \langle
				\hat\Xi_{{\bf k} \xi}^\dag
			\rangle
\right)
-
\\
\nonumber
\frac{1}{N_c}
\sum_{\mathbf{kk}'}\left[
	\bar \Gamma_{1} {\rm Tr}\!
	\left(
		\langle \hat\Xi_{{\bf k} \xi}^\dag \rangle
		\partial \langle
				\hat\Xi_{{\bf k}' \xi}^{\vphantom{\dag}}
			\rangle
		+
		\langle \hat\Xi_{{\bf k}' \xi}^{\vphantom{\dag}} \rangle
		\partial \langle \hat\Xi_{{\bf k} \xi}^\dag \rangle
	\right)
	+
	\bar \Gamma_{2} {\rm Tr}\!
	\left(
		\langle \hat\Xi_{{\bf k} \xi}^\dag \rangle
		\partial \langle \hat\Xi_{{\bf k}' \xi}^\dag \rangle
		+
		\langle \hat\Xi_{{\bf k} \xi}^{\vphantom{\dag}} \rangle
		\partial
		\langle \hat\Xi_{{\bf k}' \xi}^{\vphantom{\dag}} \rangle
	\right)\right].
\end{eqnarray}
Combining the expressions for
$\partial \langle H_0 \rangle$
and
$\partial \langle H_{\rm int} \rangle$,
we derive
\begin{eqnarray}
\partial E_{\rm var}
=
\sum_{\bf k}
	{\rm Tr} \left\{
		\partial \langle \hat\Xi_{{\bf k} \xi}^{\dag}  \rangle
	\left[
		\hat{\Delta}_{ \xi}^{\vphantom{\dag}}
		-
		\frac{1}{N_c}
		\sum_{\mathbf{k}'}
			\bar \Gamma_{1}
			\langle
				\hat\Xi_{{\bf k}' \xi}^{\vphantom{\dag}}
			\rangle
			+
			\bar \Gamma_{2}
			\langle \hat\Xi_{{\bf k}' \xi}^\dag \rangle
		+
		\frac{{\cal E}_0}{4}
		\sum_{\xi'}
			\left(
				{\rm Tr}
				\langle
					\hat\Xi_{{\bf k}' \xi'}^{\vphantom{\dag}}
				\rangle
				+
				{\rm Tr}
				\langle \hat\Xi_{{\bf k}' \xi'}^\dag \rangle
			\right) \mathbb{I}_2
	\right]
	\right\}
+ {\rm c.c.}
\end{eqnarray}
If the expression in the square brackets vanishes, so does
$\partial E_{\rm var}$.
Thus, the extrema of the variation energy can be found at
\begin{eqnarray}\label{OP_Tr_cap_Gamma_0}
\hat{\Delta}_{ \xi}^{\vphantom{\dag}}
=
\frac{1}{N_c}
\sum_{\mathbf{k}'}\left(
	\bar \Gamma_{1}
	\langle \hat\Xi_{{\bf k}' \xi}^{\vphantom{\dag}} \rangle
	+
	\bar \Gamma_{2} \langle \hat\Xi_{{\bf k}' \xi}^\dag \rangle\right)
-
	\frac{{\cal E}_0}{4} \mathbb{I}_2
	\sum_{\xi'}
		{\rm Tr}
		\langle
			\hat\Xi_{{\bf k}' \xi'}^{\vphantom{\dag}}
			+
			\hat\Xi_{{\bf k}' \xi'}^\dag
		\rangle.
\end{eqnarray}
To account for finite bias, we can shift
$\hat \Delta_\xi$
by
$(e \Phi/2)\mathbb{I}_2$
\begin{eqnarray}
%%%%%%%%%%%%%%%%%%%%%%%%%%%%%%%%%%%%%%%%%%%%%%%%%%
\label{OP_Tr_cap_Gamma}
%%%%%%%%%%%%%%%%%%%%%%%%%%%%%%%%%%%%%%%%%%%%%%%%%%
\hat{\Delta}_{ \xi}^{\vphantom{\dag}}
\rightarrow
\hat{\Delta}_{ \xi}^{\vphantom{\dag}} - \frac{e\Phi}{2} \mathbb{I}_2.
\end{eqnarray}

Formula~\eqref{OP_Tr_cap_Gamma_0},
with substitution
rule~\eqref{OP_Tr_cap_Gamma},
may be viewed as a self-consistency equation in our mean field
approximation. Yet it is more convenient to invert them, expressing
$\sum_{\bf k} \langle \hat\Xi_{{\bf k} \xi}^{\vphantom{\dag}} \rangle$
and
$\sum_{\bf k} \langle \hat\Xi_{{\bf k} \xi}^\dag \rangle$
as functions of
$\hat \Delta^{\vphantom{\dag}}_\xi$
and
$\hat \Delta^\dag_\xi$.
In this case only one non-linear term emerges in the self-consistency equation.

As an initial step toward this objective, we define
\begin{eqnarray}
%%%%%%%%%%%%%%%%%%%%%%%%%%%%%%%%%%%%%%%%%%%%%%%%%%
\label{D_defin}
%%%%%%%%%%%%%%%%%%%%%%%%%%%%%%%%%%%%%%%%%%%%%%%%%%
\hat{\cal D}_\xi
=
\hat\Delta_\xi
+
\frac{{\cal E}_0}{4}
\frac{1}{N_c}
\sum_{{\bf k}'\xi'}
	{\rm Tr}\,
	\langle
		\hat\Xi_{{\bf k}' \xi'}^{\vphantom{\dag}}
		+
		\hat\Xi_{{\bf k}' \xi'}^\dag
	\rangle
\hat{\mathbb{I}}_2
- \frac{e \Phi}{2} \hat{\mathbb{I}}_2.
\end{eqnarray}
According to
Eqs.~(\ref{OP_Tr_cap_Gamma_0})
and~\eqref{OP_Tr_cap_Gamma}, matrices
$\hat{\cal D}_\xi^{\vphantom{\dag}}$,
$\hat{\cal D}_\xi^\dag$
satisfy the following relations
\begin{eqnarray}
%%%%%%%%%%%%%%%%%%%%%%%%%%%%%%%%%%%%%%%%%%%%%%%%%%
\label{D_VS_Xi}
%%%%%%%%%%%%%%%%%%%%%%%%%%%%%%%%%%%%%%%%%%%%%%%%%%
\hat{\cal D}_\xi^{\vphantom{\dag}}
=
\frac{1}{N_c}
\sum_{{\bf k}'}\left(
	\bar\Gamma_{1}
	\langle \hat\Xi_{{\bf k}' \xi}^{\vphantom{\dag}} \rangle
	+
	\bar\Gamma_{2} \langle \hat\Xi_{{\bf k}' \xi}^\dag \rangle\right),
\quad
\hat{\cal D}_\xi^\dag
=
\frac{1}{N_c}
\sum_{{\bf k}'}\left(
	\bar\Gamma_{1} \langle \hat\Xi_{{\bf k}' \xi}^\dag \rangle
	+
	\bar\Gamma_{2}
	\langle \hat\Xi_{{\bf k}' \xi}^{\vphantom{\dag}}  \rangle\right).
\end{eqnarray}
Solving these equations, we obtain
\begin{eqnarray}
\frac{1}{N_c}
\sum_{{\bf k}'}
	\langle \hat\Xi_{{\bf k}' \xi}^{\vphantom{\dag}} \rangle
=
\frac{1}{\bar \Gamma_1^2 - \bar \Gamma_2^2}
\left(
	\bar \Gamma_1 \hat{\cal D}_\xi^{\vphantom{\dag}}
	-
	\bar \Gamma_2 \hat{\cal D}_\xi^\dag
\right),
\quad
\frac{1}{N_c}
\sum_{{\bf k}'}
	\langle \hat\Xi_{{\bf k}' \xi}^\dag \rangle
=
\frac{1}{\bar \Gamma_1^2 - \bar \Gamma_2^2}
\left(
	\bar \Gamma_1 \hat{\cal D}_\xi^\dag
	-
	\bar \Gamma_2 \hat{\cal D}_\xi^{\vphantom{\dag}}
\right).
\end{eqnarray}
Substituting definition~(\ref{D_defin}) in the latter equation, we derive after a straightforward algebra
\begin{eqnarray}
%%%%%%%%%%%%%%%%%%%%%%%%%%%%%%%%%%%%%%%%%%%%%%%%%%
\label{Xi_VS_Delta2}
%%%%%%%%%%%%%%%%%%%%%%%%%%%%%%%%%%%%%%%%%%%%%%%%%%
\frac{(\bar \Gamma_1 + \bar \Gamma_2)}{N_c}
\sum_{{\bf k}'}
	\langle \hat\Xi_{{\bf k}' \xi}^{\vphantom{\dag}} \rangle
=
\frac{1}{\bar \Gamma_1 - \bar \Gamma_2}
\left(
	\bar \Gamma_1 \hat \Delta_\xi^{\vphantom{\dag}}
	-
	\bar \Gamma_2 \hat\Delta_\xi^\dag
\right)
+
\frac{{\cal E}_0}{4}
\frac{1}{N_c}
\sum_{{\bf k}'\xi'}
	{\rm Tr}\,
	\langle
		\hat\Xi_{{\bf k}' \xi'}^{\vphantom{\dag}}
		+
		\hat\Xi_{{\bf k}' \xi'}^\dag
	\rangle
\hat{\mathbb{I}}_2
-
\frac{e \Phi}{2}
\hat{\mathbb{I}}_2,
\end{eqnarray}
Our task is not complete yet: the trace of
$\langle \hat\Xi \rangle$
must be eliminated in the right-hand side of the latter formula. To express this trace in terms of the order parameter matrix, we take trace of both sides of Eq.~(\ref{Xi_VS_Delta2})
\begin{eqnarray}
%%%%%%%%%%%%%%%%%%%%%%%%%%%%%%%%%%%%%%%%%%%%%%%%%%
\label{Xi_VS_Delta_Tr1}
%%%%%%%%%%%%%%%%%%%%%%%%%%%%%%%%%%%%%%%%%%%%%%%%%%
\frac{(\bar \Gamma_1 + \bar \Gamma_2)}{N_c}
\sum_{{\bf k}'}
	{\rm Tr}\, \langle \hat\Xi_{{\bf k}' \xi}^{\vphantom{\dag}} \rangle
=
\frac{1}{\bar \Gamma_1 - \bar \Gamma_2}
\left(
	\bar \Gamma_1 {\rm Tr}\, \hat \Delta_\xi^{\vphantom{\dag}}
	-
	\bar \Gamma_2 {\rm Tr}\, \hat\Delta_\xi^\dag
\right)
+
\frac{{\cal E}_0}{2}
\frac{1}{N_c}
\sum_{{\bf k}' \xi'}
	{\rm Tr}\,
	\langle
		\hat\Xi_{{\bf k}' \xi'}^{\vphantom{\dag}}
		+
		\hat\Xi_{{\bf k}' \xi'}^\dag
	\rangle
-
e \Phi.
\end{eqnarray}
Performing summation over
$\xi$,
we derive
\begin{eqnarray}
%%%%%%%%%%%%%%%%%%%%%%%%%%%%%%%%%%%%%%%%%%%%%%%%%%
\label{Xi_VS_Delta_Tr2}
%%%%%%%%%%%%%%%%%%%%%%%%%%%%%%%%%%%%%%%%%%%%%%%%%%
\frac{(\bar \Gamma_1 + \bar \Gamma_2)}{N_c}
\sum_{{\bf k}' \xi'}
	{\rm Tr}\, \langle
		\hat\Xi_{{\bf k}'\xi'}^{\vphantom{\dag}}
		+
		\hat\Xi_{{\bf k}'\xi'}^\dag
	\rangle
=
\sum_{\xi'}
	{\rm Tr}\, \left(
		\hat \Delta_{\xi'}^{\vphantom{\dag}}
		+
		\hat\Delta_{\xi'}^\dag
	\right)
+
2{ \cal E}_0
\frac{1}{N_c}
\sum_{{\bf k}' \xi'}
	{\rm Tr}\,
	\langle
		\hat\Xi_{{\bf k}' \xi'}^{\vphantom{\dag}}
		+
		\hat\Xi_{{\bf k}' \xi'}^\dag
	\rangle
-
 4 e \Phi.
\end{eqnarray}
Therefore
\begin{eqnarray}
%%%%%%%%%%%%%%%%%%%%%%%%%%%%%%%%%%%%%%%%%%%%%%%%%%
\label{Xi_VS_Delta_Tr3_prime}
%%%%%%%%%%%%%%%%%%%%%%%%%%%%%%%%%%%%%%%%%%%%%%%%%%
\frac{1}{N_c}
\sum_{{\bf k}' \xi'}
	{\rm Tr}\, \langle
		\hat\Xi_{{\bf k}'\xi'}^{\vphantom{\dag}}
		+
		\hat\Xi_{{\bf k}'\xi'}^\dag
	\rangle
=
\frac{4 e \Phi}{2{\cal E}_0 - \bar \Gamma_1 - \bar \Gamma_2}
-
\frac{1}{2{\cal E}_0 - \bar \Gamma_1 - \bar \Gamma_2}
\sum_{\xi'}
	{\rm Tr}\, \left(
		\hat \Delta_{\xi'}^{\vphantom{\dag}}
		+
		\hat\Delta_{\xi'}^\dag
	\right).
\end{eqnarray}
We can use Eq.~(\ref{diff_rho_Xi}) to express
$\rho_{10} - \rho_{20}$
in terms of the order parameter as follows
\begin{eqnarray}
%%%%%%%%%%%%%%%%%%%%%%%%%%%%%%%%%%%%%%%%%%%%%%%%%%
\label{n_VS_Delta}
%%%%%%%%%%%%%%%%%%%%%%%%%%%%%%%%%%%%%%%%%%%%%%%%%%
\frac{\rho_{10} - \rho_{20}}{N_c}
=
- \frac{1}{ N_c}
\sum_{{\bf k}' \xi'}
	{\rm Tr}\, \langle
		\hat\Xi_{{\bf k}'\xi'}^{\vphantom{\dag}}
		+
		\hat\Xi_{{\bf k}'\xi'}^\dag
	\rangle
=
-
\frac{4 e \Phi}{2{\cal E}_0 - \bar \Gamma_1 - \bar \Gamma_2}
+
\frac{1}{2{\cal E}_0 - \bar \Gamma_1 - \bar \Gamma_2}
\sum_{\xi'}
	{\rm Tr}\, \left(
		\hat \Delta_{\xi'}^{\vphantom{\dag}}
		+
		\hat\Delta_{\xi'}^\dag
	\right).
\end{eqnarray}
Substituting
expression~(\ref{Xi_VS_Delta_Tr3_prime})
for the trace into the right-hand side of
Eq.~(\ref{Xi_VS_Delta2}),
we establish
\begin{eqnarray}
%%%%%%%%%%%%%%%%%%%%%%%%%%%%%%%%%%%%%%%%%%%%%%%%%%
\label{Xi_VS_Delta_Tr4}
%%%%%%%%%%%%%%%%%%%%%%%%%%%%%%%%%%%%%%%%%%%%%%%%%%
&&\frac{1}{N_c}
\sum_{{\bf k}'}
	\langle \hat\Xi_{{\bf k}' \xi}^{\vphantom{\dag}} \rangle
=
\frac{1}{\bar \Gamma_1^2 - \bar \Gamma_2^2}
\left(
	\bar \Gamma_1 \hat \Delta_\xi^{\vphantom{\dag}}
	-
	\bar \Gamma_2 \hat\Delta_\xi^\dag
\right)
-
\\
\nonumber
&&\left\{
	\frac{{\cal E}_0/4}{
		(\bar \Gamma_1 + \bar \Gamma_2)
		(2{\cal E}_0 - \bar \Gamma_1 - \bar \Gamma_2)
	}
	\sum_{\xi'}
		{\rm Tr}\, \left(
			\hat \Delta_{\xi'}^{\vphantom{\dag}}
			+
			\hat\Delta_{\xi'}^\dag
		\right)
	-
	\frac{{\cal E}_0 e \Phi}{
		(\bar \Gamma_1 + \bar \Gamma_2)
		(2{\cal E}_0 - \bar \Gamma_1 - \bar \Gamma_2)
	}
\right\}
\hat{\mathbb{I}}_2
-
\frac{e \Phi}{2(\bar \Gamma_1 + \bar \Gamma_2)}
\hat{\mathbb{I}}_2,
\end{eqnarray}
Thus
\begin{eqnarray}
%%%%%%%%%%%%%%%%%%%%%%%%%%%%%%%%%%%%%%%%%%%%%%%%%%
\label{Xi_VS_Delta_Tr_APP}
%%%%%%%%%%%%%%%%%%%%%%%%%%%%%%%%%%%%%%%%%%%%%%%%%%
\nonumber
\frac{1}{N_c}
\sum_{{\bf k}'}
	\langle \hat\Xi_{{\bf k}' \xi}^{\vphantom{\dag}} \rangle
&=&
\frac{1}{\bar \Gamma_1^2 - \bar \Gamma_2^2}
\left(
	\bar \Gamma_1 \hat \Delta_\xi^{\vphantom{\dag}}
	-
	\bar \Gamma_2 \hat\Delta_\xi^\dag
\right)
-
\frac{{\cal E}_0}{
	4(\bar \Gamma_1 + \bar \Gamma_2)
	(2{\cal E}_0 - \bar \Gamma_1 - \bar \Gamma_2)
}
\sum_{\xi'}
	{\rm Tr}\, \left(
		\hat \Delta_{\xi'}^{\vphantom{\dag}}
		+
		\hat\Delta_{\xi'}^\dag
	\right)
\hat{\mathbb{I}}_2\\
&+&
\frac{e \Phi}
{2({2\cal E}_0 - \bar \Gamma_1 - \bar \Gamma_2)}
\hat{\mathbb{I}}_2.
\end{eqnarray}
This is the desired relation between the anomalous averages and the order parameter.

We can use Eq.~(\ref{Hellmann_Feyn_result}) and replace
$\frac{1}{N_c}\sum_{{\bf k}'} 	\langle \hat\Xi_{{\bf k}' \xi}^{\vphantom{\dag}} \rangle$
by the function of
$\hat{\Delta}_\xi^{\vphantom{\dag}}$
and
$\hat{\Delta}_\xi^\dag$.
As a result we obtain non-linear equations, which can be solved to calculate the order parameter matrices
$\hat{\Delta}_\xi^{\vphantom{\dag}}$
and
$\hat{\Delta}_\xi^\dag$.
If we assume that the order parameter is Hermitian,
$\hat{\Delta}_\xi^{\vphantom{\dag}} = \hat{\Delta}_\xi^\dag$,
then, such equations read as
\begin{eqnarray}
%%%%%%%%%%%%%%%%%%%%%%%%%%%%%%%%%%%%%%%%%%%%%%%%%%
\label{sefl_cons_Hermit}
%%%%%%%%%%%%%%%%%%%%%%%%%%%%%%%%%%%%%%%%%%%%%%%%%%
\frac{1}{2N_c} \sum_{\bf k}
	\frac{\hat \Delta_{ \xi}^{\vphantom{\dag}} }
	{\sqrt{
		\varepsilon_{\bf k}^2
		+ 		
		\hat \Delta_{ \xi}^2
	}}
=
\frac{\hat \Delta_\xi^{\vphantom{\dag}}}{\bar \Gamma_1 + \bar \Gamma_2}
-
\frac{{\cal E}_0}{
	2(\bar \Gamma_1 + \bar \Gamma_2)
	(2{\cal E}_0 - \bar \Gamma_1 - \bar \Gamma_2)
}
\hat{\mathbb{I}}_2
\sum_{\xi'}
	{\rm Tr}\, \hat \Delta_{\xi'}^{\vphantom{\dag}}
+
\frac{e \Phi }
{2(2{\cal E}_0 - \bar \Gamma_1 - \bar \Gamma_2)}
\hat{\mathbb{I}}_2.
\end{eqnarray}
\end{widetext}
As the order parameter is Hermitian, we diagonalize it by a suitable
unitary transformation. Thus, instead of two matrix equations (one per
valley), four coupled scalar equations emerge. Thus, after summation over
$\mathbf{k}$
and evident transformations we obtain from
Eq.~\eqref{sefl_cons_Hermit}
the self-consistency
equation~\eqref{sefl_cons_gaps_log}.

\section{Effective constants for the Hartree-Fock calculations of Jung et al.}
%%%%%%%%%%%%%%%%%%%%%%%%%%%%%%%%%%%%%%%%%%%%%%%%%%
\label{Hartree_Fock_appendix}
%%%%%%%%%%%%%%%%%%%%%%%%%%%%%%%%%%%%%%%%%%%%%%%%%%

In this Appendix we provide a detailed explanation of how model parameters
discussed in
Sec.~\ref{sec::discussion}
were extracted from the Hartree-Fock numerical data of
Ref.~\onlinecite{Jung2011},
cited below as I.

Scrutinizing Figs.~5 and~6 of I, we establish that two transitions occur
at external field of
$\sim 2.5$\,mV/nm
and
$\sim 7.5$\,mV/nm.
Multiplying these values by the inter-layer distance $d$, we find
$e\Phi^* = 0.84$\,meV,
$e\Phi_{\rm C} = 2.5$\,meV.
Similarly, one can establish that, between these transitions, the gap
reaches its maximum value when the field is
$\sim 5$\,mV/nm.
Thus,
$e\Phi_{\rm max} = 1.7$\,meV.

Using  Eq.~(\ref{ePhi_transitions}),
we obtain
\begin{eqnarray}
%%%%%%%%%%%%%%%%%%%%%%%%%%%%%%%%%%%%%%%%%%%%%%%%%%
\label{Jung_ratio}
%%%%%%%%%%%%%%%%%%%%%%%%%%%%%%%%%%%%%%%%%%%%%%%%%%
\frac{{\cal E}_0}{\bar{\Gamma}_1 + \bar{\Gamma}_2}
=
\frac{e\Phi^*}{\Delta_0}
=
2.5 \times 10^{-2},
\end{eqnarray}
where
$\Delta_0 = 33$\,meV
was used, see Fig.~6 of I. That is,
inequality~(\ref{large_E0})
is drastically violated, making $\Lambda$ and
$\lambda_{\rm FE}$
negative. As we
explained in
Sec.~\ref{FE_state_stability},
negative
$\lambda_{\rm FE}$
implies that the FE phase is metastable at vanishing bias. This conclusion
is in agreement with data in Figs.~5 and~6 of I, which both show the FE
branch extended to zero external field.

In the zero bias limit, the FE phase is characterized by
$\Delta_{\rm FE} = 25$\,meV,
visible in Fig.~6 of I. This, in conjunction with our
Eq.~(\ref{metastable_FE_OP}),
allows us to extract $\Lambda$:
\begin{eqnarray}
\Lambda = \frac{1}{4} \ln \left( \frac{\Delta_{\rm FE}}{\Delta_0} \right)
= - 6.9 \times 10^{-2}.
\end{eqnarray}
Note that, unlike
inequality~(\ref{Lambda_range}),
restricting possible values of positive $\Lambda$, negative $\Lambda$ range
stretches from $-\infty$ to zero.

With
Eq.~(\ref{Jung_ratio})
in mind, one can approximate
\begin{eqnarray}
%%%%%%%%%%%%%%%%%%%%%%%%%%%%%%%%%%%%%%%%%%%%%%%%%%
\label{Lambda_Jung}
%%%%%%%%%%%%%%%%%%%%%%%%%%%%%%%%%%%%%%%%%%%%%%%%%%
|\Lambda|
\approx
\frac{{\cal E}_0}{\nu_0 ( \bar{\Gamma}_1 + \bar{\Gamma}_2)^2},
\end{eqnarray}
and further derive
\begin{eqnarray}
%%%%%%%%%%%%%%%%%%%%%%%%%%%%%%%%%%%%%%%%%%%%%%%%%%
\label{gamma_Jung}
%%%%%%%%%%%%%%%%%%%%%%%%%%%%%%%%%%%%%%%%%%%%%%%%%%
\gamma=(\bar{\Gamma}_1+\bar{\Gamma}_2)\nu_0 \approx 0.36,
\\
%%%%%%%%%%%%%%%%%%%%%%%%%%%%%%%%%%%%%%%%%%%%%%%%%%
\label{eps_Jung}
%%%%%%%%%%%%%%%%%%%%%%%%%%%%%%%%%%%%%%%%%%%%%%%%%%
\bar{\varepsilon} = \nu_0 {\cal E}_0 \approx \Lambda \gamma^2
\approx 8.9 \times 10^{-3}.
\end{eqnarray}
This value of $\gamma$ indicates that the Hartree-Fock model of~I is in the
weak-coupling limit.
Estimate~(\ref{gamma_Jung})
is of the same order as the value in our
Eq.~(\ref{Lambda_range}),
but they are not identical. Given quite dissimilar Coulomb interaction
models employed in I and in the present paper, certain disparity between
the two is very much expected.

Comparing~(\ref{eps_Jung})
with
Eq.~(\ref{epsilon_bar}),
one cannot help but notice that the values for
$\bar{\varepsilon}$
are very much apart: our model's estimate for
$\bar{\varepsilon}$
is almost two orders of magnitude larger than the value that follows from
the data in I. The reason for such a discrepancy is unclear.

If we use the familiar mean field expression for the order parameter
$\Delta_0 = W e^{-2/\gamma}$,
where $W$ is a kind of ultraviolet cutoff, we find
$W = 8.5$\,eV.
This pre-exponential factor is very large, much larger than the two-band
theory cutoff of
$2t_0$.
Thus, within the two-band model, such a value of $W$ cannot be justified.

Our
Eq.~(\ref{energy_linearized_approximation})
predicts that, when the zero-bias FE state is metastable and
$|\Lambda|$
is small, the energy difference
$\Delta E_{\rm FE} > 0$,
separating the FE state and the ground state, is
$\Delta E_{\rm FE} \approx 8 |\Lambda| E_{\rm cond}$,
where
$E_{\rm cond} = N_c \nu_0 \Delta_0^2$
is the condensation energy. Thus, one can write
$\Delta E_{\rm FE}/E_{\rm cond} \approx 8 |\Lambda| \approx 0.56$.
We can separately extract
$E_{\rm cond} \approx 5.17 \times 10^{-8}$\,eV
from table~II of paper~I, and
$\Delta E_{\rm FE} \approx 3 \times 10^{-8}$\,eV
from Fig.~5 of the same paper. Thus, numerical data allows us to calculate
$\Delta E_{\rm FE}/E_{\rm cond} \approx 0.58$,
in good agreement with the result derived from
Eq.~(\ref{energy_linearized_approximation}).

%\bibliographystyle{apsrevlong_no_issn_url}
%\bibliography{half_met_long}

\begin{thebibliography}{41}
\expandafter\ifx\csname natexlab\endcsname\relax\def\natexlab#1{#1}\fi
\expandafter\ifx\csname bibnamefont\endcsname\relax
  \def\bibnamefont#1{#1}\fi
\expandafter\ifx\csname bibfnamefont\endcsname\relax
  \def\bibfnamefont#1{#1}\fi
\expandafter\ifx\csname citenamefont\endcsname\relax
  \def\citenamefont#1{#1}\fi

\bibitem[{\citenamefont{Rozhkov et~al.}(2016)\citenamefont{Rozhkov, Sboychakov,
  Rakhmanov, and Nori}}]{bilayer_review2016}
\bibinfo{author}{\bibfnamefont{A.}~\bibnamefont{Rozhkov}},
  \bibinfo{author}{\bibfnamefont{A.}~\bibnamefont{Sboychakov}},
  \bibinfo{author}{\bibfnamefont{A.}~\bibnamefont{Rakhmanov}},
  \bibnamefont{and} \bibinfo{author}{\bibfnamefont{F.}~\bibnamefont{Nori}},
  {``}\bibinfo{title}{Electronic properties of graphene-based bilayer
  systems},{''} \bibinfo{journal}{Phys. Rep.} \textbf{\bibinfo{volume}{648}},
  \bibinfo{pages}{1 } (\bibinfo{year}{2016}).

\bibitem[{\citenamefont{Kotov et~al.}(2012)\citenamefont{Kotov, Uchoa, Pereira,
  Guinea, and Castro~Neto}}]{Kotov2012RevModPhys}
\bibinfo{author}{\bibfnamefont{V.~N.} \bibnamefont{Kotov}},
  \bibinfo{author}{\bibfnamefont{B.}~\bibnamefont{Uchoa}},
  \bibinfo{author}{\bibfnamefont{V.~M.} \bibnamefont{Pereira}},
  \bibinfo{author}{\bibfnamefont{F.}~\bibnamefont{Guinea}}, \bibnamefont{and}
  \bibinfo{author}{\bibfnamefont{A.~H.} \bibnamefont{Castro~Neto}},
  {``}\bibinfo{title}{Electron-Electron Interactions in Graphene: Current
  Status and Perspectives},{''} \bibinfo{journal}{Rev. Mod. Phys.}
  \textbf{\bibinfo{volume}{84}}, \bibinfo{pages}{1067} (\bibinfo{year}{2012}).

\bibitem[{\citenamefont{Bao et~al.}(2012)\citenamefont{Bao, Velasco, Zhang,
  Jing, Standley, Smirnov, Bockrath, MacDonald, and Lau}}]{Bao2012}
\bibinfo{author}{\bibfnamefont{W.}~\bibnamefont{Bao}},
  \bibinfo{author}{\bibfnamefont{J.}~\bibnamefont{Velasco}},
  \bibinfo{author}{\bibfnamefont{F.}~\bibnamefont{Zhang}},
  \bibinfo{author}{\bibfnamefont{L.}~\bibnamefont{Jing}},
  \bibinfo{author}{\bibfnamefont{B.}~\bibnamefont{Standley}},
  \bibinfo{author}{\bibfnamefont{D.}~\bibnamefont{Smirnov}},
  \bibinfo{author}{\bibfnamefont{M.}~\bibnamefont{Bockrath}},
  \bibinfo{author}{\bibfnamefont{A.~H.} \bibnamefont{MacDonald}},
  \bibnamefont{and} \bibinfo{author}{\bibfnamefont{C.~N.} \bibnamefont{Lau}},
  {``}\bibinfo{title}{Evidence for a spontaneous gapped state in ultraclean
  bilayer graphene},{''} \bibinfo{journal}{PNAS}
  \textbf{\bibinfo{volume}{109}}, \bibinfo{pages}{10802}
  (\bibinfo{year}{2012}).

\bibitem[{\citenamefont{Martin et~al.}(2010)\citenamefont{Martin, Feldman,
  Weitz, Allen, and Yacoby}}]{Martin2010}
\bibinfo{author}{\bibfnamefont{J.}~\bibnamefont{Martin}},
  \bibinfo{author}{\bibfnamefont{B.~E.} \bibnamefont{Feldman}},
  \bibinfo{author}{\bibfnamefont{R.~T.} \bibnamefont{Weitz}},
  \bibinfo{author}{\bibfnamefont{M.~T.} \bibnamefont{Allen}}, \bibnamefont{and}
  \bibinfo{author}{\bibfnamefont{A.}~\bibnamefont{Yacoby}},
  {``}\bibinfo{title}{Local Compressibility Measurements of Correlated States
  in Suspended Bilayer Graphene},{''} \bibinfo{journal}{Phys. Rev. Lett.}
  \textbf{\bibinfo{volume}{105}}, \bibinfo{pages}{256806}
  (\bibinfo{year}{2010}).

\bibitem[{\citenamefont{Weitz et~al.}(2010)\citenamefont{Weitz, Allen, Feldman,
  Martin, and Yacoby}}]{Weitz2010}
\bibinfo{author}{\bibfnamefont{R.~T.} \bibnamefont{Weitz}},
  \bibinfo{author}{\bibfnamefont{M.~T.} \bibnamefont{Allen}},
  \bibinfo{author}{\bibfnamefont{B.~E.} \bibnamefont{Feldman}},
  \bibinfo{author}{\bibfnamefont{J.}~\bibnamefont{Martin}}, \bibnamefont{and}
  \bibinfo{author}{\bibfnamefont{A.}~\bibnamefont{Yacoby}},
  {``}\bibinfo{title}{Broken-Symmetry States in Doubly Gated Suspended Bilayer
  Graphene},{''} \bibinfo{journal}{Science} \textbf{\bibinfo{volume}{330}},
  \bibinfo{pages}{812} (\bibinfo{year}{2010}).

\bibitem[{\citenamefont{Mayorov et~al.}(2011)\citenamefont{Mayorov, Elias,
  Mucha-Kruczynski, Gorbachev, Tudorovskiy, Zhukov, Morozov, Katsnelson,
  Fal'ko, Geim et~al.}}]{Mayorov2011}
\bibinfo{author}{\bibfnamefont{A.~S.} \bibnamefont{Mayorov}},
  \bibinfo{author}{\bibfnamefont{D.~C.} \bibnamefont{Elias}},
  \bibinfo{author}{\bibfnamefont{M.}~\bibnamefont{Mucha-Kruczynski}},
  \bibinfo{author}{\bibfnamefont{R.~V.} \bibnamefont{Gorbachev}},
  \bibinfo{author}{\bibfnamefont{T.}~\bibnamefont{Tudorovskiy}},
  \bibinfo{author}{\bibfnamefont{A.}~\bibnamefont{Zhukov}},
  \bibinfo{author}{\bibfnamefont{S.~V.} \bibnamefont{Morozov}},
  \bibinfo{author}{\bibfnamefont{M.~I.} \bibnamefont{Katsnelson}},
  \bibinfo{author}{\bibfnamefont{V.~I.} \bibnamefont{Fal'ko}},
  \bibinfo{author}{\bibfnamefont{A.~K.} \bibnamefont{Geim}},
  \bibnamefont{et~al.}, {``}\bibinfo{title}{Interaction-Driven Spectrum
  Reconstruction in Bilayer Graphene},{''} \bibinfo{journal}{Science}
  \textbf{\bibinfo{volume}{333}}, \bibinfo{pages}{860} (\bibinfo{year}{2011}).

\bibitem[{\citenamefont{Freitag
  et~al.}(2012{\natexlab{a}})\citenamefont{Freitag, Trbovic, Weiss, and
  Sch{\"o}nenberger}}]{Freitag2012}
\bibinfo{author}{\bibfnamefont{F.}~\bibnamefont{Freitag}},
  \bibinfo{author}{\bibfnamefont{J.}~\bibnamefont{Trbovic}},
  \bibinfo{author}{\bibfnamefont{M.}~\bibnamefont{Weiss}}, \bibnamefont{and}
  \bibinfo{author}{\bibfnamefont{C.}~\bibnamefont{Sch{\"o}nenberger}},
  {``}\bibinfo{title}{Spontaneously Gapped Ground State in Suspended Bilayer
  Graphene},{''} \bibinfo{journal}{Phys. Rev. Lett.}
  \textbf{\bibinfo{volume}{108}}, \bibinfo{pages}{076602}
  (\bibinfo{year}{2012}{\natexlab{a}}).

\bibitem[{\citenamefont{Freitag
  et~al.}(2012{\natexlab{b}})\citenamefont{Freitag, Weiss, Maurand, Trbovic,
  and Sch{\"o}nenberger}}]{Freitag20122053}
\bibinfo{author}{\bibfnamefont{F.}~\bibnamefont{Freitag}},
  \bibinfo{author}{\bibfnamefont{M.}~\bibnamefont{Weiss}},
  \bibinfo{author}{\bibfnamefont{R.}~\bibnamefont{Maurand}},
  \bibinfo{author}{\bibfnamefont{J.}~\bibnamefont{Trbovic}}, \bibnamefont{and}
  \bibinfo{author}{\bibfnamefont{C.}~\bibnamefont{Sch{\"o}nenberger}},
  {``}\bibinfo{title}{Homogeneity of bilayer graphene},{''}
  \bibinfo{journal}{Solid State Communications} \textbf{\bibinfo{volume}{152}},
  \bibinfo{pages}{2053 } (\bibinfo{year}{2012}{\natexlab{b}}).

\bibitem[{\citenamefont{Veligura et~al.}(2012)\citenamefont{Veligura, van
  Elferen, Tombros, Maan, Zeitler, and van Wees}}]{veligura2012}
\bibinfo{author}{\bibfnamefont{A.}~\bibnamefont{Veligura}},
  \bibinfo{author}{\bibfnamefont{H.~J.} \bibnamefont{van Elferen}},
  \bibinfo{author}{\bibfnamefont{N.}~\bibnamefont{Tombros}},
  \bibinfo{author}{\bibfnamefont{J.~C.} \bibnamefont{Maan}},
  \bibinfo{author}{\bibfnamefont{U.}~\bibnamefont{Zeitler}}, \bibnamefont{and}
  \bibinfo{author}{\bibfnamefont{B.~J.} \bibnamefont{van Wees}},
  {``}\bibinfo{title}{Transport gap in suspended bilayer graphene at zero
  magnetic field},{''} \bibinfo{journal}{Phys. Rev. B}
  \textbf{\bibinfo{volume}{85}}, \bibinfo{pages}{155412}
  (\bibinfo{year}{2012}).

\bibitem[{\citenamefont{Velasco~Jr. et~al.}(2012)\citenamefont{Velasco~Jr.,
  Jing, Bao, Lee, Kratz, Aji, Bockrath, Lau, Varma, Stillwell
  et~al.}}]{Velasco2012}
\bibinfo{author}{\bibfnamefont{J.}~\bibnamefont{Velasco~Jr.}},
  \bibinfo{author}{\bibfnamefont{L.}~\bibnamefont{Jing}},
  \bibinfo{author}{\bibfnamefont{W.}~\bibnamefont{Bao}},
  \bibinfo{author}{\bibfnamefont{Y.}~\bibnamefont{Lee}},
  \bibinfo{author}{\bibfnamefont{P.}~\bibnamefont{Kratz}},
  \bibinfo{author}{\bibfnamefont{V.}~\bibnamefont{Aji}},
  \bibinfo{author}{\bibfnamefont{M.}~\bibnamefont{Bockrath}},
  \bibinfo{author}{\bibfnamefont{C.}~\bibnamefont{Lau}},
  \bibinfo{author}{\bibfnamefont{C.}~\bibnamefont{Varma}},
  \bibinfo{author}{\bibfnamefont{R.}~\bibnamefont{Stillwell}},
  \bibnamefont{et~al.}, {``}\bibinfo{title}{Transport spectroscopy of
  symmetry-broken insulating states in bilayer graphene},{''}
  \bibinfo{journal}{Nat. Nanotechnol.} \textbf{\bibinfo{volume}{7}},
  \bibinfo{pages}{156} (\bibinfo{year}{2012}).

\bibitem[{\citenamefont{Freitag et~al.}(2013)\citenamefont{Freitag, Weiss,
  Maurand, Trbovic, and Sch\"onenberger}}]{freitag2013}
\bibinfo{author}{\bibfnamefont{F.}~\bibnamefont{Freitag}},
  \bibinfo{author}{\bibfnamefont{M.}~\bibnamefont{Weiss}},
  \bibinfo{author}{\bibfnamefont{R.}~\bibnamefont{Maurand}},
  \bibinfo{author}{\bibfnamefont{J.}~\bibnamefont{Trbovic}}, \bibnamefont{and}
  \bibinfo{author}{\bibfnamefont{C.}~\bibnamefont{Sch\"onenberger}},
  {``}\bibinfo{title}{Spin symmetry of the bilayer graphene ground state},{''}
  \bibinfo{journal}{Phys. Rev. B} \textbf{\bibinfo{volume}{87}},
  \bibinfo{pages}{161402} (\bibinfo{year}{2013}).

\bibitem[{\citenamefont{Zhou et~al.}(2021)\citenamefont{Zhou, Xie, Ghazaryan,
  Holder, Ehrets, Spanton, Taniguchi, Watanabe, Berg, Serbyn
  et~al.}}]{trilayer_quarter2021exper}
\bibinfo{author}{\bibfnamefont{H.}~\bibnamefont{Zhou}},
  \bibinfo{author}{\bibfnamefont{T.}~\bibnamefont{Xie}},
  \bibinfo{author}{\bibfnamefont{A.}~\bibnamefont{Ghazaryan}},
  \bibinfo{author}{\bibfnamefont{T.}~\bibnamefont{Holder}},
  \bibinfo{author}{\bibfnamefont{J.~R.} \bibnamefont{Ehrets}},
  \bibinfo{author}{\bibfnamefont{E.~M.} \bibnamefont{Spanton}},
  \bibinfo{author}{\bibfnamefont{T.}~\bibnamefont{Taniguchi}},
  \bibinfo{author}{\bibfnamefont{K.}~\bibnamefont{Watanabe}},
  \bibinfo{author}{\bibfnamefont{E.}~\bibnamefont{Berg}},
  \bibinfo{author}{\bibfnamefont{M.}~\bibnamefont{Serbyn}},
  \bibnamefont{et~al.}, {``}\bibinfo{title}{Half- and quarter-metals in
  rhombohedral trilayer graphene},{''} \bibinfo{journal}{Nature}
  \textbf{\bibinfo{volume}{598}}, \bibinfo{pages}{429} (\bibinfo{year}{2021}).

\bibitem[{\citenamefont{de~la Barrera et~al.}(2022)\citenamefont{de~la Barrera,
  Sergio, Aronson, Zheng, Watanabe, Taniguchi, Ma, Jarillo-Herrero, and
  Ashoori}}]{ab_frac2022exper}
\bibinfo{author}{\bibnamefont{de~la Barrera}},
  \bibinfo{author}{\bibfnamefont{C.}~\bibnamefont{Sergio}},
  \bibinfo{author}{\bibfnamefont{S.}~\bibnamefont{Aronson}},
  \bibinfo{author}{\bibfnamefont{Z.}~\bibnamefont{Zheng}},
  \bibinfo{author}{\bibfnamefont{K.}~\bibnamefont{Watanabe}},
  \bibinfo{author}{\bibfnamefont{T.}~\bibnamefont{Taniguchi}},
  \bibinfo{author}{\bibfnamefont{Q.}~\bibnamefont{Ma}},
  \bibinfo{author}{\bibfnamefont{P.}~\bibnamefont{Jarillo-Herrero}},
  \bibnamefont{and} \bibinfo{author}{\bibfnamefont{R.}~\bibnamefont{Ashoori}},
  {``}\bibinfo{title}{Cascade of isospin phase transitions in Bernal-stacked
  bilayer graphene at zero magnetic field},{''} \bibinfo{journal}{Nat. Phys.}
  \textbf{\bibinfo{volume}{18}}, \bibinfo{pages}{771} (\bibinfo{year}{2022}).

\bibitem[{\citenamefont{Seiler et~al.}(2022)\citenamefont{Seiler, Geisenhof,
  Winterer, Watanabe, Taniguchi, Xu, Zhang, and Weitz}}]{Seiler2022}
\bibinfo{author}{\bibfnamefont{A.~M.} \bibnamefont{Seiler}},
  \bibinfo{author}{\bibfnamefont{F.~R.} \bibnamefont{Geisenhof}},
  \bibinfo{author}{\bibfnamefont{F.}~\bibnamefont{Winterer}},
  \bibinfo{author}{\bibfnamefont{K.}~\bibnamefont{Watanabe}},
  \bibinfo{author}{\bibfnamefont{T.}~\bibnamefont{Taniguchi}},
  \bibinfo{author}{\bibfnamefont{T.}~\bibnamefont{Xu}},
  \bibinfo{author}{\bibfnamefont{F.}~\bibnamefont{Zhang}}, \bibnamefont{and}
  \bibinfo{author}{\bibfnamefont{R.~T.} \bibnamefont{Weitz}},
  {``}\bibinfo{title}{Quantum cascade of correlated phases in trigonally warped
  bilayer graphene},{''} \bibinfo{journal}{Nature}
  \textbf{\bibinfo{volume}{608}}, \bibinfo{pages}{298} (\bibinfo{year}{2022}).

\bibitem[{\citenamefont{Zhou et~al.}(2022)\citenamefont{Zhou, Holleis, Saito,
  Cohen, Huynh, Patterson, Yang, Taniguchi, Watanabe, and
  Young}}]{zhou2022isospin}
\bibinfo{author}{\bibfnamefont{H.}~\bibnamefont{Zhou}},
  \bibinfo{author}{\bibfnamefont{L.}~\bibnamefont{Holleis}},
  \bibinfo{author}{\bibfnamefont{Y.}~\bibnamefont{Saito}},
  \bibinfo{author}{\bibfnamefont{L.}~\bibnamefont{Cohen}},
  \bibinfo{author}{\bibfnamefont{W.}~\bibnamefont{Huynh}},
  \bibinfo{author}{\bibfnamefont{C.~L.} \bibnamefont{Patterson}},
  \bibinfo{author}{\bibfnamefont{F.}~\bibnamefont{Yang}},
  \bibinfo{author}{\bibfnamefont{T.}~\bibnamefont{Taniguchi}},
  \bibinfo{author}{\bibfnamefont{K.}~\bibnamefont{Watanabe}}, \bibnamefont{and}
  \bibinfo{author}{\bibfnamefont{A.~F.} \bibnamefont{Young}},
  {``}\bibinfo{title}{Isospin magnetism and spin-polarized superconductivity in
  Bernal bilayer graphene},{''} \bibinfo{journal}{Science}
  \textbf{\bibinfo{volume}{375}}, \bibinfo{pages}{774} (\bibinfo{year}{2022}).

\bibitem[{\citenamefont{McCann et~al.}(2007)\citenamefont{McCann, Abergel, and
  Fal'ko}}]{MCCANN2007110}
\bibinfo{author}{\bibfnamefont{E.}~\bibnamefont{McCann}},
  \bibinfo{author}{\bibfnamefont{D.~S.} \bibnamefont{Abergel}},
  \bibnamefont{and} \bibinfo{author}{\bibfnamefont{V.~I.}
  \bibnamefont{Fal'ko}}, {``}\bibinfo{title}{Electrons in bilayer
  graphene},{''} \bibinfo{journal}{Solid State Commun.}
  \textbf{\bibinfo{volume}{143}}, \bibinfo{pages}{110} (\bibinfo{year}{2007}),
  \bibinfo{note}{exploring graphene}.

\bibitem[{\citenamefont{Nandkishore and
  Levitov}(2010{\natexlab{a}})}]{Nandkishore2010}
\bibinfo{author}{\bibfnamefont{R.}~\bibnamefont{Nandkishore}} \bibnamefont{and}
  \bibinfo{author}{\bibfnamefont{L.}~\bibnamefont{Levitov}},
  {``}\bibinfo{title}{Dynamical Screening and Excitonic Instability in Bilayer
  Graphene},{''} \bibinfo{journal}{Phys. Rev. Lett.}
  \textbf{\bibinfo{volume}{104}}, \bibinfo{pages}{156803}
  (\bibinfo{year}{2010}{\natexlab{a}}).

\bibitem[{\citenamefont{Nandkishore and
  Levitov}(2010{\natexlab{b}})}]{Nandkishore2010b}
\bibinfo{author}{\bibfnamefont{R.}~\bibnamefont{Nandkishore}} \bibnamefont{and}
  \bibinfo{author}{\bibfnamefont{L.}~\bibnamefont{Levitov}},
  {``}\bibinfo{title}{Quantum anomalous Hall state in bilayer graphene},{''}
  \bibinfo{journal}{Phys. Rev. B} \textbf{\bibinfo{volume}{82}},
  \bibinfo{pages}{115124} (\bibinfo{year}{2010}{\natexlab{b}}).

\bibitem[{\citenamefont{Vafek}(2010)}]{vafek_rg2010}
\bibinfo{author}{\bibfnamefont{O.}~\bibnamefont{Vafek}},
  {``}\bibinfo{title}{Interacting fermions on the honeycomb bilayer: From weak
  to strong coupling},{''} \bibinfo{journal}{Phys. Rev. B}
  \textbf{\bibinfo{volume}{82}}, \bibinfo{pages}{205106}
  (\bibinfo{year}{2010}).

\bibitem[{\citenamefont{Vafek and Yang}(2010)}]{vafek_nemat_rg2010}
\bibinfo{author}{\bibfnamefont{O.}~\bibnamefont{Vafek}} \bibnamefont{and}
  \bibinfo{author}{\bibfnamefont{K.}~\bibnamefont{Yang}},
  {``}\bibinfo{title}{Many-body instability of Coulomb interacting bilayer
  graphene: Renormalization group approach},{''} \bibinfo{journal}{Phys. Rev.
  B} \textbf{\bibinfo{volume}{81}}, \bibinfo{pages}{041401}
  (\bibinfo{year}{2010}).

\bibitem[{\citenamefont{Lemonik et~al.}(2010)\citenamefont{Lemonik, Aleiner,
  Toke, and Fal'ko}}]{Lemonik2010}
\bibinfo{author}{\bibfnamefont{Y.}~\bibnamefont{Lemonik}},
  \bibinfo{author}{\bibfnamefont{I.~L.} \bibnamefont{Aleiner}},
  \bibinfo{author}{\bibfnamefont{C.}~\bibnamefont{Toke}}, \bibnamefont{and}
  \bibinfo{author}{\bibfnamefont{V.~I.} \bibnamefont{Fal'ko}},
  {``}\bibinfo{title}{Spontaneous symmetry breaking and Lifshitz transition in
  bilayer graphene},{''} \bibinfo{journal}{Phys. Rev. B}
  \textbf{\bibinfo{volume}{82}}, \bibinfo{pages}{201408}
  (\bibinfo{year}{2010}).

\bibitem[{\citenamefont{Jung et~al.}(2011)\citenamefont{Jung, Zhang, and
  MacDonald}}]{Jung2011}
\bibinfo{author}{\bibfnamefont{J.}~\bibnamefont{Jung}},
  \bibinfo{author}{\bibfnamefont{F.}~\bibnamefont{Zhang}}, \bibnamefont{and}
  \bibinfo{author}{\bibfnamefont{A.~H.} \bibnamefont{MacDonald}},
  {``}\bibinfo{title}{Lattice theory of pseudospin ferromagnetism in bilayer
  graphene: Competing interaction-induced quantum Hall states},{''}
  \bibinfo{journal}{Phys. Rev. B} \textbf{\bibinfo{volume}{83}},
  \bibinfo{pages}{115408} (\bibinfo{year}{2011}).

\bibitem[{\citenamefont{Cvetkovic et~al.}(2012)\citenamefont{Cvetkovic,
  Throckmorton, and Vafek}}]{cvetkovic_multi2012}
\bibinfo{author}{\bibfnamefont{V.}~\bibnamefont{Cvetkovic}},
  \bibinfo{author}{\bibfnamefont{R.~E.} \bibnamefont{Throckmorton}},
  \bibnamefont{and} \bibinfo{author}{\bibfnamefont{O.}~\bibnamefont{Vafek}},
  {``}\bibinfo{title}{Electronic multicriticality in bilayer graphene},{''}
  \bibinfo{journal}{Phys. Rev. B} \textbf{\bibinfo{volume}{86}},
  \bibinfo{pages}{075467} (\bibinfo{year}{2012}).

\bibitem[{\citenamefont{Rakhmanov et~al.}(2012)\citenamefont{Rakhmanov,
  Rozhkov, Sboychakov, and Nori}}]{aa_graph_prl2012}
\bibinfo{author}{\bibfnamefont{A.~L.} \bibnamefont{Rakhmanov}},
  \bibinfo{author}{\bibfnamefont{A.~V.} \bibnamefont{Rozhkov}},
  \bibinfo{author}{\bibfnamefont{A.~O.} \bibnamefont{Sboychakov}},
  \bibnamefont{and} \bibinfo{author}{\bibfnamefont{F.}~\bibnamefont{Nori}},
  {``}\bibinfo{title}{Instabilities of the $AA$-Stacked Graphene Bilayer},{''}
  \bibinfo{journal}{Phys. Rev. Lett.} \textbf{\bibinfo{volume}{109}},
  \bibinfo{pages}{206801} (\bibinfo{year}{2012}).

\bibitem[{\citenamefont{Kharitonov}(2012)}]{haritonov_afm2012}
\bibinfo{author}{\bibfnamefont{M.}~\bibnamefont{Kharitonov}},
  {``}\bibinfo{title}{Antiferromagnetic state in bilayer graphene},{''}
  \bibinfo{journal}{Phys. Rev. B} \textbf{\bibinfo{volume}{86}},
  \bibinfo{pages}{195435} (\bibinfo{year}{2012}).

\bibitem[{\citenamefont{Yuan et~al.}(2013)\citenamefont{Yuan, Xu, Wang, Zhou,
  Gao, and Zhang}}]{bilayer_half-metal2013numeric_MF}
\bibinfo{author}{\bibfnamefont{J.}~\bibnamefont{Yuan}},
  \bibinfo{author}{\bibfnamefont{D.-H.} \bibnamefont{Xu}},
  \bibinfo{author}{\bibfnamefont{H.}~\bibnamefont{Wang}},
  \bibinfo{author}{\bibfnamefont{Y.}~\bibnamefont{Zhou}},
  \bibinfo{author}{\bibfnamefont{J.-H.} \bibnamefont{Gao}}, \bibnamefont{and}
  \bibinfo{author}{\bibfnamefont{F.-C.} \bibnamefont{Zhang}},
  {``}\bibinfo{title}{Possible half-metallic phase in bilayer graphene:
  Calculations based on mean-field theory applied to a two-layer Hubbard
  model},{''} \bibinfo{journal}{Phys. Rev. B} \textbf{\bibinfo{volume}{88}},
  \bibinfo{pages}{201109} (\bibinfo{year}{2013}).

\bibitem[{\citenamefont{Baima et~al.}(2018)\citenamefont{Baima, Mauri, and
  Calandra}}]{baima2018dft_half_met_graphene}
\bibinfo{author}{\bibfnamefont{J.}~\bibnamefont{Baima}},
  \bibinfo{author}{\bibfnamefont{F.}~\bibnamefont{Mauri}}, \bibnamefont{and}
  \bibinfo{author}{\bibfnamefont{M.}~\bibnamefont{Calandra}},
  {``}\bibinfo{title}{Field-effect-driven half-metallic multilayer
  graphene},{''} \bibinfo{journal}{Phys. Rev. B} \textbf{\bibinfo{volume}{98}},
  \bibinfo{pages}{075418} (\bibinfo{year}{2018}).

\bibitem[{\citenamefont{Brey and Fertig}(2013)}]{aa_graph_BreyFertig2013}
\bibinfo{author}{\bibfnamefont{L.}~\bibnamefont{Brey}} \bibnamefont{and}
  \bibinfo{author}{\bibfnamefont{H.~A.} \bibnamefont{Fertig}},
  {``}\bibinfo{title}{Gapped phase in {$AA$}-stacked bilayer graphene},{''}
  \bibinfo{journal}{Phys. Rev. B} \textbf{\bibinfo{volume}{87}},
  \bibinfo{pages}{115411} (\bibinfo{year}{2013}).

\bibitem[{\citenamefont{Sboychakov et~al.}(2013)\citenamefont{Sboychakov,
  Rakhmanov, Rozhkov, and Nori}}]{sboychakov2013AA}
\bibinfo{author}{\bibfnamefont{A.~O.} \bibnamefont{Sboychakov}},
  \bibinfo{author}{\bibfnamefont{A.~L.} \bibnamefont{Rakhmanov}},
  \bibinfo{author}{\bibfnamefont{A.~V.} \bibnamefont{Rozhkov}},
  \bibnamefont{and} \bibinfo{author}{\bibfnamefont{F.}~\bibnamefont{Nori}},
  {``}\bibinfo{title}{Metal-insulator transition and phase separation in doped
  $AA$-stacked graphene bilayer},{''} \bibinfo{journal}{Phys. Rev. B}
  \textbf{\bibinfo{volume}{87}}, \bibinfo{pages}{121401}
  (\bibinfo{year}{2013}).

\bibitem[{\citenamefont{Sboychakov et~al.}(2021)\citenamefont{Sboychakov,
  Rakhmanov, Rozhkov, and Nori}}]{aa_quarter_met2021}
\bibinfo{author}{\bibfnamefont{A.~O.} \bibnamefont{Sboychakov}},
  \bibinfo{author}{\bibfnamefont{A.~L.} \bibnamefont{Rakhmanov}},
  \bibinfo{author}{\bibfnamefont{A.~V.} \bibnamefont{Rozhkov}},
  \bibnamefont{and} \bibinfo{author}{\bibfnamefont{F.}~\bibnamefont{Nori}},
  {``}\bibinfo{title}{Bilayer graphene can become a fractional metal},{''}
  \bibinfo{journal}{Phys. Rev. B} \textbf{\bibinfo{volume}{103}},
  \bibinfo{pages}{L081106} (\bibinfo{year}{2021}).

\bibitem[{\citenamefont{Geisenhof et~al.}(2022)\citenamefont{Geisenhof,
  Winterer, Seiler, Lenz, Zhang, and Weitz}}]{Geisenhof2022}
\bibinfo{author}{\bibfnamefont{F.~R.} \bibnamefont{Geisenhof}},
  \bibinfo{author}{\bibfnamefont{F.}~\bibnamefont{Winterer}},
  \bibinfo{author}{\bibfnamefont{A.~M.} \bibnamefont{Seiler}},
  \bibinfo{author}{\bibfnamefont{J.}~\bibnamefont{Lenz}},
  \bibinfo{author}{\bibfnamefont{F.}~\bibnamefont{Zhang}}, \bibnamefont{and}
  \bibinfo{author}{\bibfnamefont{R.~T.} \bibnamefont{Weitz}},
  {``}\bibinfo{title}{Impact of Electric Field Disorder on Broken-Symmetry
  States in Ultraclean Bilayer Graphene},{''} \bibinfo{journal}{Nano Lett.}
  \textbf{\bibinfo{volume}{22}}, \bibinfo{pages}{7378} (\bibinfo{year}{2022}).

\bibitem[{\citenamefont{Sboychakov et~al.}(2023)\citenamefont{Sboychakov,
  Rozhkov, and Rakhmanov}}]{ab_supercond2023sboychakov}
\bibinfo{author}{\bibfnamefont{A.~O.} \bibnamefont{Sboychakov}},
  \bibinfo{author}{\bibfnamefont{A.~V.} \bibnamefont{Rozhkov}},
  \bibnamefont{and} \bibinfo{author}{\bibfnamefont{A.~L.}
  \bibnamefont{Rakhmanov}}, {``}\bibinfo{title}{Triplet superconductivity and
  spin density wave in biased AB bilayer graphene},{''} \bibinfo{journal}{Phys.
  Rev. B} \textbf{\bibinfo{volume}{108}}, \bibinfo{pages}{184503}
  (\bibinfo{year}{2023}).

\bibitem[{\citenamefont{Lemonik et~al.}(2012)\citenamefont{Lemonik, Aleiner,
  and Fal'ko}}]{lemonic_rg_nemat_long2012}
\bibinfo{author}{\bibfnamefont{Y.}~\bibnamefont{Lemonik}},
  \bibinfo{author}{\bibfnamefont{I.}~\bibnamefont{Aleiner}}, \bibnamefont{and}
  \bibinfo{author}{\bibfnamefont{V.~I.} \bibnamefont{Fal'ko}},
  {``}\bibinfo{title}{Competing nematic, antiferromagnetic, and spin-flux
  orders in the ground state of bilayer graphene},{''} \bibinfo{journal}{Phys.
  Rev. B} \textbf{\bibinfo{volume}{85}}, \bibinfo{pages}{245451}
  (\bibinfo{year}{2012}).

\bibitem[{\citenamefont{Min et~al.}(2008)\citenamefont{Min, Borghi, Polini, and
  MacDonald}}]{min_pseudo_fm2008}
\bibinfo{author}{\bibfnamefont{H.}~\bibnamefont{Min}},
  \bibinfo{author}{\bibfnamefont{G.}~\bibnamefont{Borghi}},
  \bibinfo{author}{\bibfnamefont{M.}~\bibnamefont{Polini}}, \bibnamefont{and}
  \bibinfo{author}{\bibfnamefont{A.~H.} \bibnamefont{MacDonald}},
  {``}\bibinfo{title}{Pseudospin magnetism in graphene},{''}
  \bibinfo{journal}{Phys. Rev. B} \textbf{\bibinfo{volume}{77}},
  \bibinfo{pages}{041407} (\bibinfo{year}{2008}).

\bibitem[{\citenamefont{Rozhkov et~al.}(2025)\citenamefont{Rozhkov, Sboychakov,
  and Rakhmanov}}]{rozhkov2025ab_su4}
\bibinfo{author}{\bibfnamefont{A.~V.} \bibnamefont{Rozhkov}},
  \bibinfo{author}{\bibfnamefont{A.~O.} \bibnamefont{Sboychakov}},
  \bibnamefont{and} \bibinfo{author}{\bibfnamefont{A.~L.}
  \bibnamefont{Rakhmanov}}, {``}\bibinfo{title}{Ordered states in AB bilayer
  graphene in a SU(4)-symmetric model},{''} \bibinfo{journal}{Phys. Rev. B}
  \textbf{\bibinfo{volume}{111}}, \bibinfo{pages}{205421}
  (\bibinfo{year}{2025}).

\bibitem[{\citenamefont{Rozhkov et~al.}(2023)\citenamefont{Rozhkov, Sboychakov,
  and Rakhmanov}}]{rozhkov2023aa_su4}
\bibinfo{author}{\bibfnamefont{A.~V.} \bibnamefont{Rozhkov}},
  \bibinfo{author}{\bibfnamefont{A.~O.} \bibnamefont{Sboychakov}},
  \bibnamefont{and} \bibinfo{author}{\bibfnamefont{A.~L.}
  \bibnamefont{Rakhmanov}}, {``}\bibinfo{title}{Ordering in the SU(4)-symmetric
  model of AA bilayer graphene},{''} \bibinfo{journal}{Phys. Rev. B}
  \textbf{\bibinfo{volume}{108}}, \bibinfo{pages}{205153}
  (\bibinfo{year}{2023}).

\bibitem[{\citenamefont{Zhang et~al.}(2011)\citenamefont{Zhang, Jung, Fiete,
  Niu, and MacDonald}}]{zhand2011hall_state_classific}
\bibinfo{author}{\bibfnamefont{F.}~\bibnamefont{Zhang}},
  \bibinfo{author}{\bibfnamefont{J.}~\bibnamefont{Jung}},
  \bibinfo{author}{\bibfnamefont{G.~A.} \bibnamefont{Fiete}},
  \bibinfo{author}{\bibfnamefont{Q.}~\bibnamefont{Niu}}, \bibnamefont{and}
  \bibinfo{author}{\bibfnamefont{A.~H.} \bibnamefont{MacDonald}},
  {``}\bibinfo{title}{Spontaneous Quantum Hall States in Chirally Stacked
  Few-Layer Graphene Systems},{''} \bibinfo{journal}{Phys. Rev. Lett.}
  \textbf{\bibinfo{volume}{106}}, \bibinfo{pages}{156801}
  (\bibinfo{year}{2011}).

\bibitem[{\citenamefont{Rakhmanov et~al.}(2023)\citenamefont{Rakhmanov,
  Rozhkov, Sboychakov, and Nori}}]{rakhmanov2023ab_FraM}
\bibinfo{author}{\bibfnamefont{A.~L.} \bibnamefont{Rakhmanov}},
  \bibinfo{author}{\bibfnamefont{A.~V.} \bibnamefont{Rozhkov}},
  \bibinfo{author}{\bibfnamefont{A.~O.} \bibnamefont{Sboychakov}},
  \bibnamefont{and} \bibinfo{author}{\bibfnamefont{F.}~\bibnamefont{Nori}},
  {``}\bibinfo{title}{Half-metal and other fractional metal phases in doped
  $AB$ bilayer graphene},{''} \bibinfo{journal}{Phys. Rev. B}
  \textbf{\bibinfo{volume}{107}}, \bibinfo{pages}{155112}
  (\bibinfo{year}{2023}).

\bibitem[{\citenamefont{Keldysh and Kopaev}(1965)}]{keldysh1965half_metal}
\bibinfo{author}{\bibfnamefont{L.~V.} \bibnamefont{Keldysh}} \bibnamefont{and}
  \bibinfo{author}{\bibfnamefont{Y.~V.} \bibnamefont{Kopaev}},
  {``}\bibinfo{title}{Possible instability of the semimetallic state toward
  Coulomb interaction},{''} \bibinfo{journal}{Sov. Phys. Solid State}
  \textbf{\bibinfo{volume}{6}}, \bibinfo{pages}{2219} (\bibinfo{year}{1965}).

\bibitem[{\citenamefont{Keldysh and
  Kopaev}(2024)}]{keldysh_kopaev_collection2024half_metal}
\bibinfo{author}{\bibfnamefont{L.~V.} \bibnamefont{Keldysh}} \bibnamefont{and}
  \bibinfo{author}{\bibfnamefont{Y.~V.} \bibnamefont{Kopaev}}, in
  \emph{\bibinfo{booktitle}{Selected Papers of Leonid V Keldysh}}
  (\bibinfo{publisher}{World Scientific}, \bibinfo{year}{2024}), pp.
  \bibinfo{pages}{41--46}.

\bibitem[{\citenamefont{Rodionov et~al.}(2025)\citenamefont{Rodionov, Rozhkov,
  Beck, Sboychakov, Kugel, and Rakhmanov}}]{rodionov2025half_metal}
\bibinfo{author}{\bibfnamefont{Y.~I.} \bibnamefont{Rodionov}},
  \bibinfo{author}{\bibfnamefont{A.~V.} \bibnamefont{Rozhkov}},
  \bibinfo{author}{\bibfnamefont{M.~E.~S.} \bibnamefont{Beck}},
  \bibinfo{author}{\bibfnamefont{A.~O.} \bibnamefont{Sboychakov}},
  \bibinfo{author}{\bibfnamefont{K.~I.} \bibnamefont{Kugel}}, \bibnamefont{and}
  \bibinfo{author}{\bibfnamefont{A.~L.} \bibnamefont{Rakhmanov}},
  {``}\bibinfo{title}{Nesting-driven ferromagnetism of itinerant
  electrons},{''} \bibinfo{journal}{Phys. Rev. B}
  \textbf{\bibinfo{volume}{112}}, \bibinfo{pages}{205129}
  (\bibinfo{year}{2025}).

\end{thebibliography}

\end{document}